\documentclass[a4paper,11pt]{article}
\pdfoutput=1 

\usepackage{jinstpub} 
\usepackage{lineno}
\usepackage{xcolor}
\usepackage[colorinlistoftodos,prependcaption]{todonotes}

\usepackage{xspace}
\newcommand{\mude}{{\sc Mu3e}\xspace}

\newcommand{\mueee}{$\mu^+ \rightarrow e^+ e^- e^+$\xspace}
\newcommand{\TiO}{${\rm TiO}_2$\xspace}
\usepackage{dashrule}

\title{\boldmath Development of the Scintillating Fiber Timing Detector for the \mude Experiment}

\author[a]{A.~Bravar,}
\author[b]{A.~Buonaura,}
\author[c]{S. Corrodi,}
\author[a]{A.~Damyanova,}
\author[a]{Y.~Demets,}
\author[c]{L.~Gerritzen,}
\author[c]{Ch.~Grab,}
\author[c]{C.~Martin~Perez,}
\author[d]{and A.~Papa }

\affiliation[a]{D\'{e}partement de physique nucl\'eaire et corpusculaire,  Universit\'e de Gen\`eve,\\
24, quai Ernest-Ansermet, 1211 Gen\`eve 4, Switzerland}
\affiliation[b]{Physik-Institut, Universit\"at Z\"urich,\\
Winterthurerstrasse 190, 8057 Zurich, Switzerland}
\affiliation[c]{Institute for Particle Physics and Astrophysics, Eidgen\"ossische Technische Hochschule Z\"urich,\\
Otto-Stern-Weg 5, 8093 Z\"urich, Switzerland}
\affiliation[d]{Laboratory for Particle Physics, Paul Scherrer Institut,\\
Forschungsstrasse 111, 5232 Villigen, Switzerland}

\emailAdd{alessandro.bravar@unige.ch}

\date{\today}

\abstract{
We present and discuss the development and performance of a compact scintillating fiber (SciFi) detector
for timing to be used in the \mude experiment at very high particle rates.
The SciFi detector is read out with multichannel silicon photomuiltipliers (SiPM) arrays at both ends
to achieve the best timing performance.
\mude is a new experiment under preparation at the Paul Scherrer Institute
to search for charged Lepton Flavor Violation
in the rare neutrinoless muon decay \mueee
using the most intense continuous surface muon beam in the world.
The \mude detector is based on thin high-voltage monolithic active silicon pixel sensors (HV-MAPS)
for very precise tracking in conjunction
with scintillating fibers and scintillating tiles coupled to SiPMs for accurate timing measurements
and it is designed to operate at very high intensities.

In order to reach a single event sensitivity of $10^{-16}$ for this rare \mueee muon decay,
all backgrounds must be rejected well below this level.
To suppress all forms of accidental background,
a very thin SciFi detector (thickness $< 0.2\%$ of a radiation length $X_0$) 
with a time resolution of 250~ps, efficiency in excess of 96\%,
and spatial resolution of $\sim 100~\mu{\rm m}$ has been developed.
In this paper we report on the development, construction, and performance of this SciFi detector.
Different scintillating fiber types have been evaluated and various assembly procedures have been tested
to achieve the best performance.

The compact size, fast response, good timing, high spatial resolution, insensitivity to magnetic fields,
and adaptable geometry
make SciFi detectors suitable for a variety of applications.
}

\keywords{Timing detectors; Scintillators, scintillation and light emission processes (solid, gas and liquid scintillators);
Detector design and construction technologies and materials}

\begin{document}
\maketitle
\flushbottom

\newpage

\setcounter{page}{1}
\section{Introduction}
\label{sec:intro}

The \mude experiment~\cite{Blondel} will search for the rare charged Lepton Flavor Violating neutrinoless muon decay
$\mu^+ \rightarrow e^+e^-e^+$
with the ultimate goal to find or exclude this process, if it occurs more than once in $10^{16}$ muon decays.
Since this process is heavily suppressed in the Standard Model of Particle Physics with a branching ratio
smaller than $10^{-54}$,
any observation above this level will be a sign for new physics.
\mude has the potential of probing for new physics at the PeV scale.

\begin{figure}[t!]
   \centering
   \includegraphics[width=0.90\textwidth]{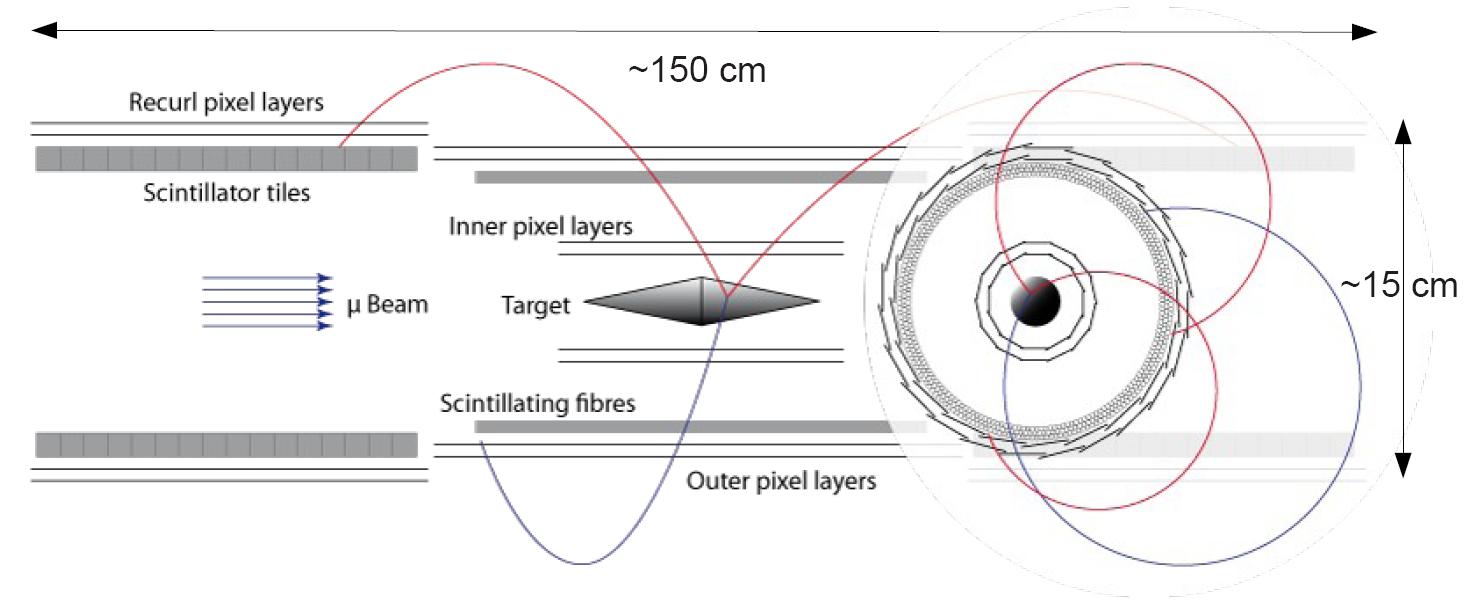}
   \caption{Schematic view of the \mude apparatus in the side and transverse projections (not to scale)~\cite{TDR}.
The phase I experiment consists of a central part and recurl stations upstream and downstream.
The central part comprises, from inside out, a hollow double cone muon stopping target,
a silicon pixel inner double layer, the scintillating fiber, and a second silicon pixel outer double layer sub-detector.
The detector is inserted in a 2~m long solenoidal magnet with a 1~T field.}
   \label{fig:experiment}
\end{figure}

The \mude experiment is in preparation at the Paul Scherrer Institute in Villingen
using the most intense continuous surface muon beam in the world (presently $1 \times 10^8~{\rm muons}/s$).
A detailed description of the \mude apparatus can be found in the Technical Design of the experiment~\cite{TDR}.
The \mude detector is based on
$50~\mu{\rm m}$ thick high-voltage monolithic active silicon pixel sensors (HV-MAPS)~\cite{HV-MAPS}
for very precise tracking in conjunction
with scintillating fibers (SciFi) and scintillating tiles coupled to silicon photomultipliers (SiPMs) for accurate timing measurements.
The detector is designed to operate at very high rates in excess of $10^8$ muon decays per second.
Positive muons are stopped in a $5~\mu{\rm m}$ thick hollow double cone target, where they decay.
The detector is inserted into a 2~m long 1~T solenoidal homogeneous magnetic field.
Figure \ref{fig:experiment} illustrates the cylindrical arrangement of the sub-detectors around the target
parallel to the magnetic field and to the beam.

To suppress all forms of combinatorial background
two sub-detector systems based on plastic scintillators provide complementary precise time measurements
for decay positrons and electrons.
The superposition of positrons and electrons can mimic the signature of the \mueee decay.
Electrons are produced for instance in the muon radiative decays
$\mu^+ \rightarrow e^+e^-e^+{\bar \nu_\mu}\nu_e$ or by Bhabba scattering of positrons in the target.
In order to achieve the required momentum resolution of $\Delta p_{eee} < 0.5~{\rm MeV}/c$ multiple Coulomb scattering,
and consequently the material budget of the detectors, has to be minimized.
A very thin SciFi detector (thickness $< 0.2\%$ of a radiation length $X_0$) 
with a time resolution of 250~ps, efficiency in excess of 96\%,
and spatial resolution of $\sim 100~\mu{\rm m}$
has been developed.
The SciFi detector is read out at both ends with multichannel SiPM arrays~\cite{SiPM},
which will detect $\mathcal{O}(15)$ photons per {\it minimum ionizing particle} (MIP) crossing. 
In addition to timing, the SciFi detector will help to reduce ambiguities in the track reconstruction
and will reliably determine the sense of rotation (i.e. the charge)
of the recurling tracks in the central region of the apparatus
using the time of flight between consecutive crossings of the SciFi detector.
Exploiting the SciFi detector alone
with a time resolution of 250~ps and a 90\% signal efficiency working point
(all three tracks from the \mueee decay are measured)
provides a survival fraction for the background  due to Bhabha electron/positron-pairs plus one Michel positron
of  $\mathcal{O}{(3 \cdot 10^{-2})}$,
well below the required sensitivity~\cite{TDR}.

The SciFi detector is cylindrical in shape,
with a radius of 61~mm and a length of 300~mm (270~mm in the acceptance region).
It is composed of 12 SciFi ribbons 300~mm long and 32.5~mm wide.
In the \mude baseline design, the ribbons are formed by staggering 3 layers of $250~\mu{\rm m}$ diameter round double-clad
scintillating fibers (type SCSF-78MJ from Kuraray~\cite{Kuraray})
to assure continuous coverage and high detection efficiency,
while minimizing the thickness of the detector.
The width of the fiber ribbons matches the size of the multichannel SiPM photo-sensors.
Incidentally, the width of the SiPM photo-sensors determines also the radius of the
detector.\footnote{The radius of a circle inscribed in a regular dodecagon with side 32.5~mm,
i.e. the width of the photo-sensor, is indeed 61~mm.}
The SciFi detector is located 5~mm below the outer double-layer silicon pixel detector (see Figure~\ref{fig:experiment}).
The space constraints in the central part of the Mu3e apparatus impose a very compact design.
Moreover, the fiber ribbons are longitudinally staggered by 10~mm because of the mechanical constraints
and to minimize gaps between the SciFi ribbons.
The SciFi arrays are coupled at both ends to multichannel SiPM arrays and will be read out
with a dedicated mixed mode ASIC, the MuTRiG~\cite{MuTRiG}.

The compact size, fast response, good timing, high spatial resolution, insensitivity to magnetic fields,
and adaptable geometry make these detectors suitable for a variety of applications.

In the following we present the assembly of the SciFi ribbons and their physical properties,
and the photo-sensors used for detecting the scintillation light from the fibers.
Then we discuss the performance of the detector like the light yield, achievable time resolution, efficiency, etc.
Different scintillating fiber materials and assembly procedures (like the number of staggered SciFi layers)
have been evaluated to achieve the best performance compatible with the \mude requirements.

\clearpage
\section{The Scintillating Fiber Ribbons}
\label{sec:ribbon}

\begin{figure}[!t]
   \centering
   \includegraphics[width=0.60\textwidth]{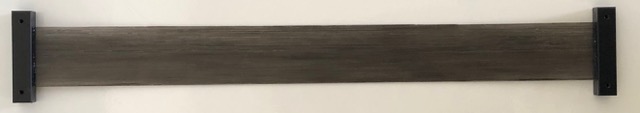}
   \hspace*{5mm}
   \includegraphics[width=0.30\textwidth]{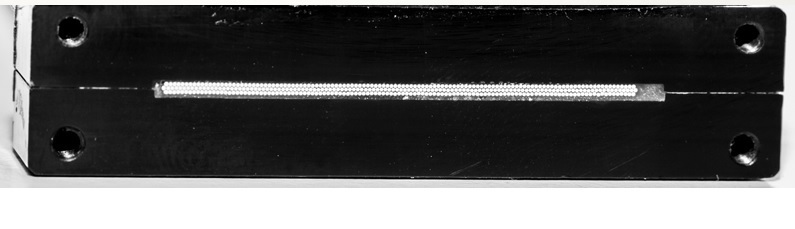}
   \caption{Full size SciFi ribbon prototype.
The ribbon is formed by staggering 3 layers of $250~\mu{\rm m}$ diameter round scintillating fibers.
The fibers are assembled with black epoxy, which gives the black color to the ribbons.
right) Socket for coupling the SciFi ribbons to the photo-sensors. The fibers are diamond polished.}
   \label{fig:scifi}
\end{figure}

In the \mude baseline design,
the SciFi ribbons are formed by staggering three layers of $250~\mu{\rm m}$ diameter round
high purity double-clad fibers from Kuraray~\cite{Kuraray}, type SCSF-78MJ.
The length of the ribbons is 300~mm with a width of 32.5~mm (there are 128 fibers in the bottom layer).
Polytec EP 601-Black epoxy is used for the assembly of the SciFi ribbons.
This two component, low viscosity, black-colored adhesive
was chosen for its handling properties.
Using a \TiO loaded adhesive is not a viable option due to the high $Z$ of titanium,
which would generate an unacceptably large amount of multiple scattering in the ribbon.
Figure~\ref{fig:scifi} shows a full size SciFi ribbon
with the end pieces glued to the fibers to couple the ribbon to the photo-sensor.

\begin{table}[b!]
   \centering
   \begin{tabular}{l l}
      \hline
      characteristic                            &  value \\
      \hline
      emission peak [nm]                   & 450   \\
      trapping efficiency [\%]             & 5.4   \\
      numerical aperture                    & 0.74  \\
      cladding thickness [\% of fiber radius]  & 3 / 3   \\
      decay time [ns]                         & 2.8    \\
      attenuation length [m]               & >~4.0  \\
      core material                            & Polystyrene (PS) \\
      inner cladding                          & Acrylic (PMMA)   \\
      outer cladding                          & Fluor-acrylic (FP) \\
      refractive index (inside out)        & 1.59~/~1.49~/~1.42  \\
      density [g/cm$^3$] (inside out)  & 1.05~/~1.19~/~1.43  \\
      \hline
   \end{tabular}
   \caption{Properties of the Kuraray's $250~\mu{\rm m}$ diameter high purity round double-clad SCSF-78MJ scintillating fibers
as quoted by the manufacturer~\cite{Kuraray}.}
   \label{tab:FibresScintillators}
\end{table}

The constraints on the material budget, the occupancy, and position resolution
require the use of the thinnest available scintillating fibers. 
Different blue-emitting $250~\mu{\rm m}$ diameter round double-clad scintillating fibers
(Kuraray fibers SCSF-78 and SCSF-81~\cite{Kuraray}, and NOL-11~\cite{NOL11}, and Bicron fibers BCF-12)
\footnote{The timing characteristics of these blue-emitting scintillating fibers are discussed in~\cite{fibers}.}
have been studied intensively in test beams (see Section~\ref{sec:setup}).
Fiber ribbons consisting of two to six staggered layers have been evaluated
using also different types of adhesives for the assembly. 
The detailed results of these studies are presented in the following Sections.
In this work only results for the high yield fibers SCSF-78 and NOL-11 are presented.
The performance of low light yield fibers (SCSF-81 and BCF-12) is worse. 
Based on their performance with respect to light yield and time resolution,
round double-clad SCSF-78MJ fibers from Kuraray were chosen.
Table~\ref{tab:FibresScintillators} summarizes the characteristics of SCSF-78MJ  scintillating fibers as quoted by the manufacturer~\cite{Kuraray}.
Novel NOL fibers, based on Nanostructured Organosilicon Luminophores~\cite{NOL11}, 
which have very short decay times ($\sim 1~{\rm ns}$),
give the best timing performance.
These fibers, however, were not available in sufficiently large quantities at the time of the construction of the SciFi detector,
but will be considered for future SciFi detector upgrades.

\subsection{Making Of the Fiber Ribbons}
\label{sec:ribbons:production}

\begin{figure}[!t]
   \centering
   \includegraphics[width=0.56\textwidth]{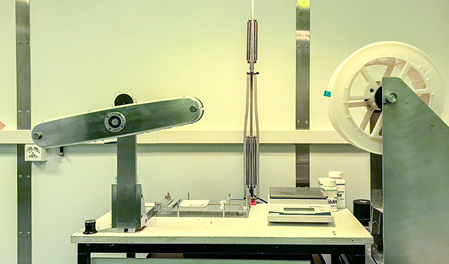}
   \hfill
   \includegraphics[width=0.42\textwidth]{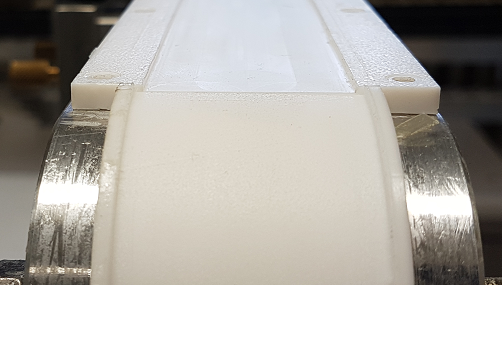}
   \caption{Winding tool for the preparation of the $32.5~{\rm mm} \times 300~{\rm mm}$ SciFi ribbons
(the fiber is not visible in the photo).
Up to 6 SciFi layers can be staggered using this tool.
A detail of the U-channel is shown on the right.}
   \label{fig:tool}
\end{figure}

\begin{figure}[b!]
   \centering
   \includegraphics[width=0.49\textwidth]{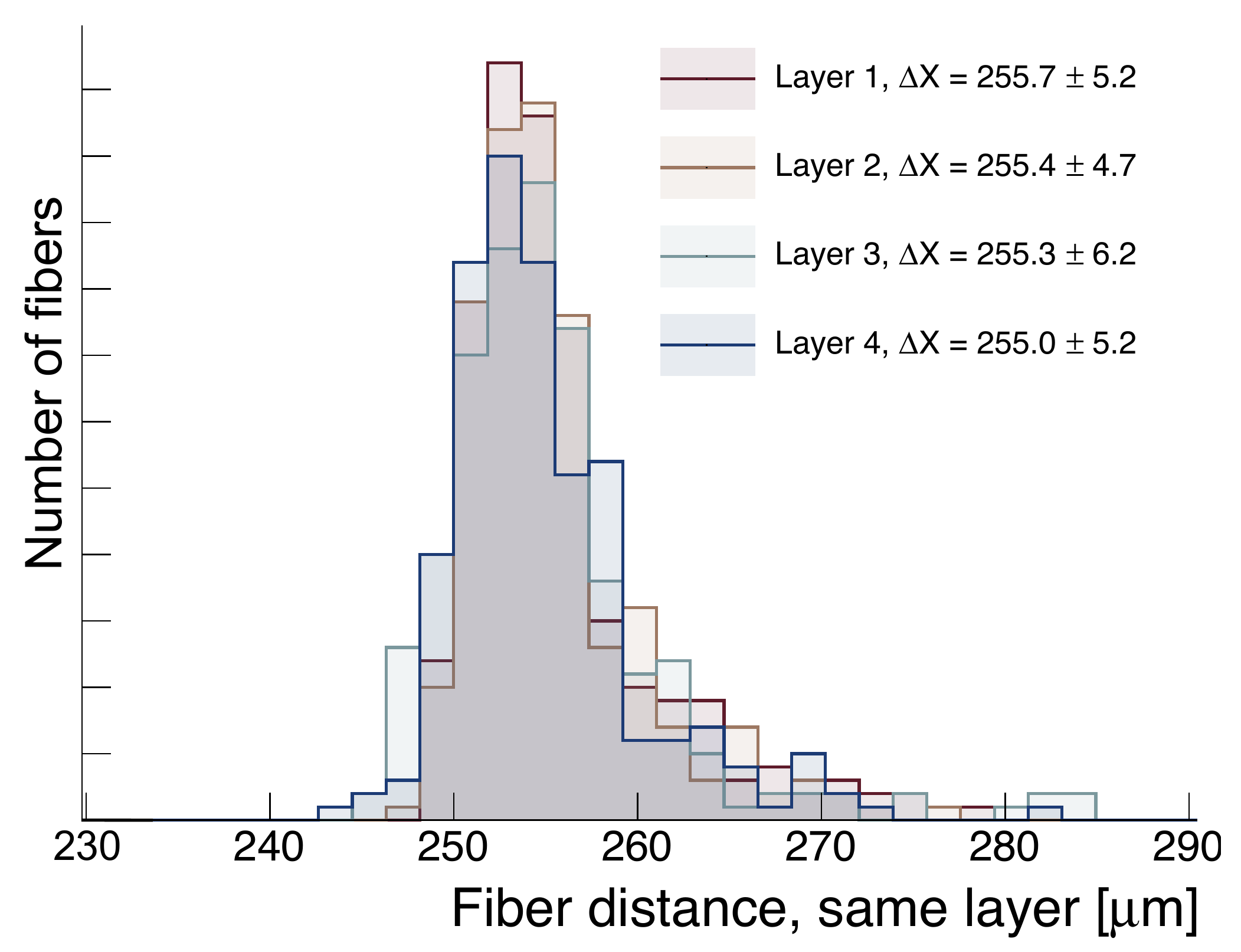}
   \includegraphics[width=0.49\textwidth]{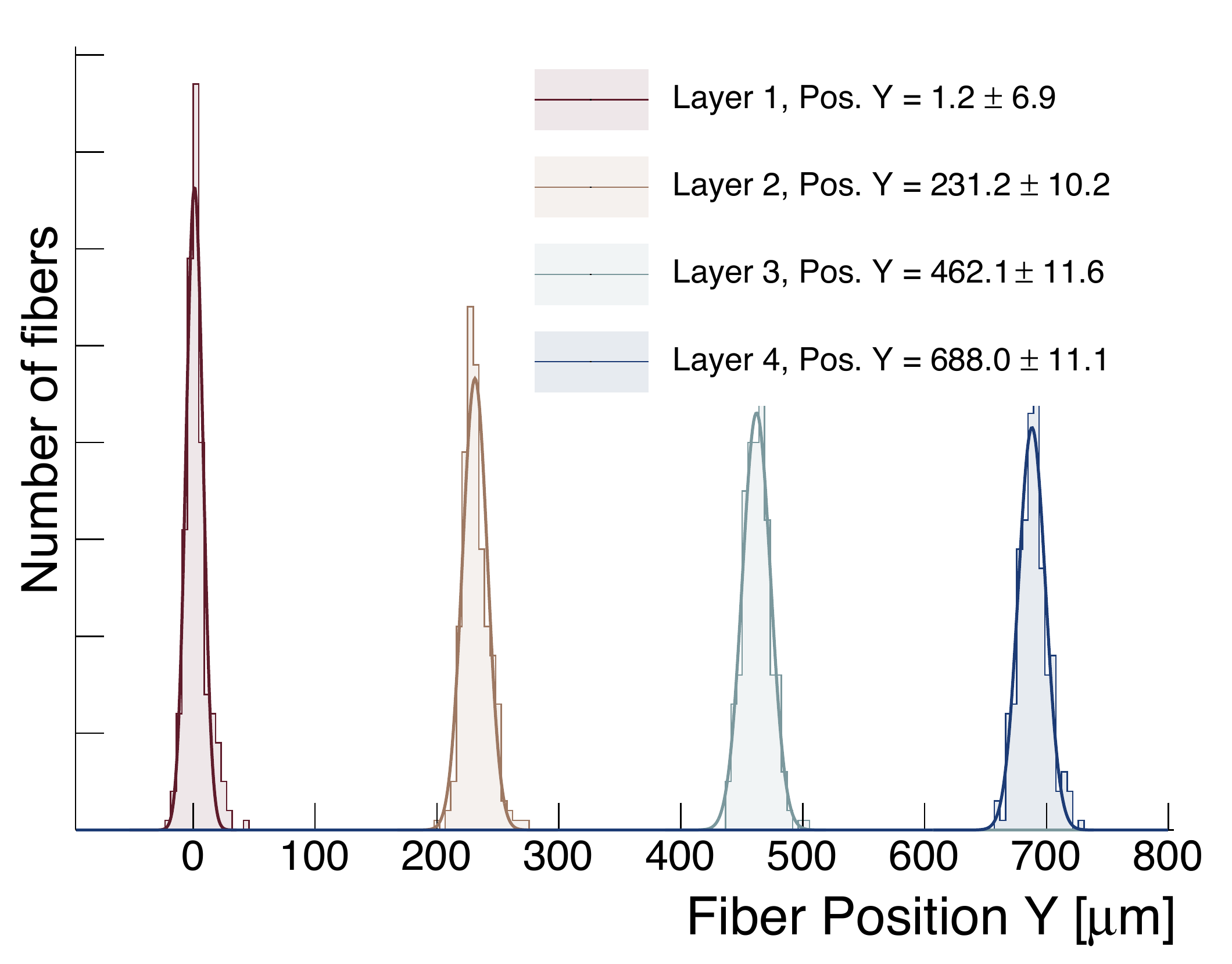}
   \caption{Metrology of a four-layer SciFi ribbon:
left) the average distance between neighboring fibers in the same layers is around $255~\mu{\rm m}$,
right) while the vertical distance between the layers is around $230~\mu{\rm m}$.}
   \label{fig:ribbonGeom}
\end{figure}

With the help of the winding tool depicted in Figure~\ref{fig:tool} (left), the fibers are assembled into ribbons consisting of two to six staggered fiber layers.
The ribbons are formed layer-by-layer.
First, a U-channel shown in Figure~\ref{fig:tool} (right) made in teflon is tightly filled with a first layer of fibers.
The depth of the U-channel is 0.5~mm.
To ease the extraction of the assembled ribbons the U-channel is coated with a thin layer of wax-based spray Trennspray P6.
The tight filling of the U-channel with fibers assures a good alignment of the fibers, better than $10~\mu{\rm m}$ (see Figure~\ref{fig:ribbonGeom}),
which is essential for good tracking resolution.
No template to position the fibers is used.
Then a thin layer of glue is applied.
The amount of glue is minimized to reduce the amount of non-active materials.
Before the glue completely dries, a second SciFi layer is added.
The bottom layer helps to stagger the second layer and determines the overall alignment of individual fibers in the ribbon.
Iterating the procedure, up to 6 layers have been staggered.
Two ribbons are manufactured at the same time.
To minimize edge effects, mainly due to the low amount of glue applied during the manufacturing process,
the assembled ribbons are slightly wider than their final width.
After removing the ribbons from the winding tool,
the superfluous edge fibers are peeled off, down to the design width of 32.5~mm.
Each ribbon is then placed in a jig and the end pieces are glued at each ribbon's end (see Figure~\ref{sec:ribbon}).
Finally the ribbons are cut to the nominal length of 300~mm and diamond polished.

\begin{figure}[t!]
   \centering
   \includegraphics[width=0.5\textwidth]{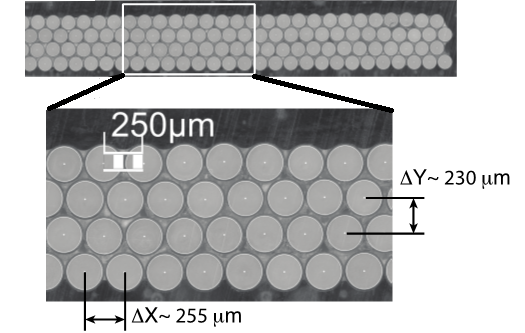}
   \caption{Diamond polished cut view of a 4-layer SciFi ribbon.
The cladding makes the fibers appear smaller in diameter.
The centers of the fibers (white dots) are identified with a pattern recognition algorithm
and their position is determined.}
   \label{fig:ribbonFront}
\end{figure}

Three different types of adhesive have been used to test the performance of the ribbons (see following Sections).
A first set of ribbons has been prepared with {\sc POLYTEC EP 601} epoxy,
which is a two component, clear, low viscosity adhesive.
The epoxy is cured at room temperature.
A second set of ribbons has been prepared by admixing ${\rm TiO}_2$ powder, 20\% by weight,
to the clear epoxy.
While the use of ${\rm TiO}_2$ loaded adhesive is the standard approach when assembling SciFi ribbons,
the material budget constraints in \mude do not allow to use high Z materials.
The last and final set of ribbons have been prepared using {\sc POLYTEC EP 601-Black} adhesive,
which is also a two component, black-colored, low viscosity adhesive.
For the final assembly of the \mude fiber ribbons, the latter is used.
The highest reduction of the optical cross-talk between the fibers has been observed with this black-colored adhesive
(Section~\ref{sec:cluster}).

\subsection{Fiber Alignment}

Figure~\ref{fig:ribbonFront} shows the cross section of a four-layer fiber ribbon prototype.
The transverse position of each fiber in the SciFi ribbon has been determined with a pattern recognition algorithm applied to this cut view. 
The fibers in a layer are spaced by $255~\mu{\rm m}$ center to center
with very good uniformity throughout the ribbon as it can be assessed from Figure~\ref{fig:ribbonGeom} (left).
The $255~\mu{\rm m}$ {\it pitch} is due to non-uniformities in the fiber's cross section,
which compensate when aligning several fibers in the same layer.
The uniform separation between the layers is $\sim 230~\mu{\rm m}$
as shown in Figure~\ref{fig:ribbonGeom} (right),
which gives an overall thickness of about $710~\mu{\rm m}$ for a three-layer ribbon.
This thickness, including the glue, corresponds to less than 0.2\% of a radiation length $X_0$.
The fill factor of the fiber ribbon is of 82\%,
which is slightly smaller compared to the tightest possible configuration of 87\%,
when the fibers are in direct contact with each other.

\begin{figure}[b!]
   \centering
   \includegraphics[width=0.49\textwidth]{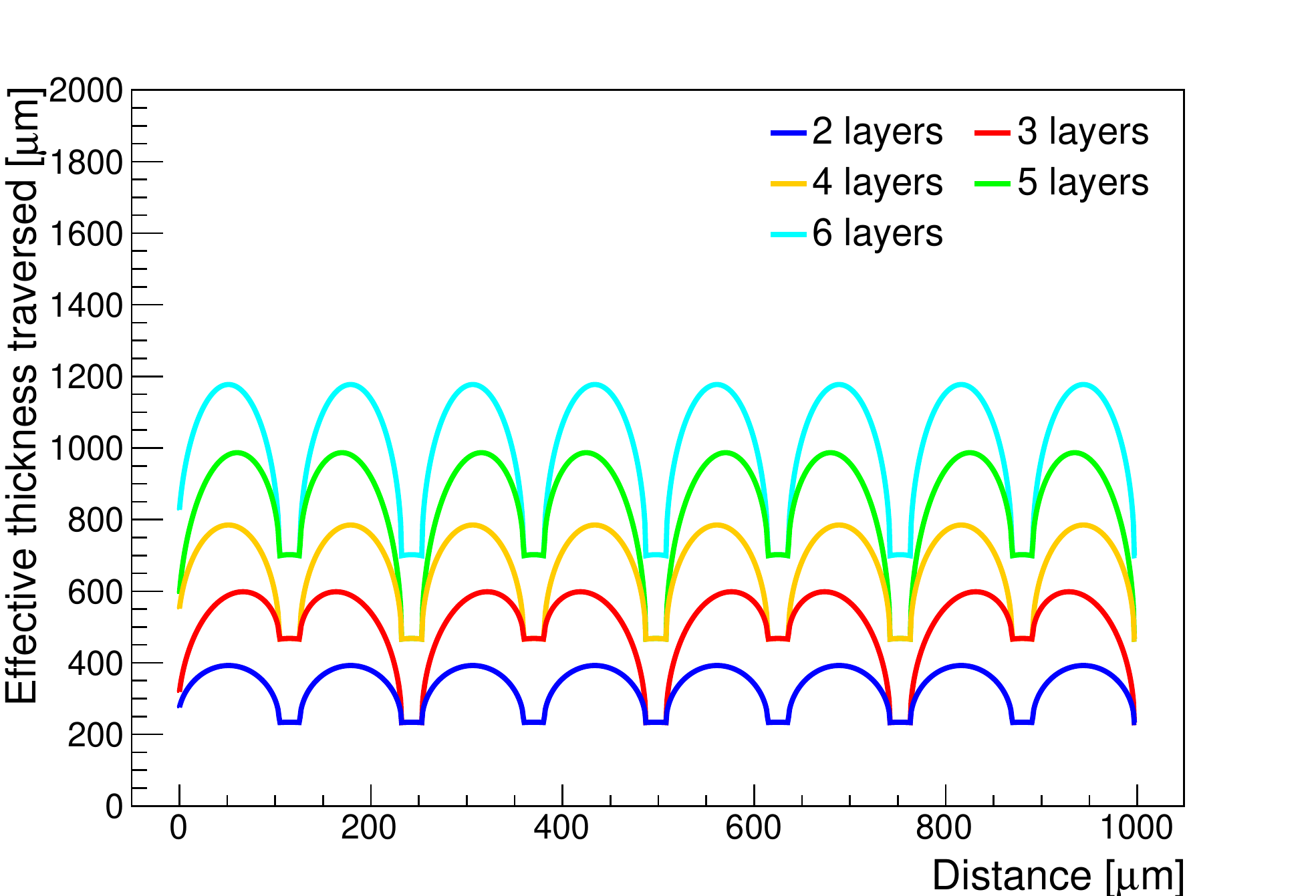}
   \includegraphics[width=0.49\textwidth]{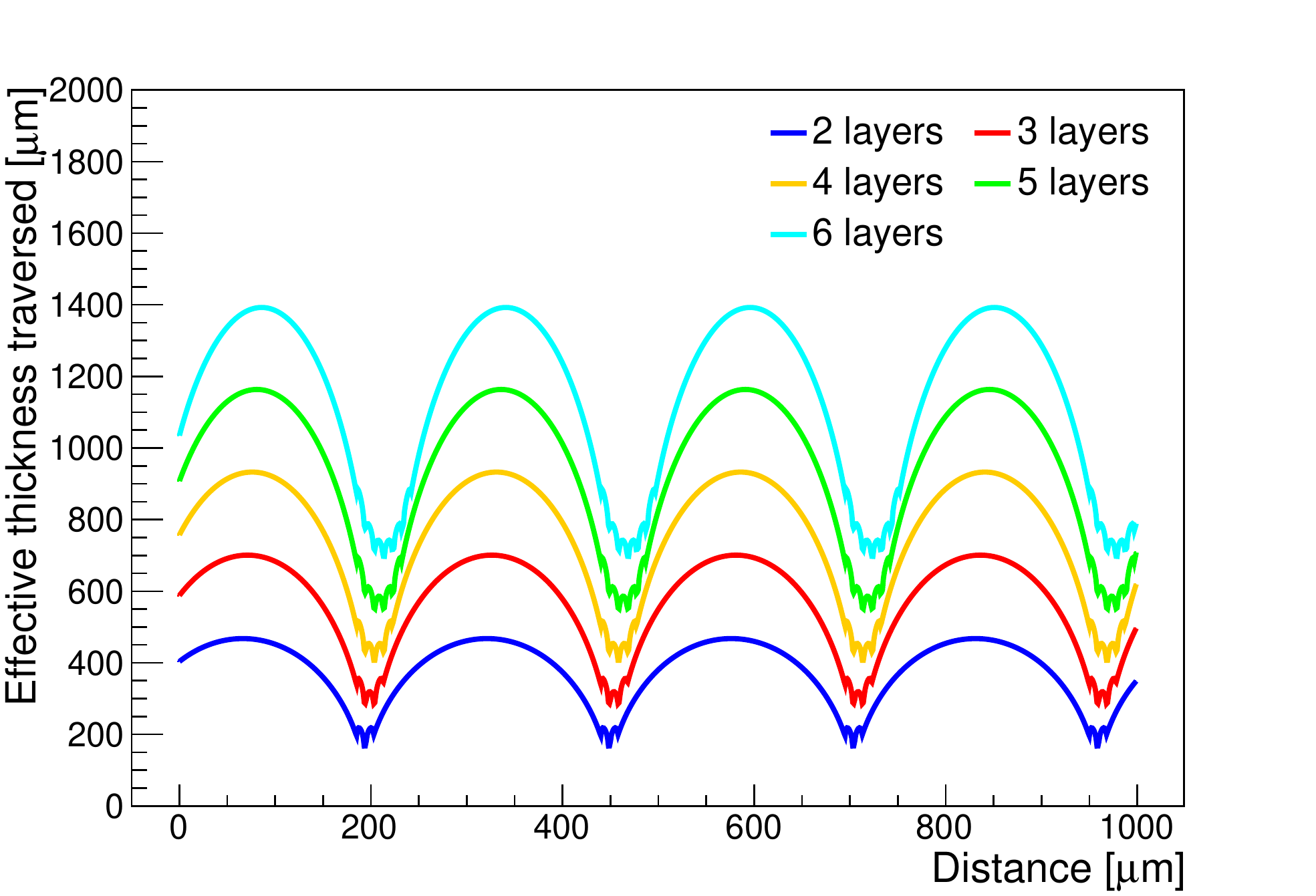}
   \caption{Effective thickness of the active scintillating fiber material traversed by particles impinging on the SciFi ribbon
at and angle of $0^\circ$ (left) and $27^\circ$ (right)
as a function of the impact position across the ribbon
for different numbers  of staggered fiber layers.
An effective diameter of $230~\mu{\rm m}$ for the scintillating fibers has been assumed excluding the cladding.}
   \label{fig:LYcrossing}
\end{figure}

The effective thickness of the active fiber scintillating material traversed by particles
impinging on the SciFi ribbon at an angle $\vartheta$ of $0^\circ$ and $27^\circ$ w.r.t. the normal to the ribbon
is shown in Figure~\ref{fig:LYcrossing} as a function of the impact position across the SciFi ribbon
for different numbers of staggered fiber layers.
An effective diameter of $230~\mu{\rm m}$ for the $250~\mu{\rm m}$ scintillating fibers has been assumed,
which excludes the cladding.
The fibers are positioned according to Figure~\ref{fig:ribbonGeom}.
In \mude, the positrons from muons decaying in the stopping target will cross the SciFi ribbons at an average angle $\vartheta = 27^\circ$
because of the bending in the magnetic field.
Since the scintillation light yield is proportional to the thickness of the traversed active material,
the number of detected photons will depend on the crossing point across the SciFi ribbon.
For a $0^\circ$ crossing angle
the average active thickness is $510~\mu{\rm m}$ for a 3-layer ribbon
and $680~\mu{\rm m}$ for a 4-layer ribbon.
The average active thickness increases with the crossing angle.
A {\it resonance} effect, emphasizing the structure of the SciFi ribbon and shown in Figure~\ref{fig:LYcrossing} (right),
occurs around $\vartheta = 30^\circ$
(incidentally this angle is close to the average crossing angle in \mude),
because the fibers are aligned on an axis forming an angle $\vartheta = 31^\circ$,
as it can be deduced from Figure~\ref{fig:ribbonFront}.
For a $27^\circ$ crossing angle the average active thickness is $580~\mu{\rm m}$ for a 3-layer ribbon
and $780~\mu{\rm m}$ for a 4-layer ribbon.
To minimize the multiple Coulomb scattering in the ribbons we had to compromise between uniformity and thickness,
and 3-layer SciFi ribbons were preferred.
Moreover, the width of the clusters for 3-layer ribbons is smaller, especially at non-zero crossing angles (see Section~\ref{sec:cluster}),
which reduces the occupancy of the detector.

\subsection{Thermal Expansion and Sagging}
\label{sec:scifiSag}

\begin{figure}[h!]
   \centering
   \includegraphics[width=0.49\textwidth]{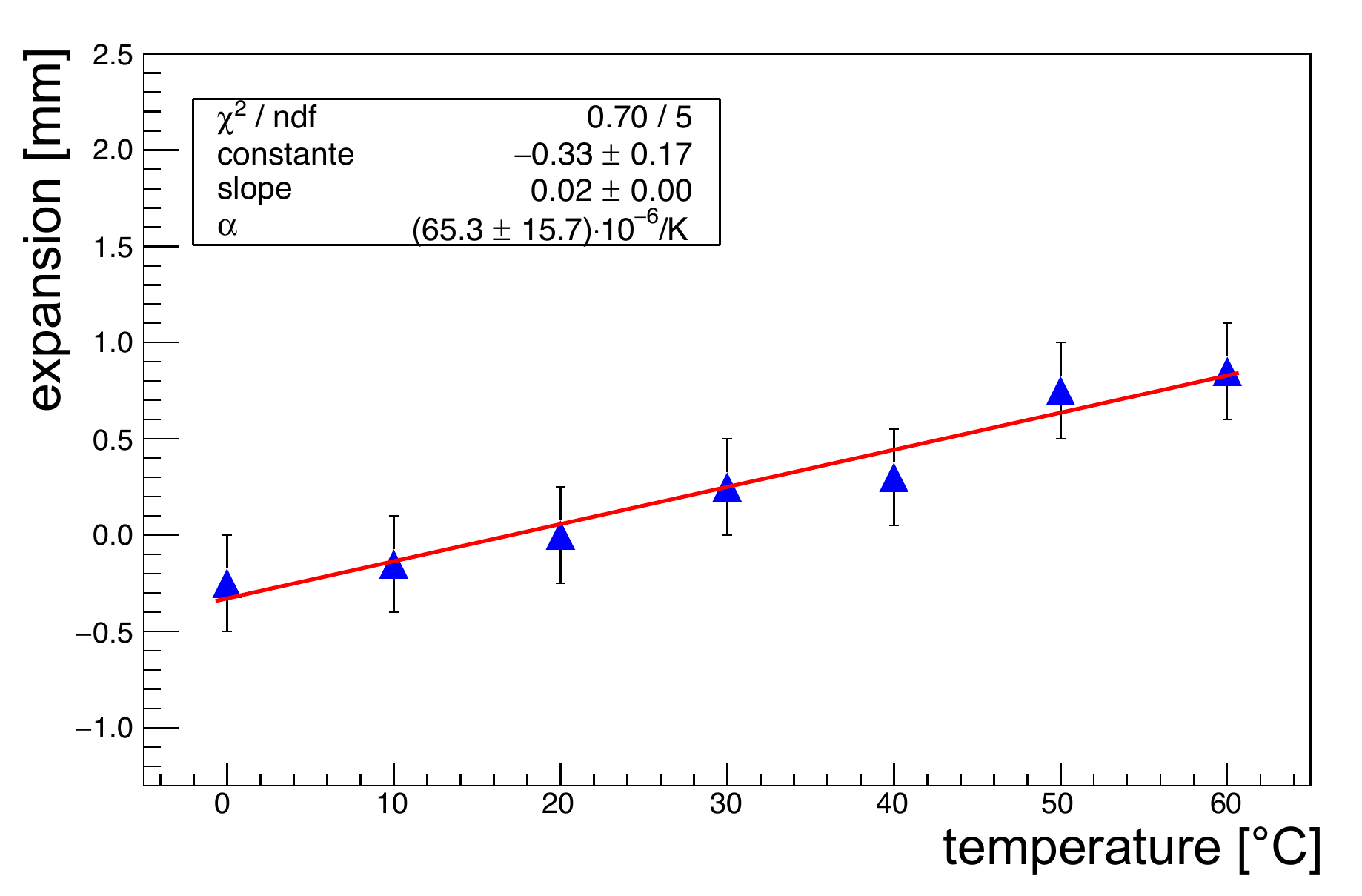}
   \includegraphics[width=0.49\textwidth]{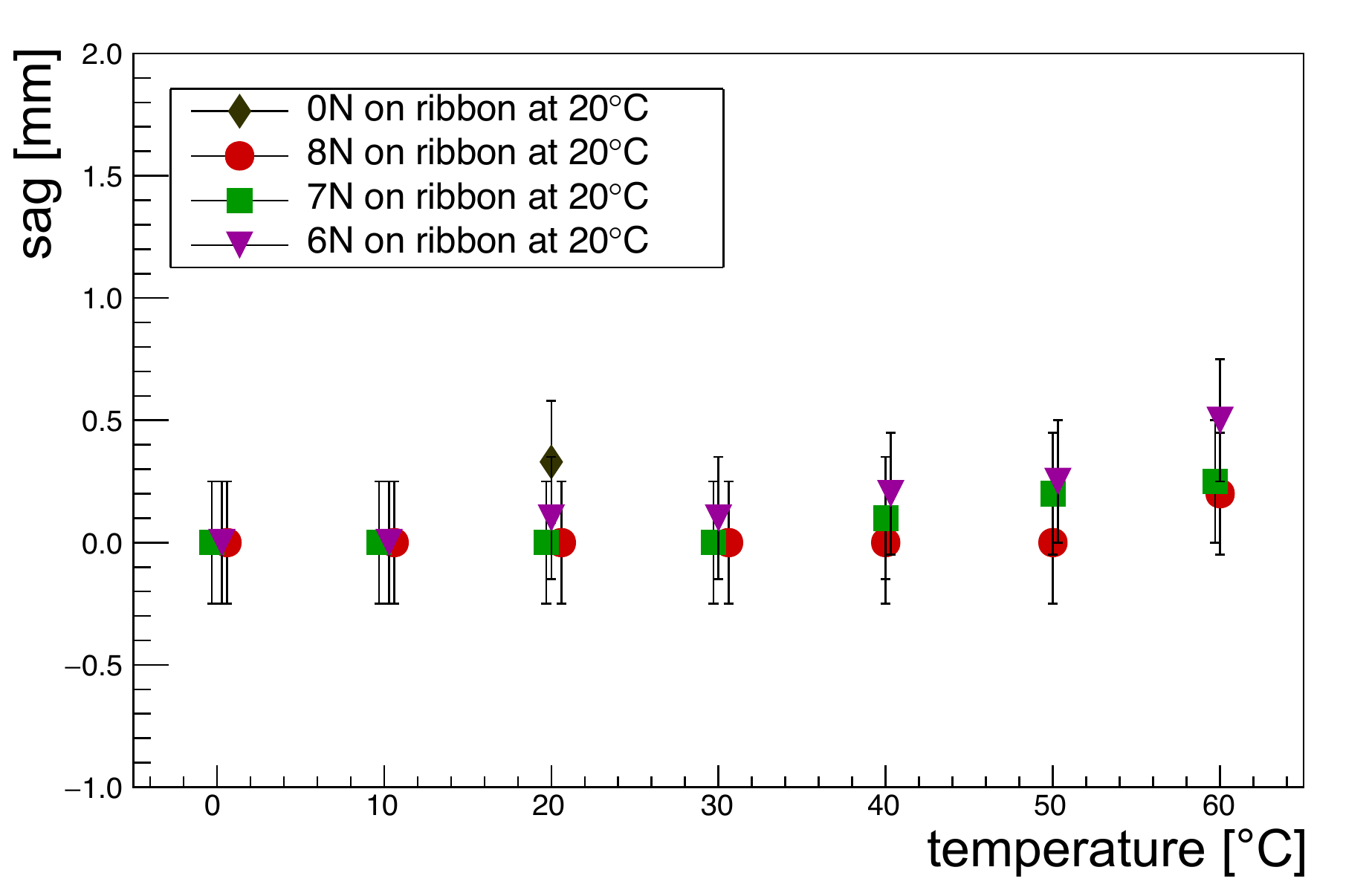}
   \caption{3-layer Scifi ribbon elongation as a function of the temperature (left) and
sag as a function of the temperature for different values of the applied tension (right).
Also shown is the sag at $20^\circ~{\rm C}$ with no tension applied.}
   \label{fig:scifiSag}
\end{figure}

The thermal expansion and sagging of the fiber ribbons has been measured
by warming a 3-layer fiber ribbon in a climate chamber
between $0^\circ~{\rm C}$ and $60^\circ~{\rm C}$.
The SciFi ribbon is spring loaded at one side with two identical springs.
The fiber ribbons will be in proximity of the silicon pixel detectors,
which can warm up to $50^\circ~{\rm C}$.
A thermal expansion coefficient of
$(65 \pm 16) \times 10^{-6} / {\rm K}$ has been measured for the 300~mm long fiber ribbons,
as shown in Figure~\ref{fig:scifiSag} (left).
An elongation of $\sim 1~{\rm mm}$ is expected for a $50^\circ~{\rm C}$ thermal excursion.
The elongation will be compensated by spring loading the ribbons (see Figure~\ref{fig:SciFiCAD} in Section~\ref{sec:mecano}).
Figure~\ref{fig:scifiSag} (right) shows the sag of the same fiber ribbon
as a function of the temperature for different values of the applied tension
(the reported value is twice the tension provided by a single spring).
A tension of 8~N is required to prevent sagging over the whole temperature range
and to guarantee the correct positioning of the SciFi detector.

\clearpage
\section{Multichannel Silicon Photo-Multiplier Arrays}
\label{sec:SiPM}

The scintillation light produced in the fibers is detected at both fiber ends with multichannel Silicon Photo-Multiplier (SiPM) arrays.
Acquiring the signals on both sides improves the time resolution, helps to distinguish between noise and signal,
and increases the detection efficiency of the whole system because of an improved pileup rejection.
Moreover, by taking the mean-time of the two time measurements,
the timing measurements is independent of the hit position
(see Figure~\ref{fig:MTz})
and thus no position dependent correction is necessary.

\begin{figure}[b!]
   \centering
   \vspace*{-3mm}
   \includegraphics[width=0.60\textwidth]{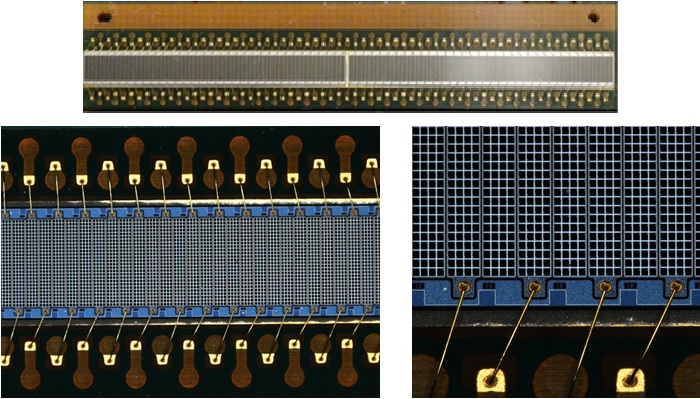}
   \hspace*{10mm}
   \includegraphics[width=0.13\textwidth]{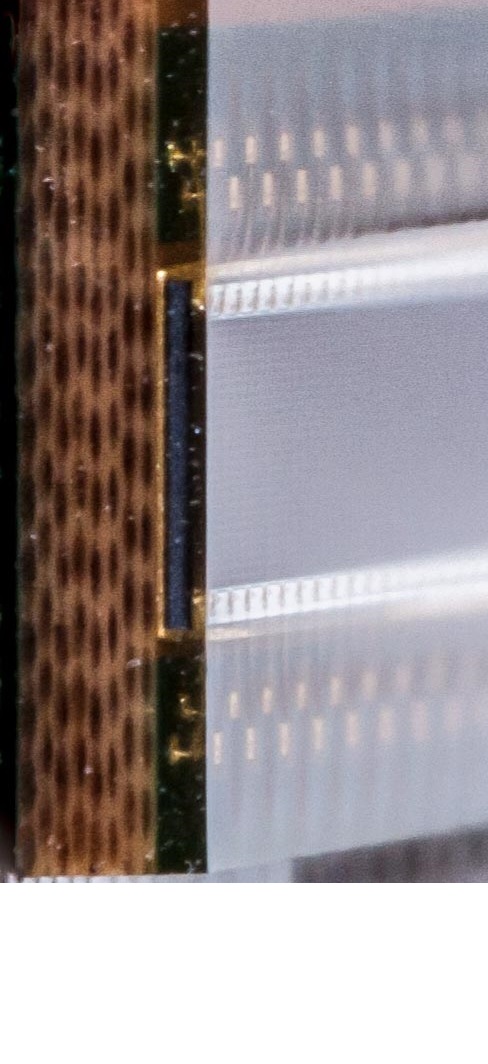}
   \caption{Photograph of a Hamamatsu S13552-HRQ SiPM column array
including a close view showing the pixel structure of the sensor.
Several pixels are arranged in vertical {\it columns} to provide the longitudinal segmentation of the sensor .
The side profile of the sensor is schematized on the right.}
   \label{fig:sipmArray}
\end{figure}

The \mude fiber detector is read out with the 128 channel Hamamatsu SiPM arrays~\cite{SiPM}, device S13552-HRQ, 
with a high quench resistance $R_Q$ of $\sim 500~{\rm k}\Omega$.
The large $R_Q$ allows to operate the SiPM arrays at high over bias voltages, and thus to increase the gain
and the photon detection efficiency (PDE) of the sensor.
The drawback is a longer decay time of the output signal and a longer recovery time.
The segmentation of the sensor is obtained by arranging the individual SiPM cells into independent
readout columns or channels (see Figure~\ref{fig:sipmArray}).
Each channel consists of 104 $57.5~\mu{\rm m} \times 62.5~\mu{\rm m}$ cells
arranged in a $4 \times 26$ grid giving a sensitive area of 
$230~\mu{\rm m} \times 1625~\mu{\rm m}$ per SiPM array column.
The cells are separated by trenches of the fifth generation Hamamatsu low-crosstalk development (LCT5).
A $20~\mu{\rm m}$ gap separates the columns,
resulting in a $250~\mu{\rm m}$ readout pitch.
Each sensor comprises 64 such channels, which share a common cathode.
Two sensors, separated by a gap of $220~\mu{\rm m}$,
form the 128 channel device shown in Figure~\ref{fig:sipmArray}.
The sensors are delivered wire bonded on a PWB with BGA soldering balls on the backside,
and are coated with a $105~\mu{\rm m}$ thick epoxy resin.
Table~\ref{tab:SiPMArrays} summarizes the most important features of the sensor.
These sensors have a gain of $3.8 \times 10^6$, a PDF of 48\%, and a cross-talk probability of 3\%
when operated at an overbias voltage $V_{ob} = V_{op} - V_{bd}$ of 3.5~V above the breakdown voltage $V_{bd}$. 
The dark count rate (DCR) of an non-irradiated sensor at $25^\circ~{\rm C}$ is $ \sim 15~{\rm kHz}$ per channel
at a 0.5 photon equivalent threshold and $\sim 1~{\rm kHz}$ at a 1.5 threshold.
In \mude the sensors will be operated at a temperature below $-10^\circ~{\rm C}$.
The cooling of the detector is required to limit the radiation damage effects.
and to reduce the DCR.
A short overview of radiation effects is given in Section~\ref{sec:rad}.
Radiation effects will be discussed in detail in a separate work~\cite{rad}
(see also~\cite{Gerritzen}).

\begin{table}[t!]
    \centering
    \begin{tabular}{l l}
    \hline
    characteristic                                                &  value \\
    \hline
    breakdown voltage                                        & $\sim 52.4~{\rm V}$ \\
    variation per sensor                                       & $\pm 300~{\rm mV}$ \\
    variation between sensors                               & $\pm 500~{\rm mV}$ \\
    temperature coefficient ($^\ast$)                    & 53.7~mV/K \\
    gain ($^\ast$)                                               & $3.8 \times 10^6$ \\
    dark count rate (0.5 ph. thr.)                          & 15~kHz \\
    direct cross-talk                                             & $< 3\%$ \\
    delayed cross-talk                                          & $< 2.5\%$ \\
    after-pulse                                                    & 0\% \\
    peak PDE ($^\ast$)                                       & 48\% \\
    max PDE wavelength ($^\ast$)                       & 450~nm \\
    mean quench resistance $R_Q$                      & $500 \pm 10~{\rm k}\Omega$ \\
    recovery time $\tau_{\rm recovery}$               & $69 \pm 2~{\rm ns}$ \\
    rise time $\tau_r$                                          & < 1~ns \\
    short decay component $\tau_{\rm short}$               & $\sim 1~{\rm ns}$ \\
    long decay component $\tau_{\rm long}$                 & $50 \pm 3~{\rm ns}$ \\
    \hline
    \end{tabular}
    \caption{Characteristics of the multichannel SiPM array, Hamamatsu device S13552-HRQ, at an overbias voltage of $V_{ob} = 3.5~{\rm V}$
and $T = 25^\circ~{\rm C}$.
Most of these values have been determined by us, while those indicated with ($^\ast$) are from~\cite{SiPM}.}
   \label{tab:SiPMArrays}
\end{table}

This sensor~\cite{SiPM} has been developed for the LHCb experiment
and matches the requirements of the \mude fiber detector.
Single photon detection capabilities, very fast intrinsic time response (the single photon jitter is $\sim 220~{\rm ps}$),
and insensitivity to magnetic field
are the key features for use in the \mude experiment.

\begin{figure}[t!]
   \centering
   \includegraphics[width=0.60\textwidth]{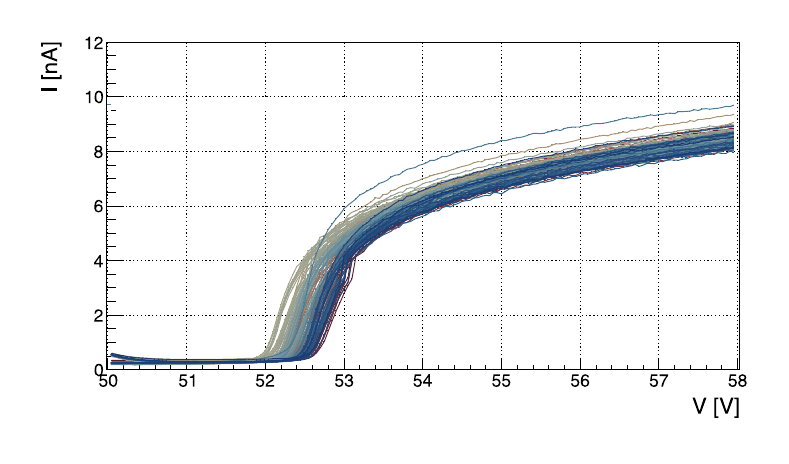}
   \vspace*{-5mm}
   \caption{$I - V$ curves for all 128 channels of one Hamamatsu S13552-HRQ SiPM array.
All breakdown voltages are comprised within $\pm 300~{\rm mV}$ of the central value of $\sim 52.4~{\rm V}$.}
  \label{fig:SiPMIVcurves}
\end{figure}

\begin{figure}[h]
   \centering
   \includegraphics[width=0.49\textwidth]{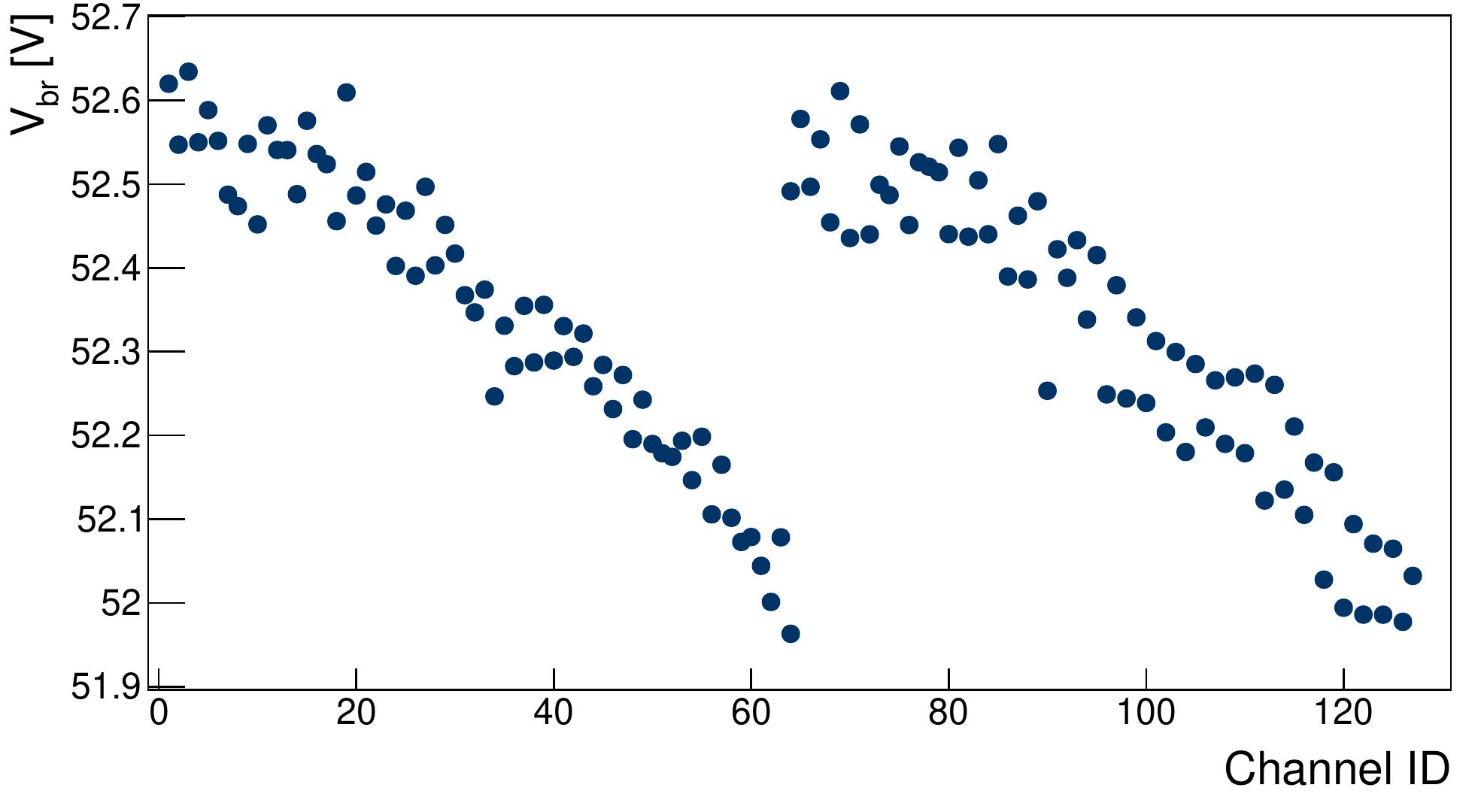}
   \includegraphics[width=0.49\textwidth]{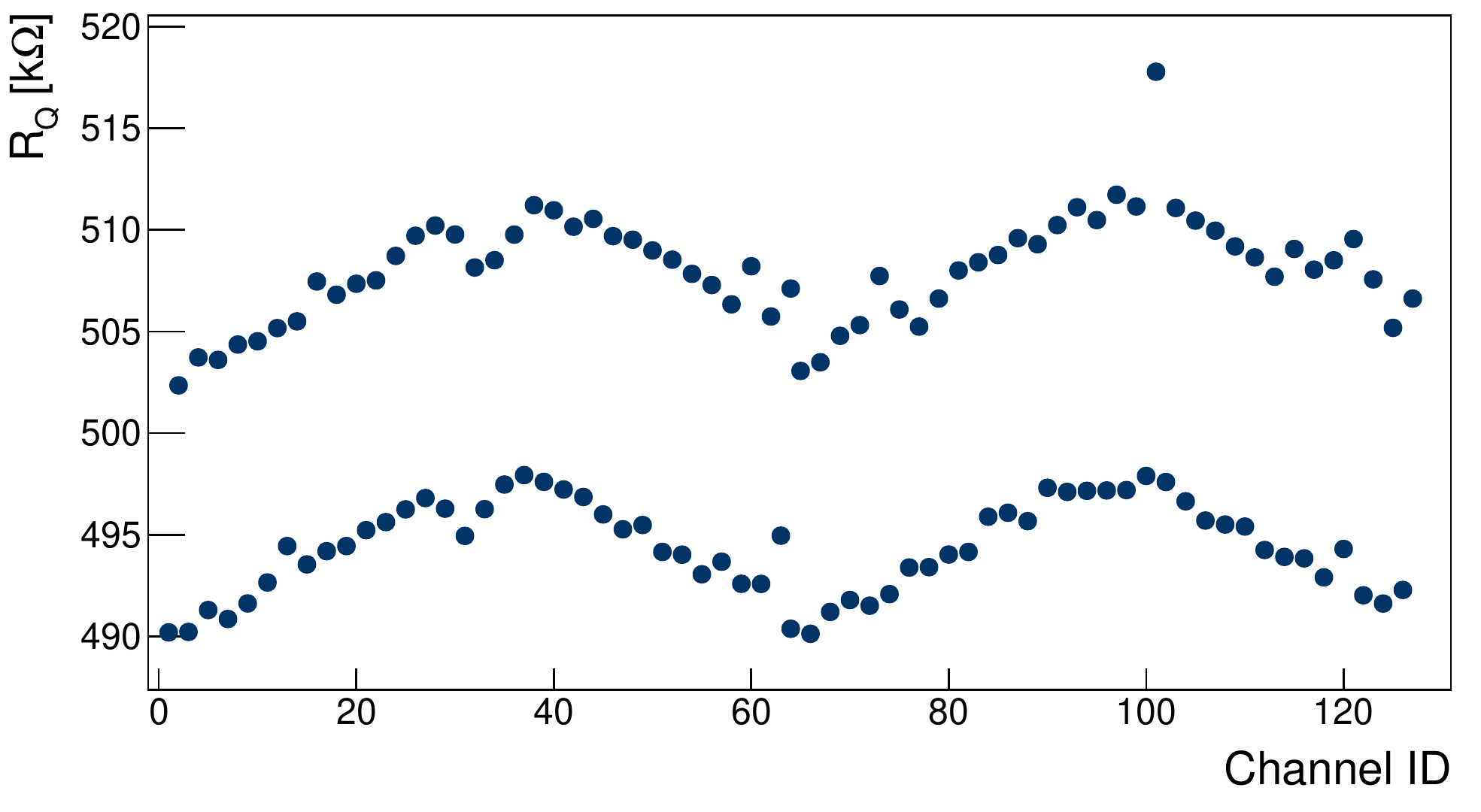}
   \vspace*{-2mm}
   \caption{Breakdown voltage $V_{bd}$ (left) and quench resistance $R_Q$ (right) measured for all 128 channels
of a Hamamatsu S13552-HRQ SiPM array.}
  \label{fig:SiPMIVcurves2}
\end{figure}

To characterize the sensors we measured the $I - V$ curves shown in Figure~\ref{fig:SiPMIVcurves}
for all 128 channels of the SiPM array.
The breakdown voltages $V_{bd}$ (Figure~\ref{fig:SiPMIVcurves2} left)
and the quench resistance $R_Q$ (Figure~\ref{fig:SiPMIVcurves2} right) are determined channel by channel
from the $I - V$ measurements.
The breakdown voltage $V_{bd}$ is comprised within $\pm 300~{\rm mV}$ of the central value of $\sim 52.4~{\rm V}$
(for the sensor in the Figure~\ref{fig:SiPMIVcurves}),
while the quench resistance $R_Q$ values are centered around $500~{\rm k}\Omega$ for the same sensor.
$V_{bd}$ drifts from 52.6~V at one end of the sensor to 52.0~V at the other end.
The repetitive pattern is due to the fact that a single device consists of two separate dies.
The alternating pattern in $R_Q$ originates from the fact that the columns are alternatingly bonded on opposite sides of the sensor.
The same pattern has been observed in all measured SiPM arrays.
$V_{bd}$ varies from sensor to sensor by $\pm 500~{\rm mV}$.
Since all channels share a common cathode, the sensor is operated at a voltage common to all channels.
This leads to small but visible gain differences between channels. 
The performance of the sensor can be further improved by adjusting the bias voltage for each channel.
This is possible, for instance,
with the MuTRiG readout ASIC (Section~\ref{sec:MuTRiG}), which allows for the fine tuning of the bias voltage
around a common value for each individual channel.

\begin{figure}[b!]
   \centering
   \includegraphics[width=0.66\textwidth]{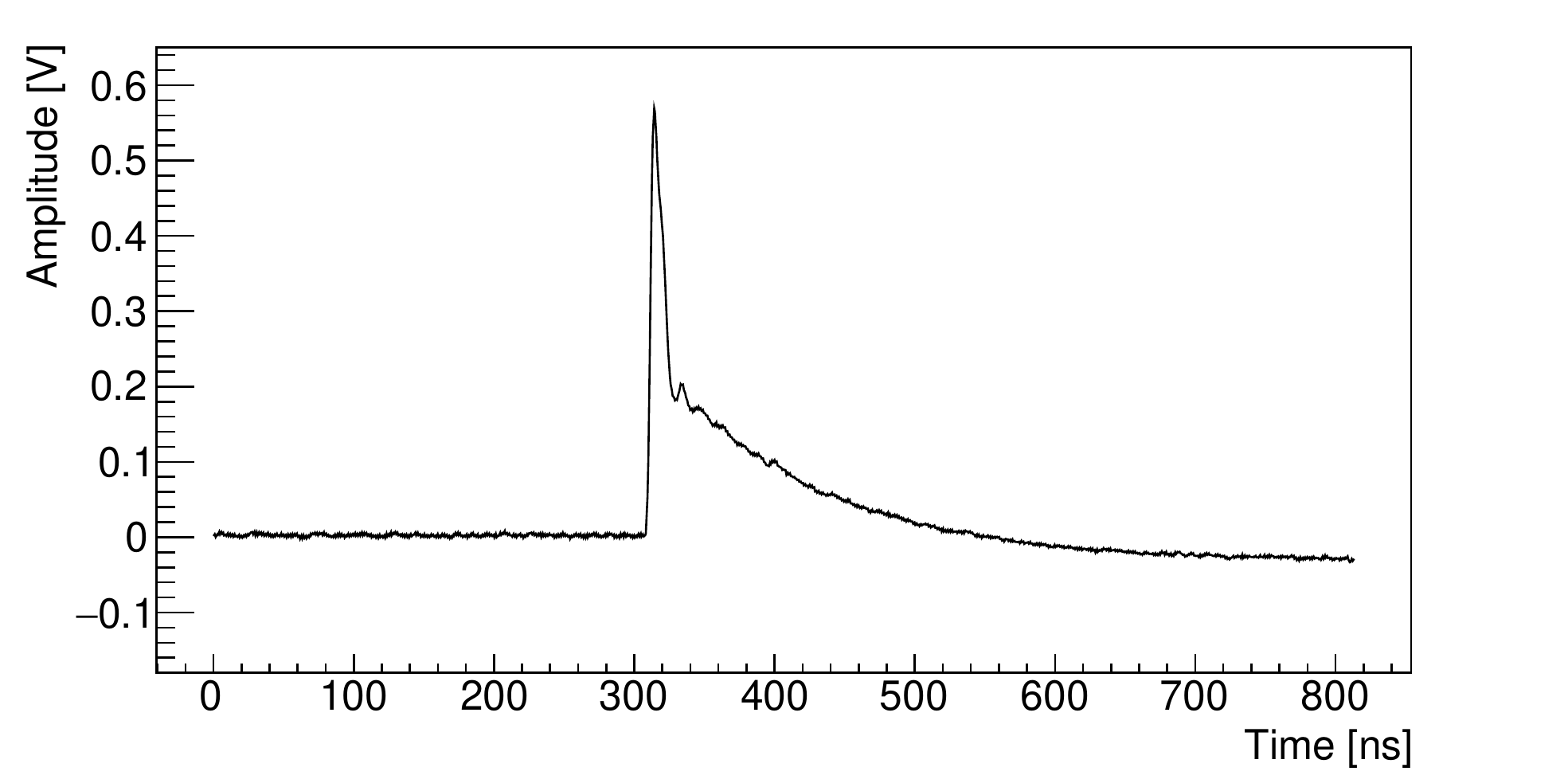}
\caption{$1~\mu{\rm s}$ long digitized waveform.
The sensor is excited with a 405~nm laser.
From the pulse shape one can extract several parameters characteristic for this SiPM,
like the rise time $\tau_r$ and the short and long components of the decay time $\tau_{\rm short}$ and $\tau_{\rm long}$,
the cross-talk between SiPM cells, afterpulsing, etc.,
which are reported in Table~\ref{tab:SiPMArrays}.
In this example, some additional peaks, due to the delayed cross-talk, are visible.}
   \label{fig:wflong}
\end{figure}

Figure~\ref{fig:wflong} shows a digitized pulse (see Section~\ref{sec:SigProc}) for one channel of the SiPM array
over a $1~\mu{\rm s}$ acquisition window.
The sensor has been excited with a 405~nm laser.
The shape of the pulse is characteristics of this SiPM: a fast rise time $\tau_r < 1~{\rm ns}$
($\tau_r$ is driven by the speed of the amplifier) and a
two component decay of the pulse with a sharp {\it kink} few ns after the main peak with a short component $\tau_{\rm short} \sim 1~{\rm ns}$
and a long component $\tau_{\rm long} \sim 50~{\rm ns}$,
which is due to the large quench resistance $R_Q$.
Sometime additional peaks from delayed cross-talk between the SiPM cells of the same channel are visible, like in the shown example.
From the analysis of the waveforms we have estimated the direct and delayed cross-talk, after pulsing, etc. (see Table~\ref{tab:SiPMArrays}).

\begin{figure}
   \centering
   	\begin{tikzpicture}[scale=2, very thick, font={\sffamily}]
	\foreach \i in {1,...,8}
		\draw[thick] (-0.115+0.25*\i, 0.8125) -- +(0.230, 0) -- +(0.230, -1.625) -- +(0, -1.625) -- +(0,0);
	\foreach \i in {9,...,11}
		\draw[thick, dotted] (-0.115+0.25*\i, 0.8125) -- +(0.230, 0) -- +(0.230, -1.625) -- +(0, -1.625) -- +(0,0);
	
	\foreach \j in {0,...,8}
		{\draw[thick, blue] (0.3+0.255*\j, 0.0) circle (0.125);
		\draw[thick, blue] (0.1725+0.255* \j, -0.22) circle (0.125);
		\draw[thick, blue] (0.1725+0.255*\j, 0.22) circle (0.125);
		};
	\foreach \j in {9,...,10}
		{\draw[thick, blue, dotted] (0.3+0.255*\j, 0.0) circle (0.125);
		\draw[thick, blue, dotted] (0.1725+0.255* \j, -0.22) circle (0.125);
		\draw[thick, blue, dotted] (0.1725+0.255*\j, 0.22) circle (0.125);
		};
		
		\draw[thick, <->] (3.35, 0.8125) -- (3.35, -0.8125);
		\node[rotate=90] at (3.45,0) {1.625~mm};

		\draw[thick, <->] (3.0, 0.3) -- (3.0, -0.3);
		\node[rotate=90] at (3.13,0) {0.7~mm};

		\draw[thick,<->] (0.25*2, -0.9125) -- (0.25*3,  -0.9125);
		\draw[thin, dotted] (0.25*2, -0.6) -- (0.25*2, -0.9225);
		\draw[thin, dotted] (0.25*3, -0.6) -- (0.25*3, -0.9225);
		\node at (0.25*2.5, -1.1025) {pitch $250~\mu{\rm m}$};

		\draw[thick,<->] (0.25*2-0.115, 0.9125) -- (0.25*2+0.115,  0.9125);
		\draw[thin, dotted] (0.25*2-0.115, 0.6) -- (0.25*2-0.115, 0.9225);
		\draw[thin, dotted] (0.25*2+0.115, 0.6) -- (0.25*2+0.115, 0.9225);
		\node at (0.25*2, 1.1) {active $230~\mu{\rm m}$};
		
		\draw[thick, blue, <->] (0.1725+0.255*6, -0.75) -- (0.1725+0.255*7, -0.75);
		\draw[thin, blue, dotted] (0.1725+0.255*6, -0.75) -- (0.1725+0.255*6, -0.22);
		\draw[thin, blue, dotted] (0.1725+0.255*7, -0.75) -- (0.1725+0.255*7, -0.22);
		\node[blue] at (0.3+0.255*6, -0.95) {fiber pitch $255~\mu{\rm m}$};

		\draw[thick, ->] (0.25*6, 1.1025) -- (0.25*5+0.115, 0.82);
		\node[anchor=west, align=left] at (0.25*6, 1.1) {inactive $20~\mu{\rm m}$};


	\end{tikzpicture}
   \vspace*{-8mm}
\caption{Mapping of the SciFi ribbon on the SiPM array.
No one to one matching is possible between the fibers and the SiPM columns
and the light signal will be spread over several columns.}
   \label{fig:fiberMapping}
\end{figure}
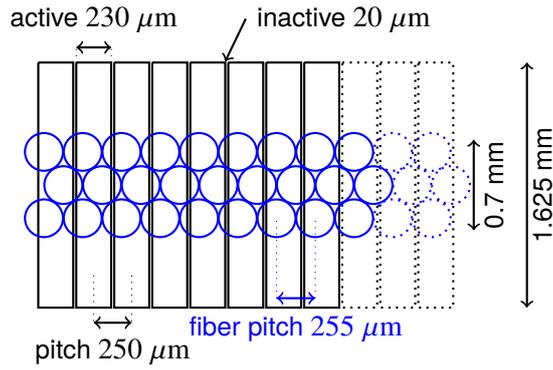

The fiber ribbons are coupled directly to the surface of the photo-sensors.
Figure~\ref{fig:fiberMapping} shows the mapping of the SciFi ribbon to the SiPM array.
No one to one matching is possible between the fibers and the SiPM columns
because of the staggering of the fibers.
The deployed mapping scheme adds an inefficiency of $(1-\varepsilon_{\text{fill-factor}}) = 8\%$ in addition to the sensor's PDE.
For easy detector assembly and maintainability, the coupling will be done by mechanical pressure provided by fixation screws
(see Figure~\ref{fig:SciFiCAD}).
No spring loading of the ribbons to the SiPMs is foreseen, and no optical interfaces will be used
to minimize the spread of the light signal in the interface between the SciFi ribbons and the sensor. 
Given the thickness of the SciFi ribbons, a smaller height SiPM array could also be used,
however we adopted this one because at the time of procurement the thickness of the SciFi ribbons (3- or 4-fiber layers) has not yet been fixed.
The geometric overlap with the 1.625~mm high sensor columns ensures the full coverage of the ribbon and of the photons,
which are emitted at an angle up to $45^\circ$ w.r.t. the fiber axis. 

\clearpage
\section{Test Setup}
\label{sec:setup}

\begin{figure}[b!]
   \centering
   \vspace*{-2mm}
   \includegraphics[width=0.9\textwidth]{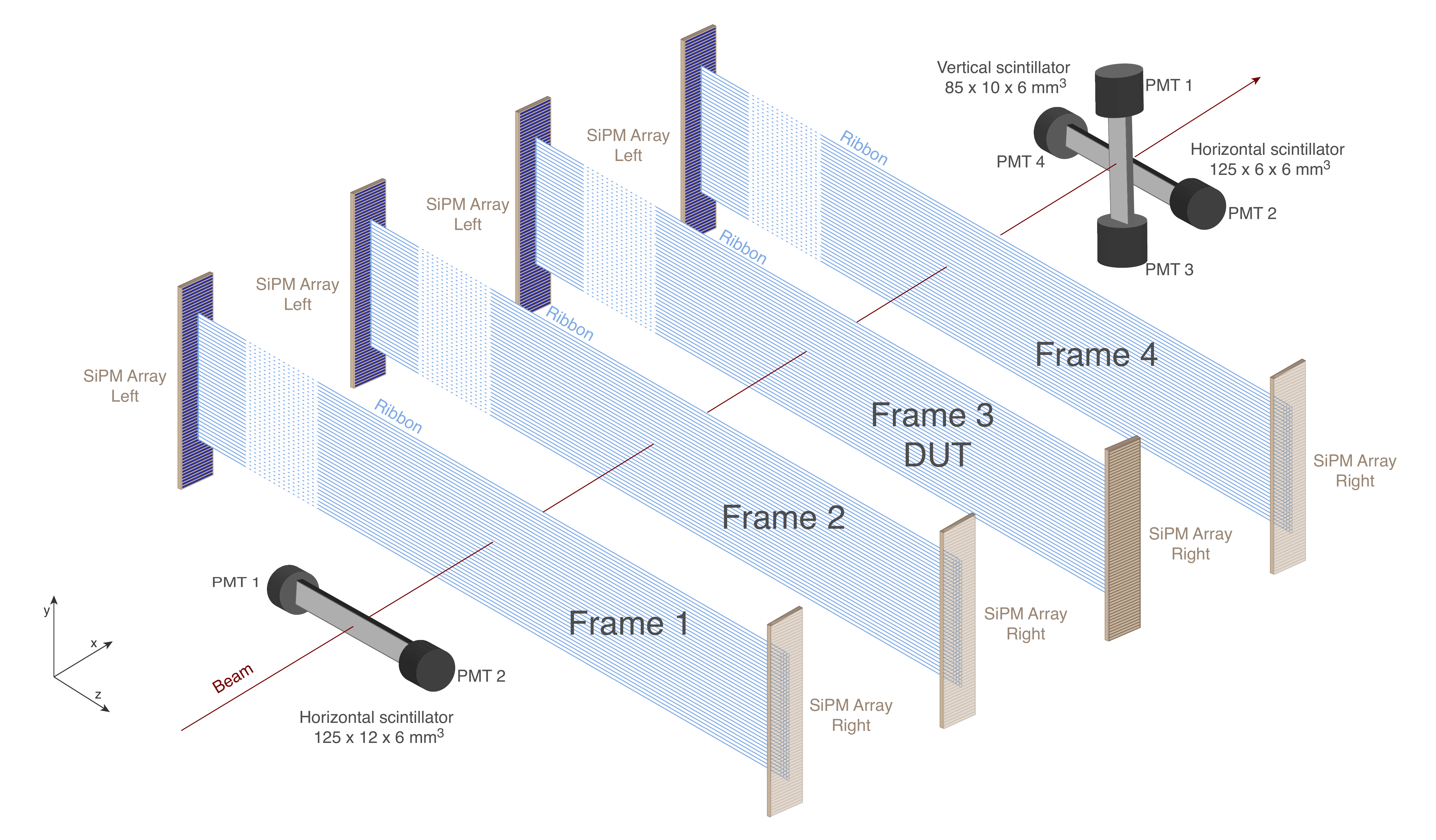}
   \caption{Schematics of the Scifi telescope comprising fours SciFi ribbons and three trigger scintillators.
The third most downstream detector is referred to as Device Under Test (DUT).}
   \label{fig:setup}
\end{figure}

The SciFi detector has been studied in extensive test beam campaigns at CERN (T9 beamline) and PSI ($\pi$M1 beamline).
Complementary measurements have been performed using a $^{90}{\rm Sr}$ $\beta$ source, as well.
When using the $\beta$ source, energetic electrons were selected by applying a high threshold on the trigger scintillator
located behind the SciFi ribbon.
The timing measurements have been performed with beams only.
To study the SciFi detector we assembled a telescope consisting of four SciFi detectors,
as illustrated in Figure~\ref{fig:setup}.
All the detectors are placed in a light tight box.
Three out of four SciFi ribbons are used for tracking, while the third most downstream one
is the Device Under Test (DUT).
The ribbons are mounted in motorized frames movable vertically for alignment and vertical scans.
The DUT can also be tilted w.r.t. the beam around the horizontal axis for angular scans.
Moreover, the whole setup is mounted on rails and can be moved transversely w.r.t. the beam for position scans (horizontal scans).
A scintillator bar of $125~{\rm mm} \times 12~{\rm mm} \times 6~{\rm mm}$ placed upstream
and a scintillator cross formed by a horizontal bar of $125~{\rm mm} \times 6~{\rm mm} \times 6~{\rm mm}$
and a vertical bar of $85~{\rm mm} \times 10~{\rm mm} \times 6~{\rm mm}$ placed downstream
are used for triggering on beam particles.
The scintillators were coupled at both ends to Hamamatsu PMTs, model H6524. 
The trigger cross provides also the external time reference for timing measurements
with a time resolution of 80~ps.
The acquisition of an event is triggered by a 3-fold coincidence between the 3 scintillator bars.

The SciFi ribbons are of the same manufacturing and size as the ones that will be used in \mude (i.e. 32.5~mm wide and 300~mm long).
Prototype fiber ribbons consisting of two to six fiber layers have been assembled using different blue-emitting scintillating fibers
and different types of adhesive, as well.
The SciFi ribbons in the telescope are coupled to 128 channel SiPM arrays, Hamamatsu device S13552 HQR, at both ends.
Not all the channels of the SiPM arrays, however, have been instrumented:
64 channels of the DUT, corresponding to a vertical extension of 16~mm,
and 32 channels for the other ribbons, corresponding to a vertical extension of 8~mm,
have been equipped with readout electronics.
In the following, if not stated otherwise, the results are presented for the \mude baseline design,
which consists of SciFi ribbons made of 3 staggered layers of $250~\mu{\rm m}$ diameter round SCSF-78MJ scintillating fibers
prepared with the {\sc Polytec EP 601-Black} epoxy.

\subsection{Test Beam Readout Electronics}
\label{sec:electro}

\begin{figure}[t!]
   \centering
   \includegraphics[width=0.45\textwidth]{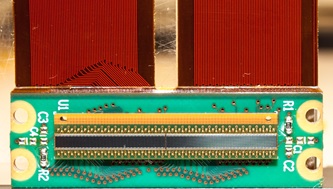}
   \hspace*{5mm}
   \includegraphics[width=0.42\textwidth]{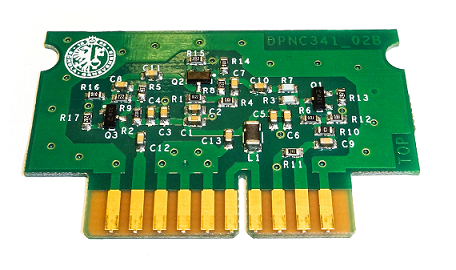}
   \caption{left) SiPM array module soldered to a carrier PCB with integrated flex print cables with 128 signal lines.
right) Transistor based fast hybrid amplifier used in test beam measurements presented in this work.}
   \label{fig:sipmflex}
\end{figure}

The SiPM array modules are soldered on a carrier PCB with integrated flex print cables (Figure~\ref{fig:sipmflex} left),
which connect each SiPM channel to a fast transistor based three-stage common emitter amplifier (Figure~\ref{fig:sipmflex} right).
A shunt resistor $R_{\text{shunt}} = 39~\Omega$ converts the SiPM's current signal into a voltage,
which is AC-coupled by a $C = 100~{\rm nF}$ capacitance to the amplification stages.
Simulations of the amplifier showed a cut-off frequency of 400~MHz, i.e. a rise time of around 1~ns.
A variable gain in the order of 35~dB, controlled by the bias voltage applied to the amplifiers,
provides single photo-electron amplitudes of $\sim 100~{\rm mV}$.
The maximal amplitude never exceeded 1~V.
Typical SiPM amplified signals show a rise time of approximately 1~ns (see Figure~\ref{fig:wflong}).

The SiPM signals are recorded with waveform digitizers based on the DRS4 circular capacitor array~\cite{DRS}.
The digitizer boards comprise 32 analog input channels (i.e. 4 DRS4 ASICs)~\cite{Damyanova}.
The inputs range between 0~V and $-1~{\rm V}$ and are inverted on the board to match the dynamic range of the DRS4 ASIC.
The baseline is shifted by 50~mV for detailed monitoring.
Each DRS board is read out via a USB 3.0 link.
The DRS4 ASIC comprises 9 independent capacitor arrays with a depth of 1024 cells each.
The $9^{\rm th}$ channel is used for the synchronization of the ASICs.
The waveforms are sampled at 5~GHz.
The sampling speed is controlled by a PLL locked to a reference clock ($\sim 2.5~{\rm MHz}$ for 5~GHz sampling).
The charges stored in the capacitor arrays are digitized at a much lower frequency of 33~MHz with 12 bit Analog-to-Digital converters (ADC).
An example of a digitized waveform is shown in Figure~\ref{fig:wfcalib}.
Each capacitor array cell is voltage calibrated
and the time width of each sampling cell is measured with a 100~MHz sinusoidal signal generated on the digitizer board.
The DRS4 capacitor arrays are synchronized with an external 50~MHz sinusoidal reference signal,
which is recorded simultaneously with data in a separate channel of the ASIC (channel 9).
The detailed description of the calibration procedure is given in~\cite{Damyanova}.

Note that the readout electronics described here is used to study the SciFi detector in test beams as presented in this work.
For the readout of the SciFi detector in \mude, the MuTRiG ASIC will be used (see Section~\ref{sec:MuTRiG}).

\subsection{The Beamline}

\begin{figure}[t!]
   \centering
   \includegraphics[width=0.8\textwidth]{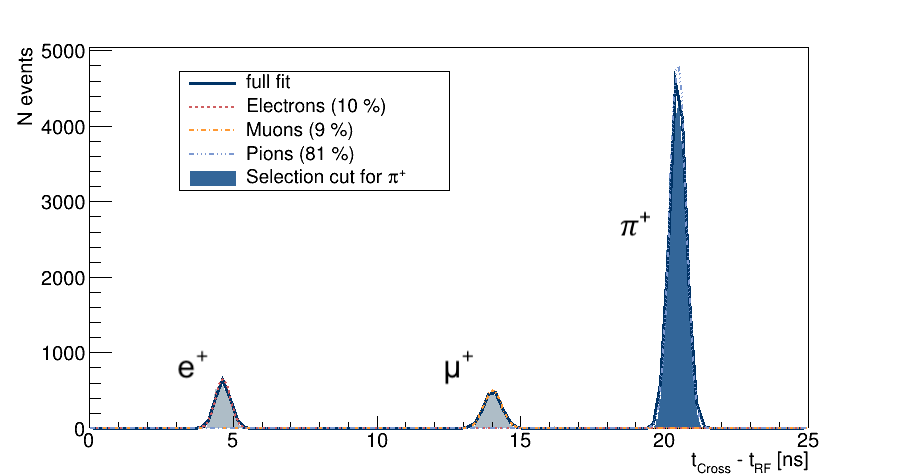}
   \caption{Time of flight particle identification based on
the time difference between the cyclotron accelerating cavities RF signal of 50.63~MHz and the trigger cross.
At $210~{\rm MeV}/c$ the beam consists of $81\%~\pi^+$, $9\%~\mu^+$, and $10\%~{\rm e}^+$.
Particle concentrations are extracted after a composite fit of three Gaussian functions.
The area shaded in dark blue represent the cut applied to select pions.
The pattern repeats each 19.74~ns.}
   \label{fig:pid}
\end{figure}

Most of the measurements have been performed in the $\pi$M1 beam line at the Paul Scherrer Institute (PSI).
The beam was set at a momentum of $210~{\rm MeV}/c$ and contained mainly positive pions ($> 85\%$).
At this momenta the ionization energy loss of pions ($\beta \gamma = 1.5$) is 20\% higher
than the energy that a minimum ionizing particle (MIP) would deposit
and is comparable to the energy loss of highly relativistic particles on the relativistic plateau.

Secondary particles from a carbon target ({\sc target M}) in a 1.8~A continuous beam of 590~MeV protons
are extracted and selected w.r.t. the $p/m$ ratio by a set of dipole and quadrupole magnets.
The average continuous beam intensity was of around $10^6$~part/s
with a spot of $15~{\rm mm} \times 10~{\rm mm}$ on the DUT (i.e. in the center of the telescope). 
Beam particles are identified with time of flight (ToF) measurements
between the production target, which is located 21~m upstream of the experimental setup,
by recording the 50.63~MHz RF sinusoidal signal of the proton cyclotron's accelerating cavities (the RF pattern has a length of 19.74~ns)
and the trigger cross.
The beam particle identification is illustrated in Figure~\ref{fig:pid},
which shows the distribution of the time difference between the trigger signal and the RF signal.
Beam particles are selected in a 3~ns window around each peak.
88\% of beam particles have passed this selection.
Particle concentrations are extracted after a composite fit of three Gaussian functions.
At $210~{\rm MeV}/c$ the beam consists of $81\%~\pi^+$, $9\%~\mu^+$, and $10\%~{\rm e}^+$.
For our studies we selected the most abundant particles, the pions.

\clearpage
\section{Signal Processing}
\label{sec:SigProc}

\begin{figure}[h!]
   \centering
   \includegraphics[width=0.70\textwidth]{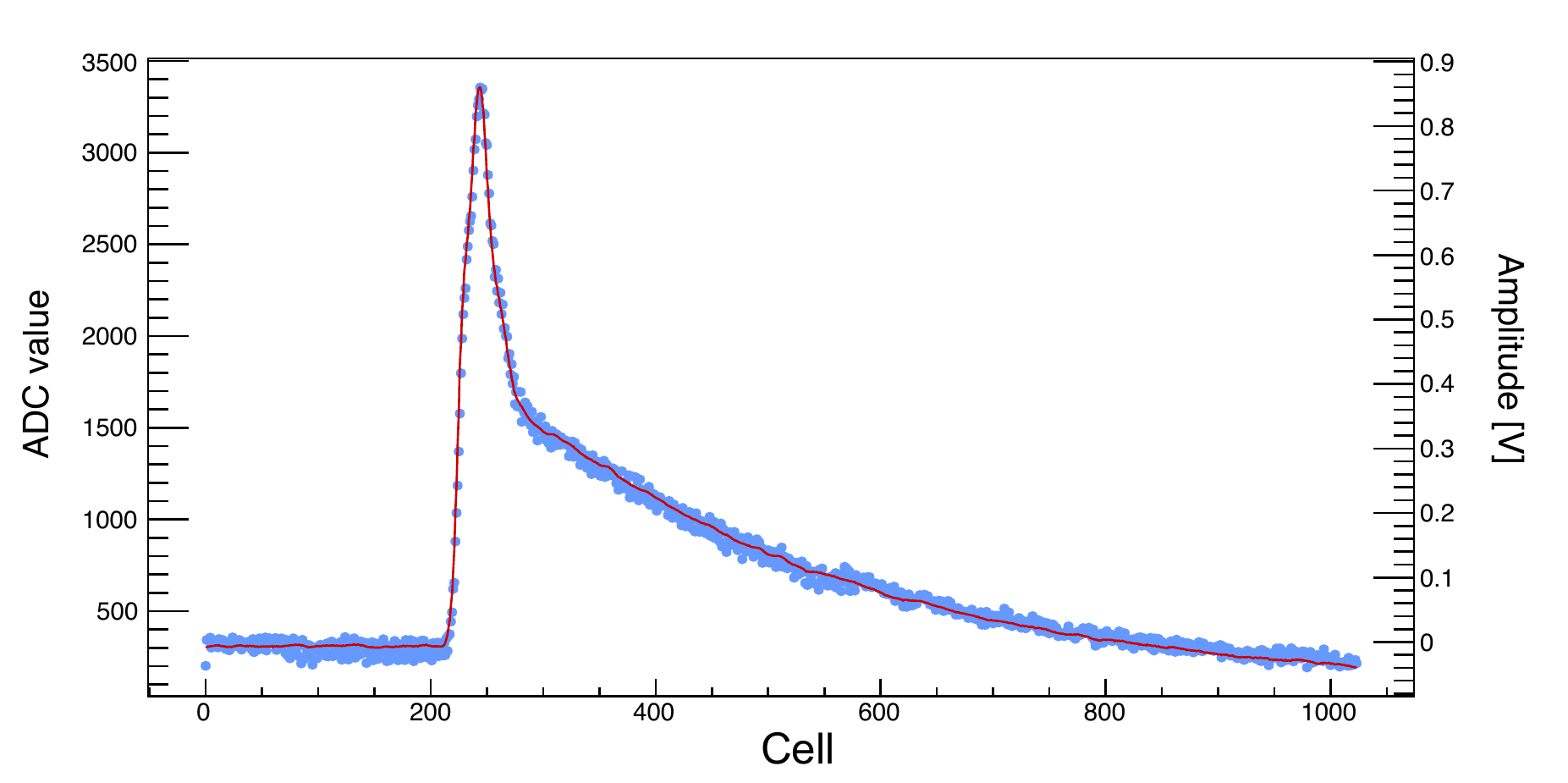}
   \includegraphics[width=0.66\textwidth]{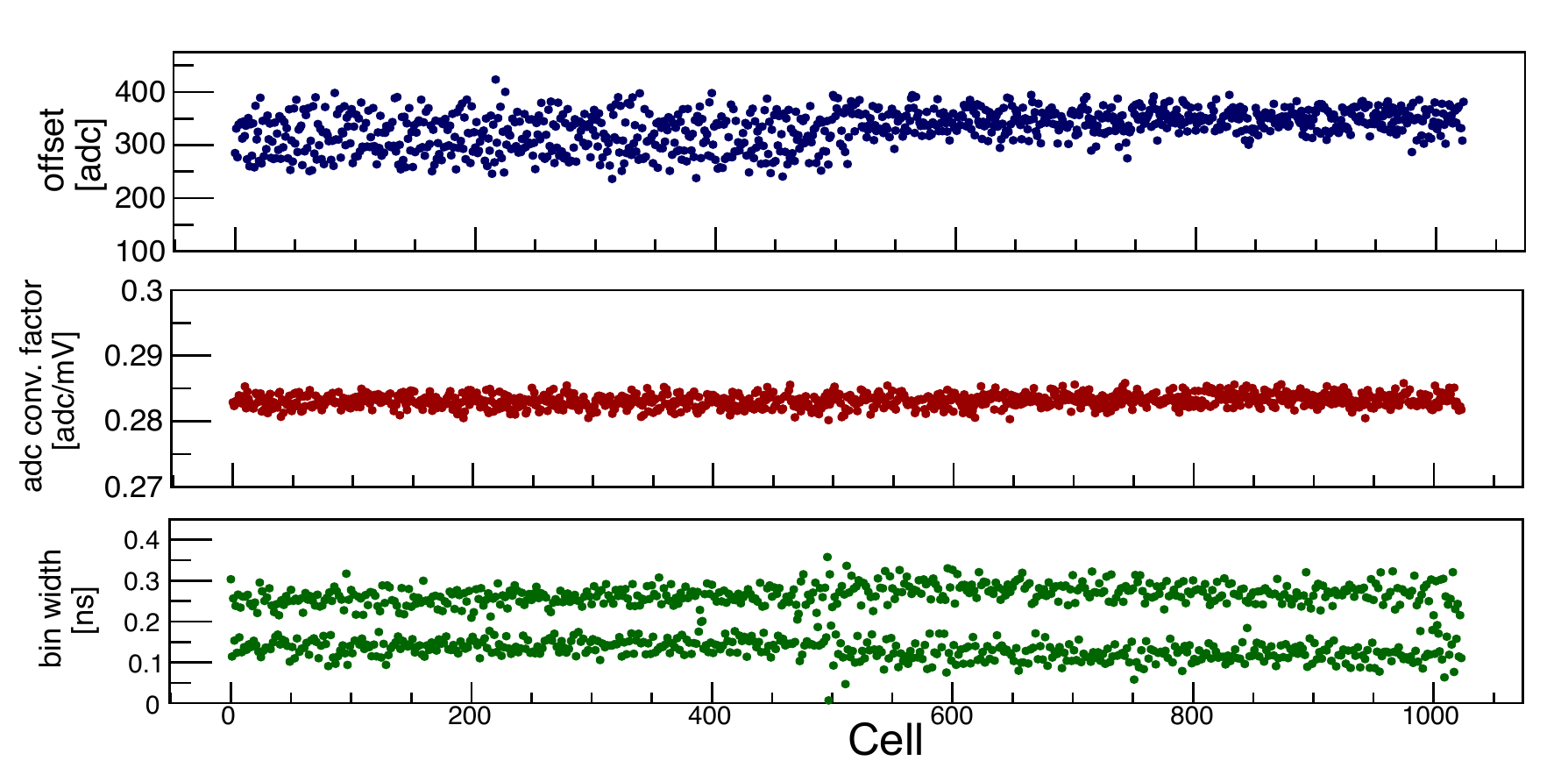}
   \caption{top) Example of a recorded waveform before and after voltage calibration.
The voltage calibrated waveform is plotted on top of the ADC values (continuous line). 
The baseline is not subtracted in order to illustrate the span of the ADC values.
bottom) Waveform calibration: baseline, ADC to voltage conversion factors, and time cell widths.}
   \label{fig:wfcalib}
\end{figure}

Data was acquired with the DRS4 based waveform digitizers as discussed in Section~\ref{sec:setup}. 
An example of a digitized waveform with 1024 samples over 205~ns
from the raw analog-to-digital converter (ADC) values to the voltage calibrated waveform is shown in Figure~\ref{fig:wfcalib}.
All information, like charge and arrival time of scintillation photons, is extracted from the waveforms.
Figure~\ref{fig:wfcalib} also shows the ADC to voltage conversion factors and the time width for each cell of the DRS4 capacitor array.

\subsection{Waveform Processing}

\begin{figure}[t!]
   \centering
   \includegraphics[width=0.75\textwidth]{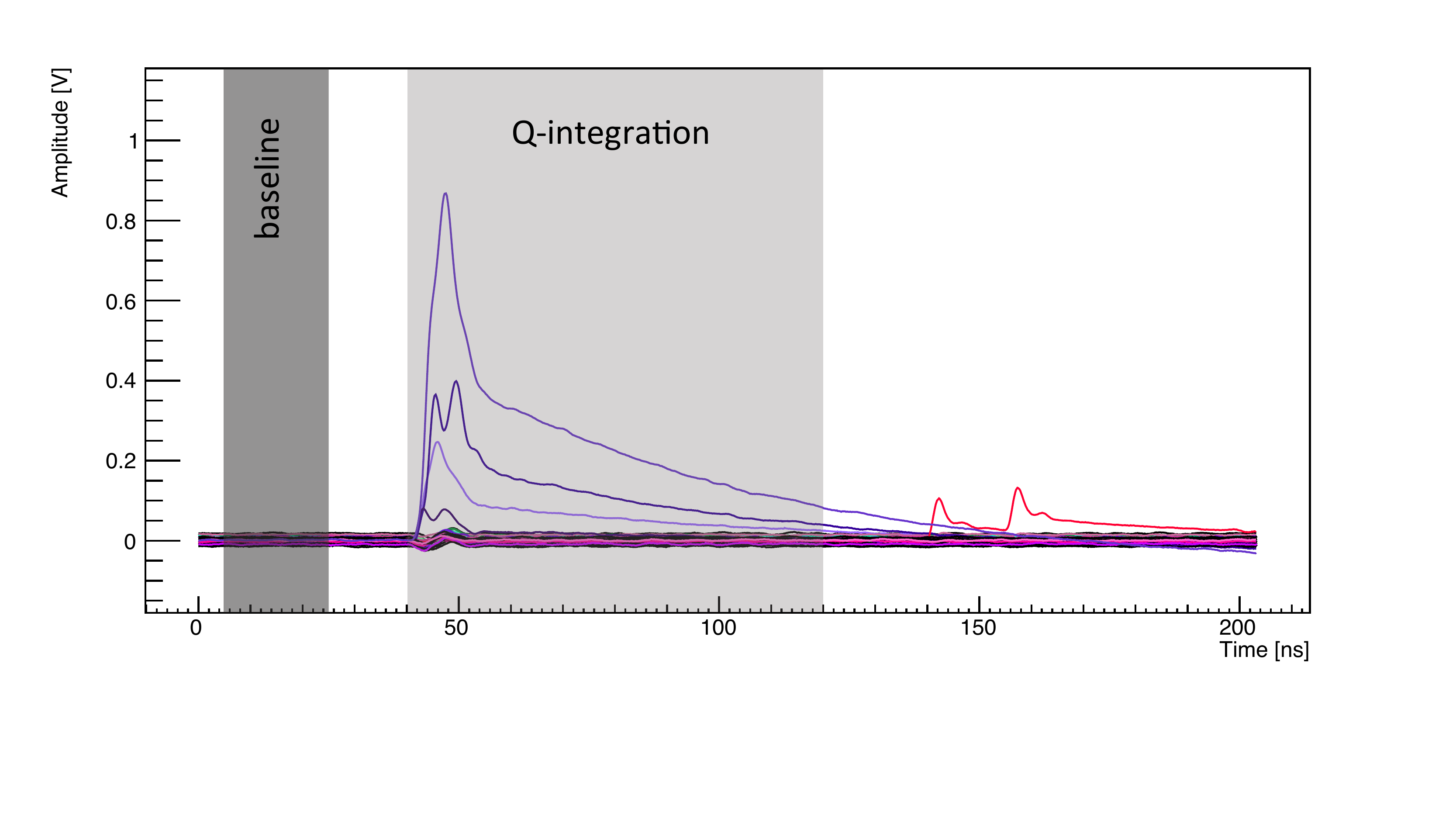}
   \caption{Example of waveforms recorded in on event for 32 consecutive channels.
Few adjacent SiPM channels (4 in this example) show a large activity (i.e. amplitude $> 100~{\rm mV}$.
Some electronic cross-talk at the level of 3\% is visible in the non active channels.
The main part of the signals is integrated over the shaded region ``Q-integration” to extract the charge,
while the area labeled ``baseline” is used for baseline determination. 
The shape of the pulse reflects the fact that not all scintillation photons arrive at the same time.
The late peaks in a single waveform are due to delayed cross-talk between SiPM cells.}
   \label{fig:wfANA}
\end{figure}

Figure~\ref{fig:wfANA} shows the voltage and time calibrated waveforms for 32 consecutive channels of a SiPM array
acquired with one 32 ch. DRS digitization board during one event.
In this example, several consecutive SiPM channels show a large activity.
The light signal generated by the crossing particle is spread over several channels of the SiPM array
as discussed in Section~\ref{sec:cluster}.   
Most of the scintillation photons arrive at the same time and generate signals with steep rising edges.
In the Figure, the signals start around 40~ns after the opening of the acquisition window.
The starting time of the signal is determined by the trigger delay,
which can be adapted such that the waveform is fully contained in the digitizer acquisition window.  
The shape of the pulses reflects the fact
that not all scintillation photons are emitted at the same time
and arrive following the de-exitation of the scintillating fiber (see Figure~\ref{fig:TDecay} and e.g.~\cite{fibers}). 
A few late peaks in the signal tails come mainly from the delayed cross-talk between SiPM cells in the same column.
Electronic cross-talk in the readout chain (mainly in the flex print circuit) at the level of 3\% 
generates the concurrent dips visible in the non active channels.
The time alignment of all channels is determined with respect to the 50~MHz sinusoidal reference signal
injected into each DRS4 ASIC (Section~\ref{sec:electro}) and the external trigger (Section~\ref{sec:Talign}).

To extract the charge generated by the scintillation photons in each SiPM channel,
the waveforms, after baseline subtraction, are integrated over a time interval (i.e. {\it gate}) of 80~ns,
which corresponds to 400 consecutive samples of the waveform,
starting around 5~ns before the start of the pulse.
At this stage the measured charge, which includes also the amplifier's gain, is expressed in arbitrary units.
Thanks to the single photon detection capabilities of the SiPMs, the charge can be normalized to the charge corresponding to one photon
without knowing the absolute amplification of the SiPM and of the electronics
and therefore it is expressed in terms of the number of detected photons or simply photo-electrons (ph.e.).
The baseline is evaluated on an event by event basis
by averaging the waveform over 100 consecutive samples (i.e. over a 20~ns interval)
between 25~ns and 5~ns before the start of the signal
and it is subtracted from the waveform event by event.

\subsection{Charge Normalization}
\label{sec:chNorm}

\begin{figure}[t!]
   \centering
   \includegraphics[width=0.49\textwidth]{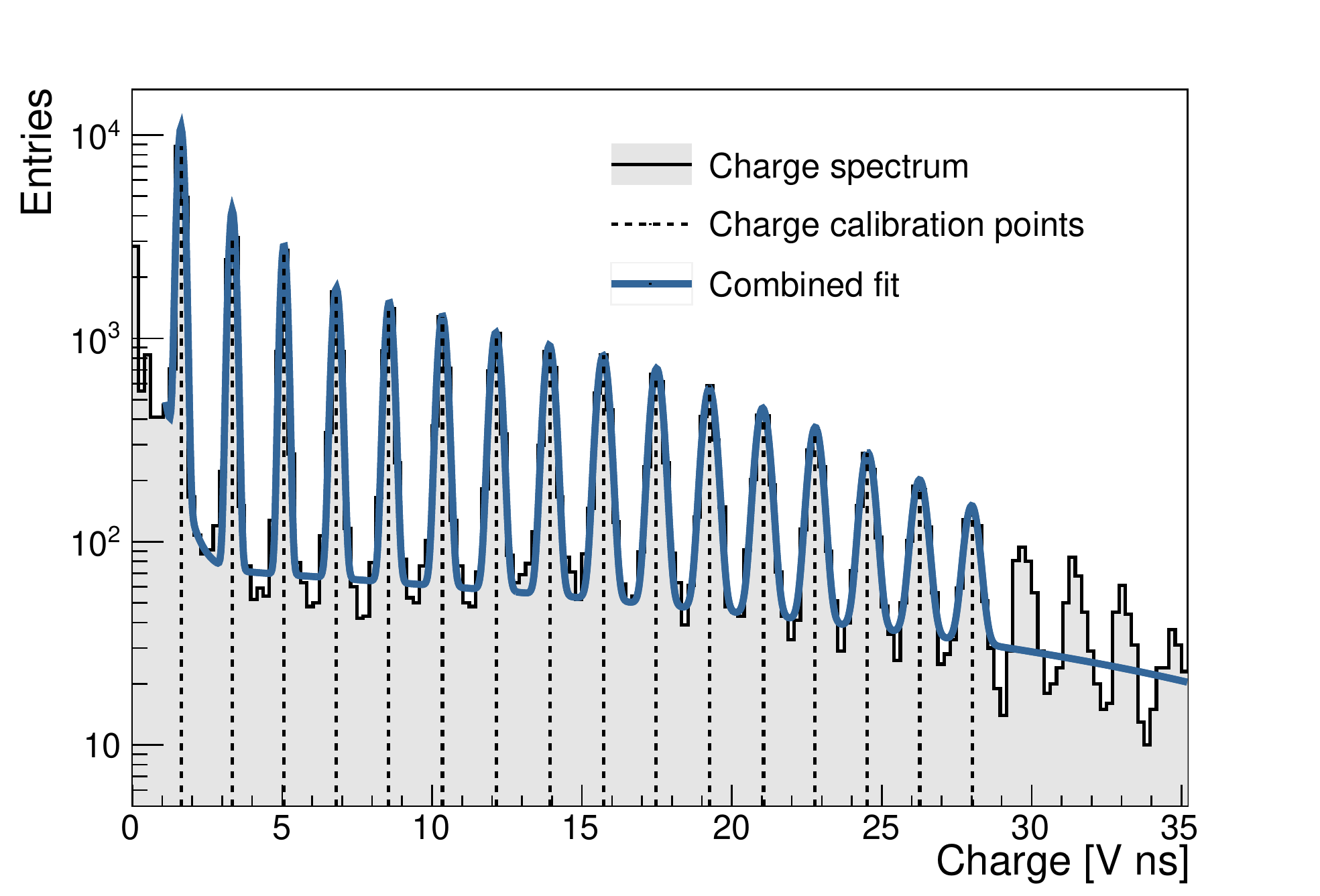}
   \includegraphics[width=0.49\textwidth]{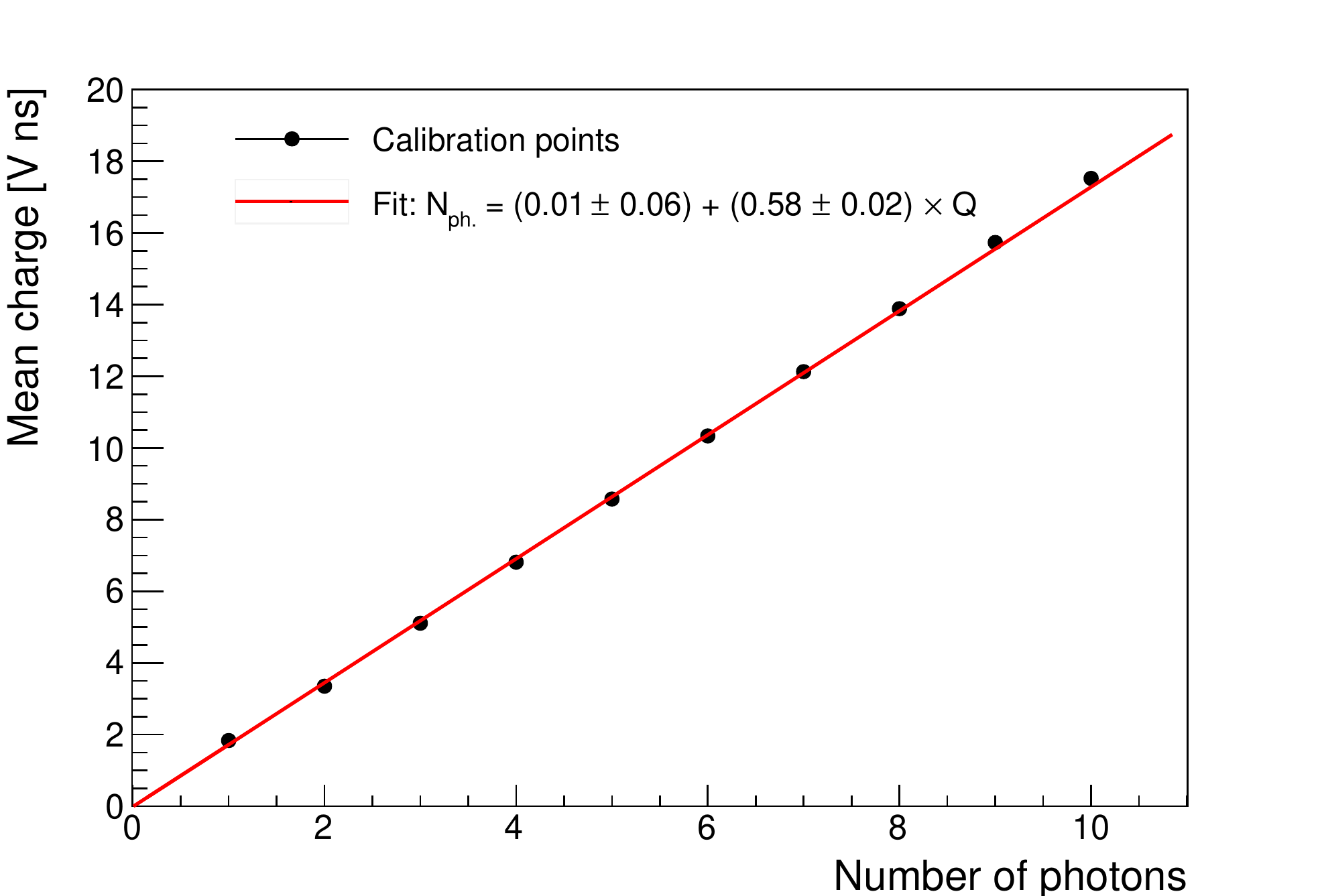}
   \caption{left) Integrated charge distribution for one channel.
Each peak corresponds to the number of detected photons in one event and it is fitted with a Gaussian.
The centroids of the fits are used as input for the photo-electron normalization.
right) Correlation between the integrated charge and the number of detected photons. 
The integrated charges is proportional to the number of detected photons.
The slope of the linear fit gives the relative gain of the channel.}
   \label{fig:calib_fit}
\end{figure}

\begin{figure}
   \centering
   \includegraphics[width=0.47\textwidth]{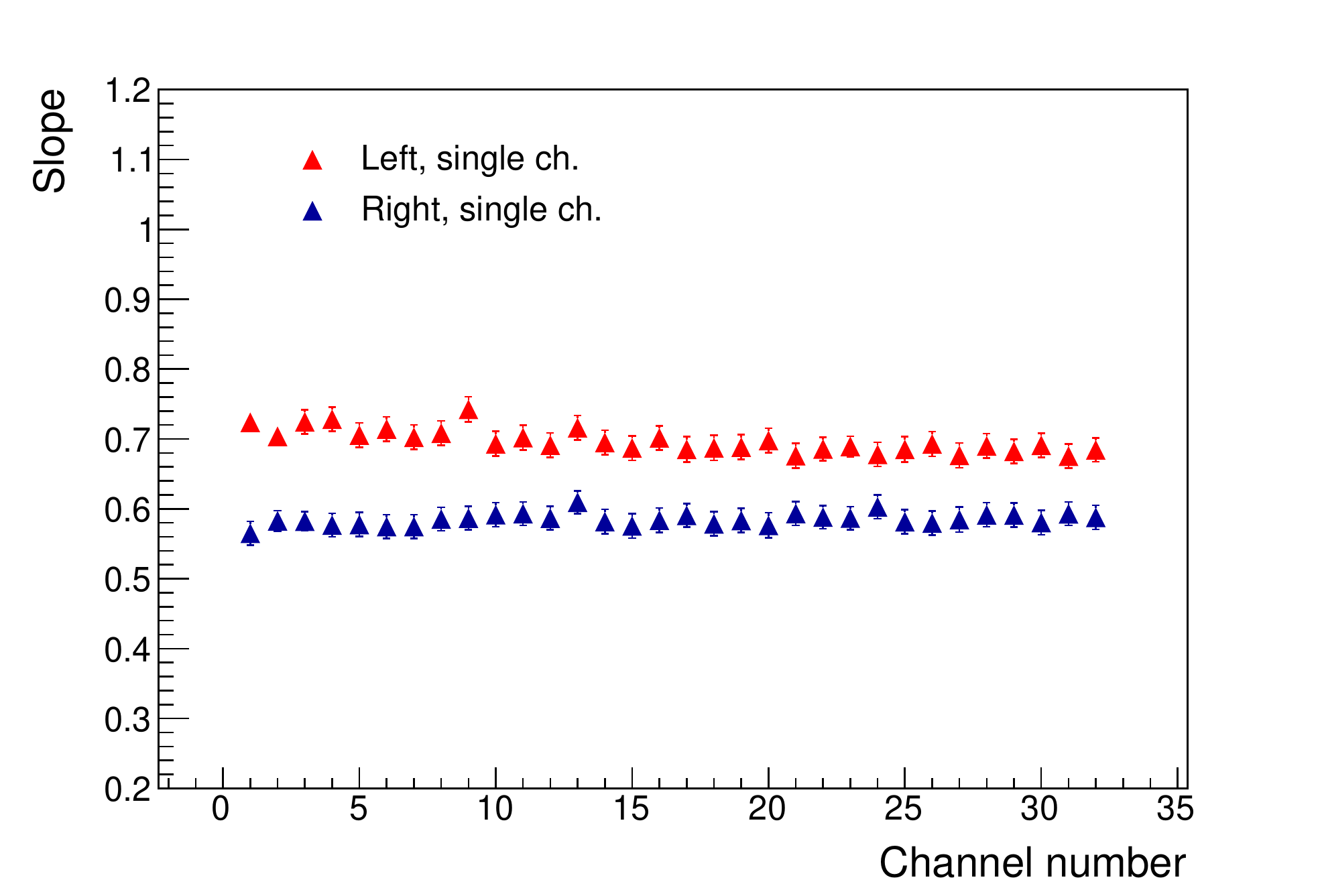}
   \includegraphics[width=0.47\textwidth]{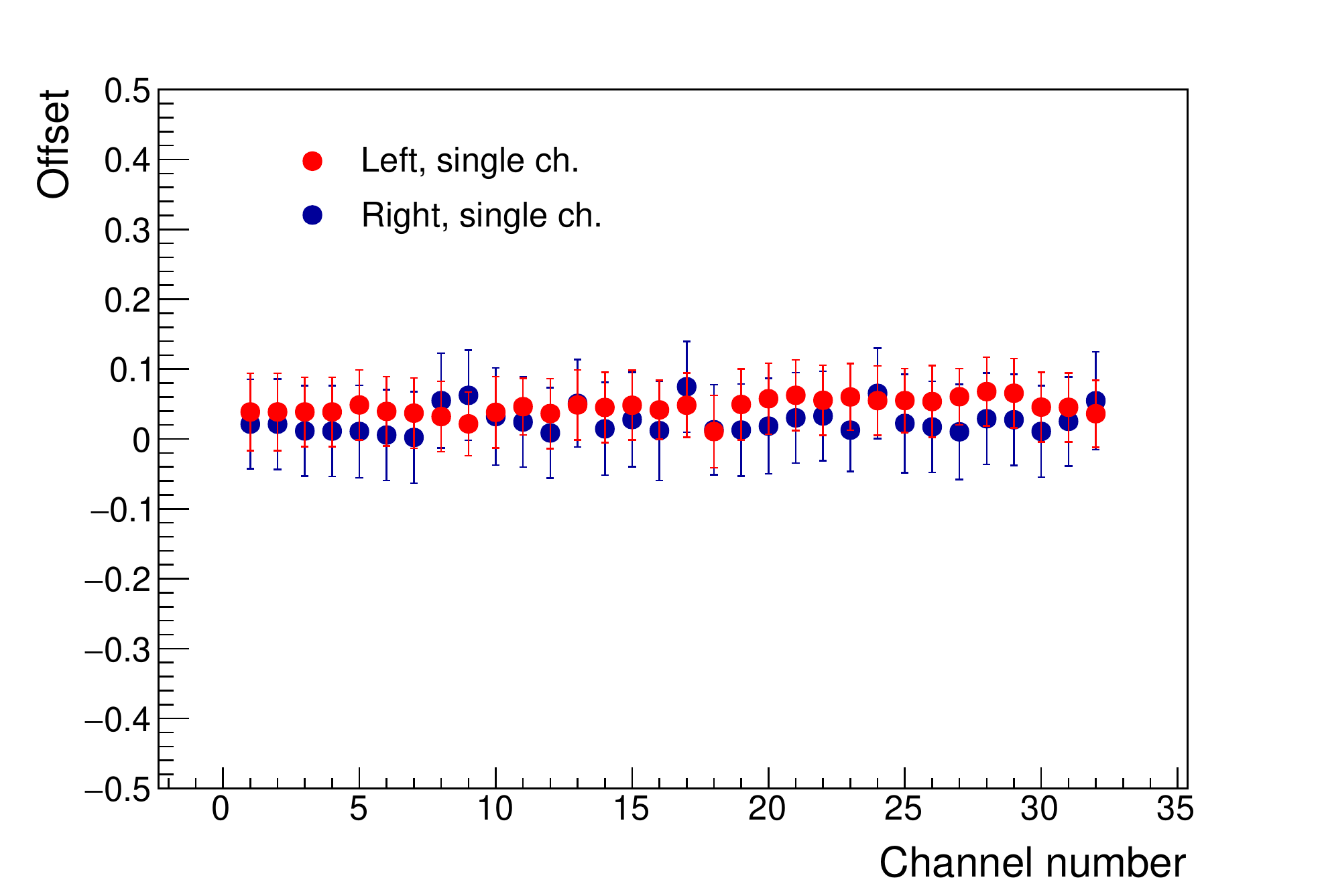}
   \caption{Slope (left) and offset (right) of the linear fit for the two SiPM arrays at the opposite ends of the SciFi ribbon ($2\times 32$ channels).
The variations are due to to differences in the gain of the amplifiers
and the fact that all channels of the SiPM array are operated at the same voltage,
while the difference between the two sensors comes from the fact
that both sensors were operated at the same bias voltage.}
   \label{fig:calib_slope}
\end{figure}

To determine the number of detected scintillation photons, the charge distribution obtained by integrating the waveforms
and shown in Figure~\ref{fig:calib_fit} (left) is normalized to the charge generated by a single photon.
Each peak above the {\it continuum}, coming mainly from the cross-talk between the SiPM cells in the same column,
corresponds to a specific number of detected photons, and it is fitted with a Gaussian.
The peak positions versus the corresponding number of photons $n_{ph}$ are displayed in Figure~\ref{fig:calib_fit} (right).
The charge per detected photon, corresponding to the combined gain of the sensor and the amplifier,
is determined by a linear interpolation,
which validates the linear response of the sensor and of the readout electronics.
In addition, the fit extrapolates to zero validating the correct subtraction of the baseline.
This procedure is applied to each channel.
The relative gain of the channels from the same SiPM array shows an overall uniformity of 8\% (Figure~\ref{fig:calib_slope}).
The small variations are due differences in the gain of the amplifiers
and to the fact that all channels of the SiPM array share the same cathode and are thus operated at the same bias voltage.
The two sensors at the opposite ends of the SciFi ribbon were also operated at the same bias voltage.

\begin{figure}[!t]
   \centering
   \includegraphics[width=0.6\textwidth]{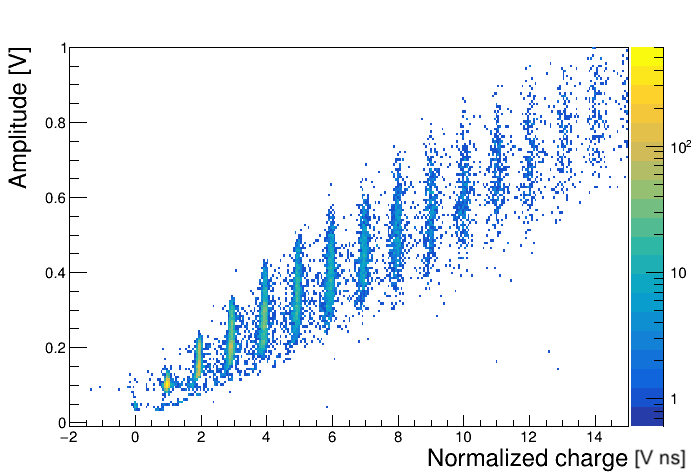}
   \caption{Charge vs maximum signal amplitude for one channels of the SiPM after gain normalization.}
   \label{fig:AmpVsCharge}
\end{figure}

Figure~\ref{fig:AmpVsCharge} shows the correlation between the integrated charge and the maximum signal amplitude
for all active channels\footnote{By active we mean all channels that have passed a threshold equivalent to 1/3~ph.e..
The amplitude of the electrical cross-talk is below this threshold and will not show up when the 1/3~ph.e. selection is applied.}
of the SiPM array from one 32 ch. DRS acquisition board before gain normalization.
The integrated charge represents better the number of detected photons in an event,
since it is based on 400 consecutive samples of the pulse and therefore it is less sensitive to amplitude fluctuations.
The spread in the amplitudes for a specific number of detected photons, which grows with the increasing number of photons,
is driven by the fluctuations in the voltage measurements.
From that one can conclude that the pulse shape is not the same for the same number of detected photons.

\subsection{Determination of Threshold Levels}

\begin{figure}[!b]
   \centering
   \includegraphics[width=0.49\textwidth]{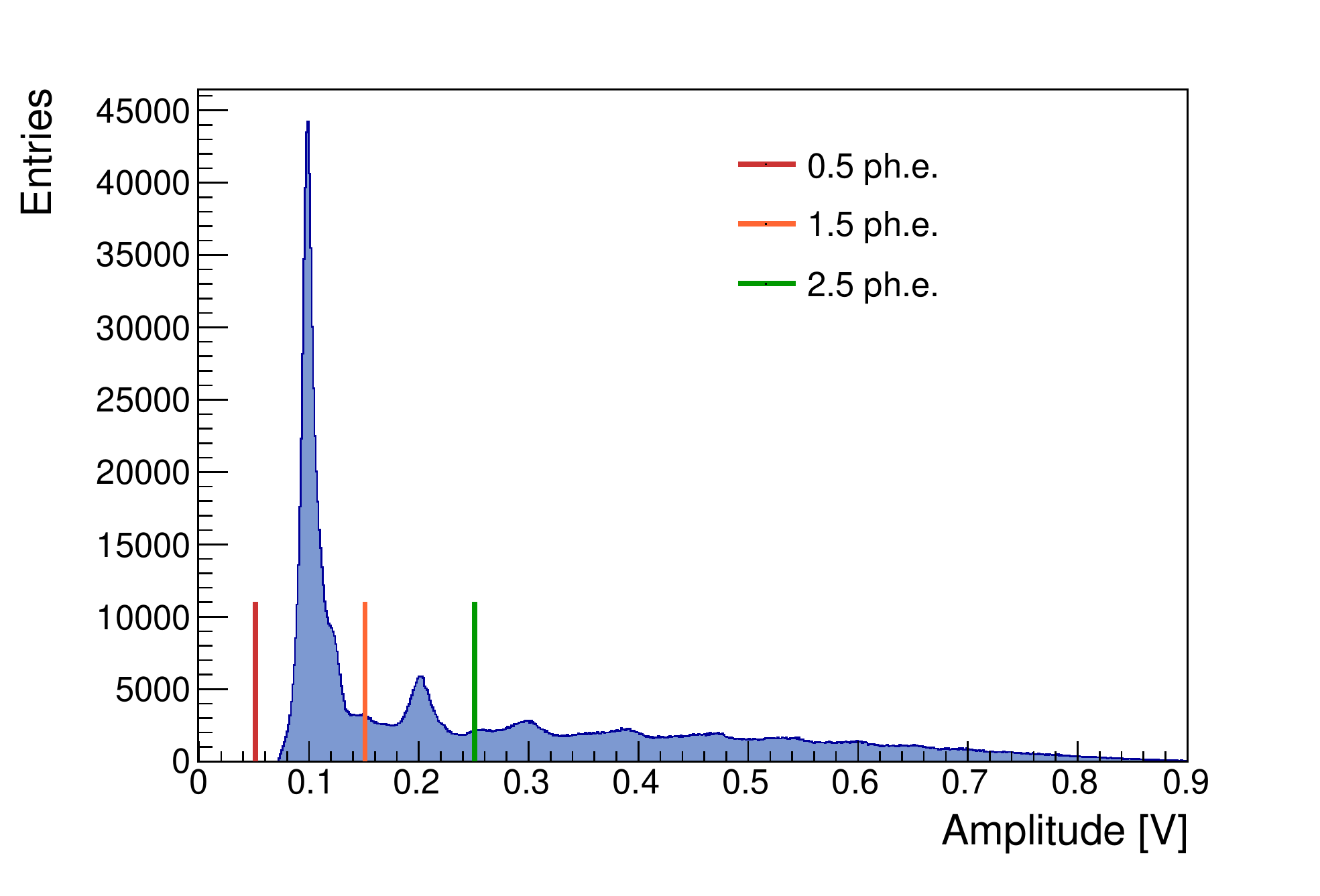}
   \includegraphics[width=0.49\textwidth]{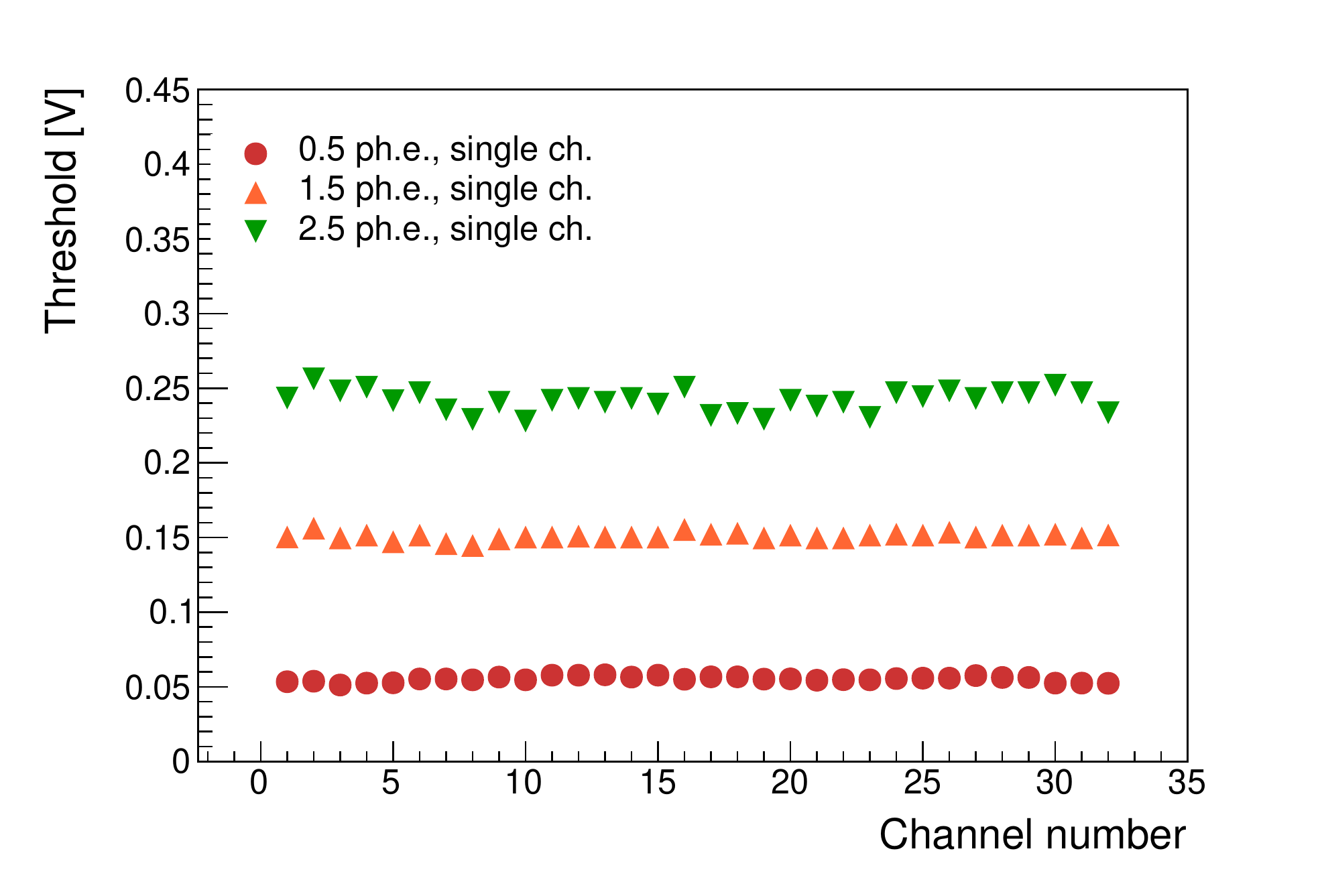}
   \caption{left) Signal maximum amplitude.
Thresholds corresponding to 0.5 ph.e. (red), 1.5 ph.e. (orange) and 2.5 ph.e. (blue) have been determined for each
channel based on this and similar plots.
right) Thresholds for individual channels.}
   \label{fig:amplitude}
\end{figure}

Hit selection is based on the signal's amplitude rather than a charge threshold,
since this information is readily available and it can be used also at trigger level.
This is also the operation mode of the MuTRiG ASIC (see Section~\ref{sec:MuTRiG}),
which will be used in the \mude experiment for SiPM signal digitization.
Figure~\ref{fig:amplitude} (left) shows the maximum amplitude distribution for 32 consecutive channels of the SiPM array after baseline subtraction.
A threshold corresponding to the 1/3~ph.e. amplitude is applied to remove noise and select only events with some {\it activity}. 
Photon peaks are clearly visible in the amplitude spectrum at low photon multiplicities
and are evenly spaced, which implies a linear response to the number of detected photons.
To select channels with a minimum number of detected photons $n_{\rm ph.e.}$,
amplitude thresholds are set between two consecutive peaks in the amplitude distribution
as illustrated in Figure~\ref{fig:amplitude} (left):
to select, for example, a channel with $n_{\rm ph.e.} \ge 2$,
the threshold is set between the 1-photon and 2-photon peaks at a threshold of 1.5 equivalent photo-electrons. 
Amplitude thresholds thus determined are shown in Figure~\ref{fig:amplitude} (right) and are evenly spaced.

\begin{figure}[!t]
   \centering
   \includegraphics[width=0.5\textwidth]{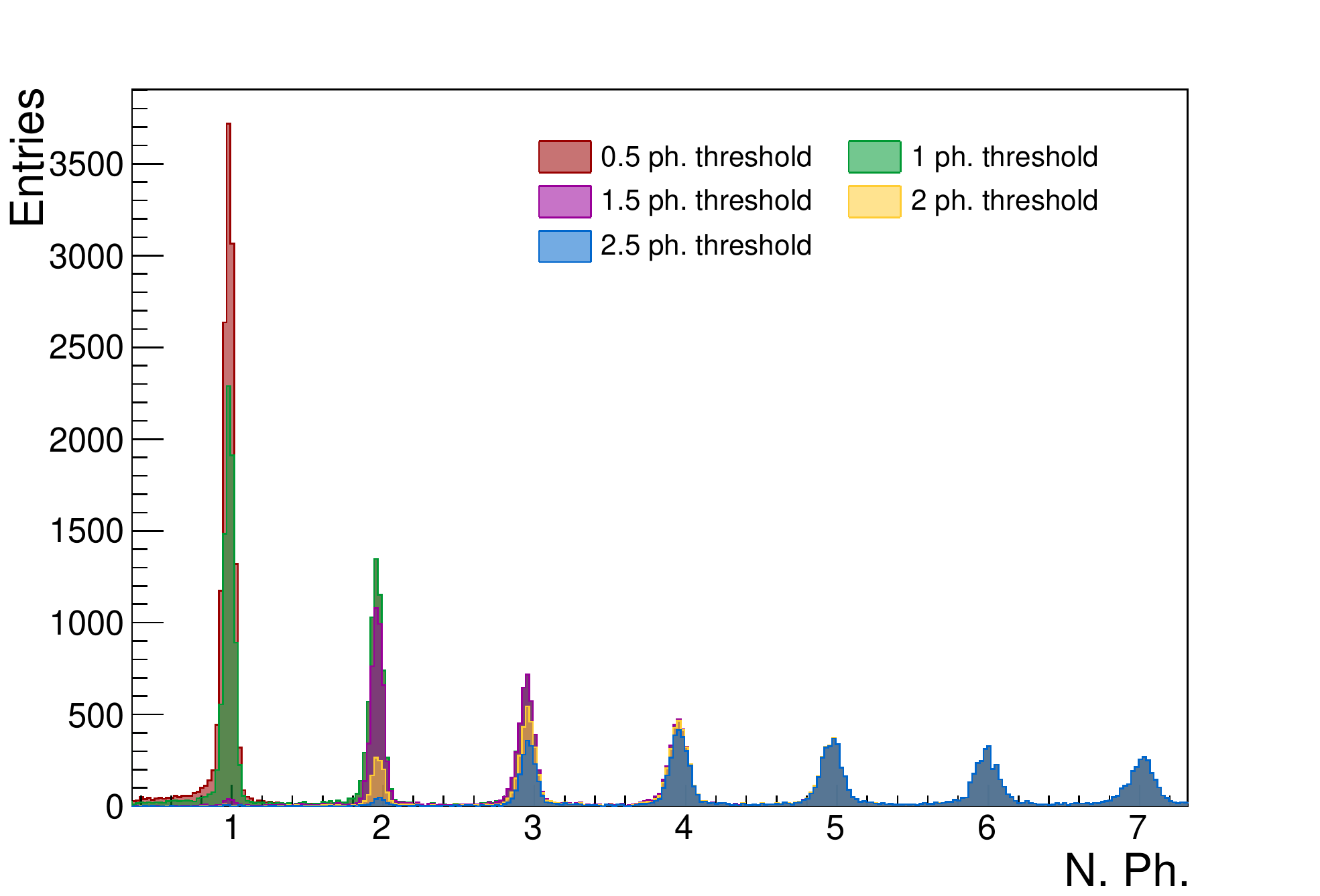}
   \caption{Effect of the amplitude thresholds on the integrated charge distribution for one channel.}
   \label{fig:QvsThr}
\end{figure}

\begin{figure}[b!]
   \centering
   \includegraphics[width=0.49\textwidth]{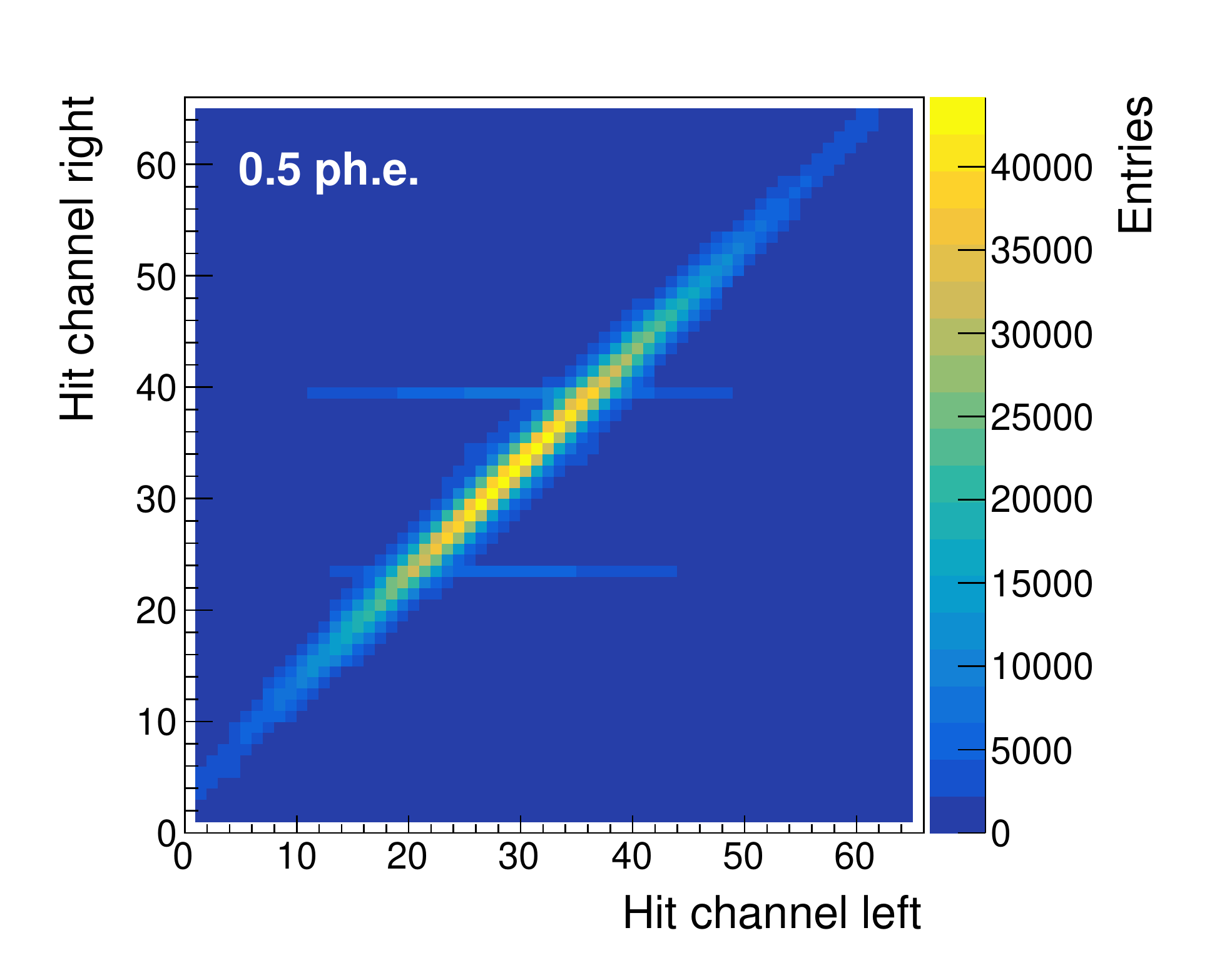}
   \includegraphics[width=0.49\textwidth]{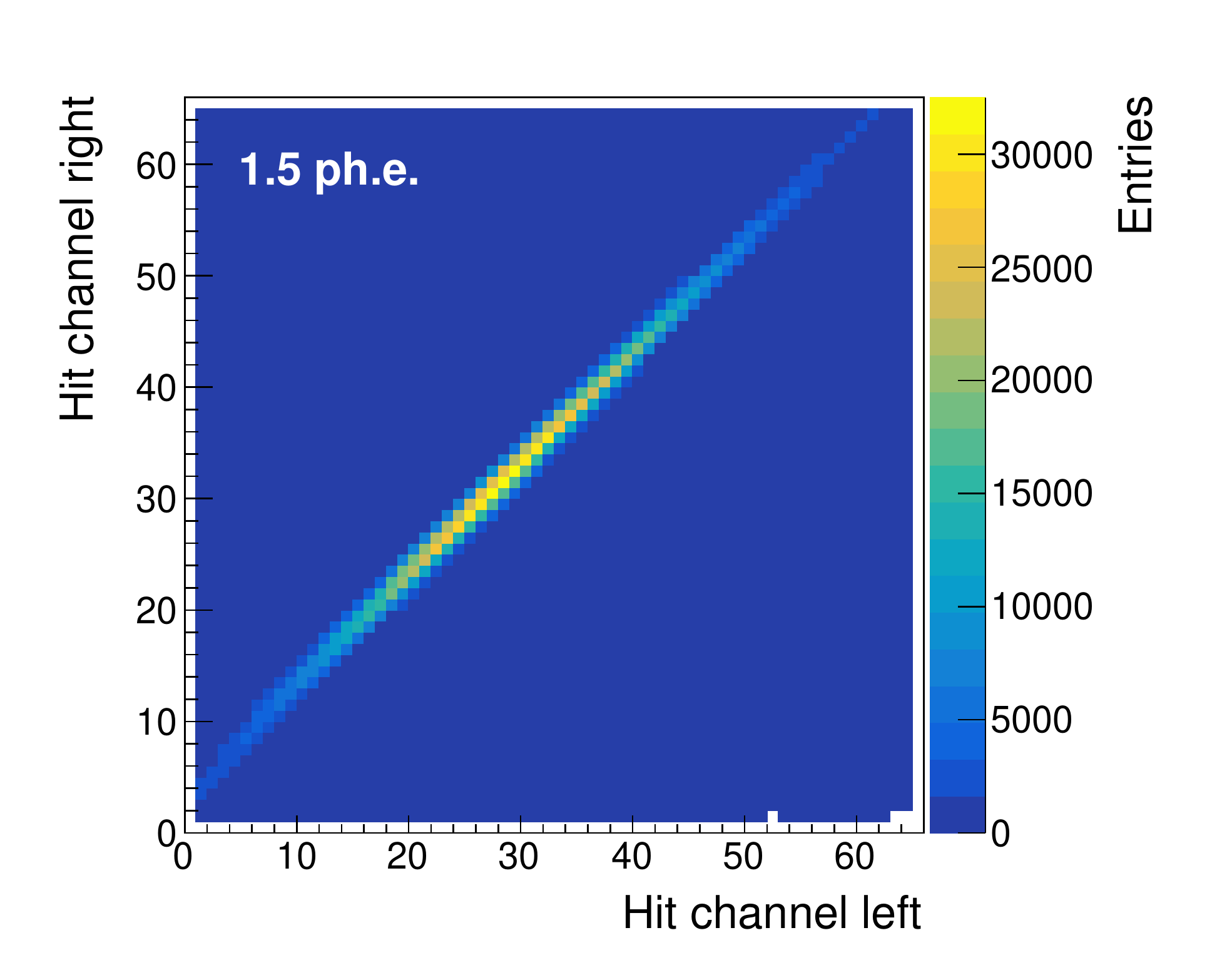}
   \caption{Correlation between the hits generated in the two SiPM arrays at each side of a ribbon exposed to the $\pi$M1 beam 
for a 0.5 ph.e (left) and 1.5 ph.e (right) threshold.
Each column represents an individual SiPM channel, $250~\mu{\rm m}$ wide, that gave a signal.
Some noise is visible at the 0.5 ph.e. threshold.}
   \label{fig:correl}
\end{figure}

To understand better the impact of the amplitude thresholds on the charge distribution of Figure~\ref{fig:calib_fit}
and the number of detected photons $n_{\rm ph.e.}$,
Figure~\ref{fig:QvsThr} shows how the amplitude thresholds bite into the charge distribution;
for instance, for a 1.5 ph.e. threshold about 15\% of 1-photon events pass the selection,
while about 15\% 2-photon events are lost.
The thresholds are particularly relevant for the detection efficiency of the detector discuss in Section~\ref{sec:efficiency}.

\subsection{Left -- Right Correlations}

The light generated by a particle crossing the SciFi ribbon propagates in the ribbon in opposite direction
and it is detected at both ends of the ribbon.
Figure~\ref{fig:correl} shows the correlation between the {\it hits} detected in the two SiPM arrays at opposite ends
for 0.5 ph.e. and 1.5 ph.e. thresholds. 
Note that the SciFi ribbons measure the hit position in the vertical direction (Figure~\ref{fig:setup}), i.e. across the ribbon.
The width of each SiPM channel is $250~\mu{\rm m}$.
The vertical extension of the beam spot impinging on the SciFi ribbon, determined by the vertical size of the trigger scintillators,
is clearly visible in the Figure.
From the width of the correlations one can estimate the cluster width (Section~\ref{sec:cluster}),
i.e. how much the optical signal is spread over several channels of the SiPM array.

\clearpage
\section{Clustering}
\label{sec:cluster}

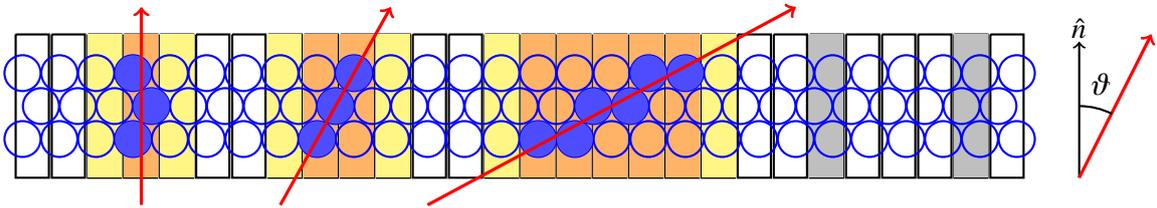
\begin{figure}[!b]
   \centering
   	\begin{tikzpicture}[scale=1.9, thick, font={\sffamily}]
	\foreach \i in {3,...,30}
	     	\draw (-0.115+0.25*\i, 0.5) -- +(0.230, 0) -- +(0.230, -1.0) -- +(0, -1.0) -- +(0,0);
	
     \fill[yellow!60!white] (-0.115+0.25*5, 0.5) -- +(0.230, 0) -- +(0.230, -1.0) -- +(0, -1.0) -- +(0,0);
     \fill[orange!60!white] (-0.115+0.25*6, 0.5) -- +(0.230, 0) -- +(0.230, -1.0) -- +(0, -1.0) -- +(0,0);
     \fill[yellow!60!white] (-0.115+0.25*7, 0.5) -- +(0.230, 0) -- +(0.230, -1.0) -- +(0, -1.0) -- +(0,0);
     
     \fill[yellow!60!white] (-0.115+0.25*10, 0.5) -- +(0.230, 0) -- +(0.230, -1.0) -- +(0, -1.0) -- +(0,0);
     \fill[orange!60!white] (-0.115+0.25*11, 0.5) -- +(0.230, 0) -- +(0.230, -1.0) -- +(0, -1.0) -- +(0,0);
     \fill[orange!60!white] (-0.115+0.25*12, 0.5) -- +(0.230, 0) -- +(0.230, -1.0) -- +(0, -1.0) -- +(0,0);
     \fill[yellow!60!white] (-0.115+0.25*13, 0.5) -- +(0.230, 0) -- +(0.230, -1.0) -- +(0, -1.0) -- +(0,0);
     
     \fill[yellow!60!white] (-0.115+0.25*16, 0.5) -- +(0.230, 0) -- +(0.230, -1.0) -- +(0, -1.0) -- +(0,0);
     \fill[orange!60!white] (-0.115+0.25*17, 0.5) -- +(0.230, 0) -- +(0.230, -1.0) -- +(0, -1.0) -- +(0,0);
     \fill[orange!60!white] (-0.115+0.25*18, 0.5) -- +(0.230, 0) -- +(0.230, -1.0) -- +(0, -1.0) -- +(0,0);
     \fill[orange!60!white] (-0.115+0.25*19, 0.5) -- +(0.230, 0) -- +(0.230, -1.0) -- +(0, -1.0) -- +(0,0);
     \fill[orange!60!white] (-0.115+0.25*20, 0.5) -- +(0.230, 0) -- +(0.230, -1.0) -- +(0, -1.0) -- +(0,0);
     \fill[orange!60!white] (-0.115+0.25*21, 0.5) -- +(0.230, 0) -- +(0.230, -1.0) -- +(0, -1.0) -- +(0,0);
     \fill[yellow!60!white] (-0.115+0.25*22, 0.5) -- +(0.230, 0) -- +(0.230, -1.0) -- +(0, -1.0) -- +(0,0);

     \fill[black!25!white] (-0.115+0.25*25, 0.5) -- +(0.230, 0) -- +(0.230, -1.0) -- +(0, -1.0) -- +(0,0);
     
     \fill[black!25!white] (-0.115+0.25*29, 0.5) -- +(0.230, 0) -- +(0.230, -1.0) -- +(0, -1.0) -- +(0,0);

	\foreach \j in {2,...,29}
		\draw[blue] (0.1725+0.255*\j, 0.23) circle (0.125);
	\foreach \j in {2,...,28}
		\draw[blue] (0.3+0.255*\j, 0.0) circle (0.125);
	\foreach \j in {2,...,29}
		\draw[blue] (0.1725+0.255* \j, -0.23) circle (0.125);
		
	     \fill[blue!70!white] (0.1725+0.255*5, -0.23) circle (0.125);
		\fill[blue!70!white] (0.3+0.255*5, 0.0) circle (0.125);
		\fill[blue!70!white] (0.1725+0.255*5, 0.23) circle (0.125);
		\draw[very thick, red, ->] (0.23+0.255*5, -0.69) -- (0.23+0.255*5, 0.69);
	     \fill[blue!70!white] (0.1725+0.255*10, -0.23) circle (0.125);
		\fill[blue!70!white] (0.3+0.255*10, 0.0) circle (0.125);
		\fill[blue!70!white] (0.1725+0.255*11, 0.23) circle (0.125);
		\draw[very thick, red, ->] (0.1725+0.255*9, -0.69) -- (0.1725+0.255*12, 0.69);
	     \fill[blue!70!white] (0.1725+0.255*16, -0.23) circle (0.125);
	      \fill[blue!70!white] (0.1725+0.255*17, -0.23) circle (0.125);
		\fill[blue!70!white] (0.3+0.255*17, 0.0) circle (0.125);
		\fill[blue!70!white] (0.3+0.255*18, 0.0) circle (0.125);
		\fill[blue!70!white] (0.1725+0.255*19, 0.23) circle (0.125);
		\fill[blue!70!white] (0.1725+0.255*20, 0.23) circle (0.125);
		\draw[very thick, red, ->] (0.1725+0.255*13, -0.69) -- (0.1725+0.255*23, 0.69);

		\draw[thick, black, ->] (8.0, -0.5) -- (8.0, 0.45);
		\draw[very thick, red, ->] (8.0, -0.5) -- (8.5, 0.5);
		\node at (8.0, 0.54) {$\hat{n}$};
		\node at (8.15, 0.15) {\large $\vartheta$};
		\draw[thick, black] (8.0, 0.0) arc (90.0 : 63.0 : 0.5);
	\end{tikzpicture}
   \vspace*{-5mm}
   \caption{Transverse view of a 3-layer SciFi ribbon mapped to a SiPM array.
A particle (red arrow) traversing the ribbon excites some fibers (blue),
which can trigger avalanches in several channels of the SiPM array (orange).
The optical cross-talk spreads the signal to neighboring channels (yellow).
The width of the cluster depends also on the particle's crossing angle (from left to right $0^\circ$, $30^\circ$, and $60^\circ$ w.r.t. the normal).
A {\it dark} count (gray), which is identical to a real signal, should be excluded from the cluster.}
   \label{fig:cluster}
\end{figure}

Particles crossing the SciFi ribbon excite few fibers (as illustrated in Figure~\ref{fig:cluster} in blue)
and produce scintillation photons, which are transported by the same fibers to the photo-sensors.
Because of the mapping of the fibers to the columns of the SiPM array (Figure~\ref{fig:fiberMapping}),
optical cross-talk between the fibers, delta rays, the photon's exit angle, 
light scattering at the optical junction between the SciFi ribbon and the photo-sensor, etc.
the scintillation light signal is spread over several adjacent SiPM columns.
This triggers avalanches in several neighboring channels of the SiPM array
(see Figure~\ref{fig:wfANA}, where several channels show a large activity).
Moreover, in \mude particles will cross the SciFi ribbons at an average angle $\vartheta$ of $27^\circ$ w.r.t the normal to the SciFi ribbon,
which will spread the light signal even further.
Due to cross-talk between the SiPM channels and dark count effects,
fake signals might appear in ``non-signal'' channels, as well.
The goal of the clustering is to identify and group together all SiPM channels excited by the crossing particles
while eliminating any unwanted contributions from accidentally triggered SiPM channels.

The first step in building a cluster is to select candidate SiPM channels, which fulfill the following constraints:
\begin{itemize}
\itemsep0em
   \item the amplitude of the selected channels should pass a predefined threshold level of 0.5 ph.e. or higher
(the same threshold is applied to all channels);
   \item the peak of the signal is matched to the trigger time within $-5$ and $+10~{\rm ns}$ \\ 
   (this requirement removes dark counts below the 0.2\% level at a DCR of 15~kHz);
\end{itemize}

\noindent followed by the cluster formation:

\begin{itemize}
\itemsep0em
   \item only directly contiguous channels among the selected ones are placed in a clusters \\
(a one channel gap to account for inefficiency and dead channels is allowed);
   \item the channel with the highest amplitude is identified as the {\it seed} for further selections;
   \item a minimum hit multiplicity, usually of two, is imposed on the cluster
   (the multiplicity or width of the cluster is a parameter used to eliminate noise events);
   \item the cluster must be matched to a track reconstructed in the SciFi telescope within $\pm 5~\sigma_{\rm res}$,
where $\sigma_{\rm res}$ is the width of the residual distribution shown in Figure~\ref{fig:residuals};
   \item finally, the cluster time is set to the earliest time stamp of the channels assigned to it. 
\end{itemize}

\subsection{SciFi Ribbon Alignment}
\label{sec:SciFi_align}

Before clusters can be matched to a track, the SciFi telescope needs to be aligned.
This is an iterative process, in which first clusters are identified without track matching,
then the SciFi detectors in the telescope are aligned, and finally clusters are matched to tracks.
For the alignment of the SciFi telescope (Figure~\ref{fig:setup}), clusters in all 4 SciFi ribbons are required.
Tracks are reconstructed across the SciFi telescope with straight lines neglecting the multiple scattering
using the clusters identified above:
selected hits are required to pass a predefined threshold of 0.5~photo-electrons,
while the {\it seed} must pass a tighter threshold of 1.5~ph.e.. 
If more than one cluster is present, the one closest to the trigger time is selected.
Edge effects are removed by rejecting the clusters reconstructed less than 4 channels away from the ribbons edges.

\begin{figure}[!b]
   \centering
   \includegraphics[width=0.49\textwidth]{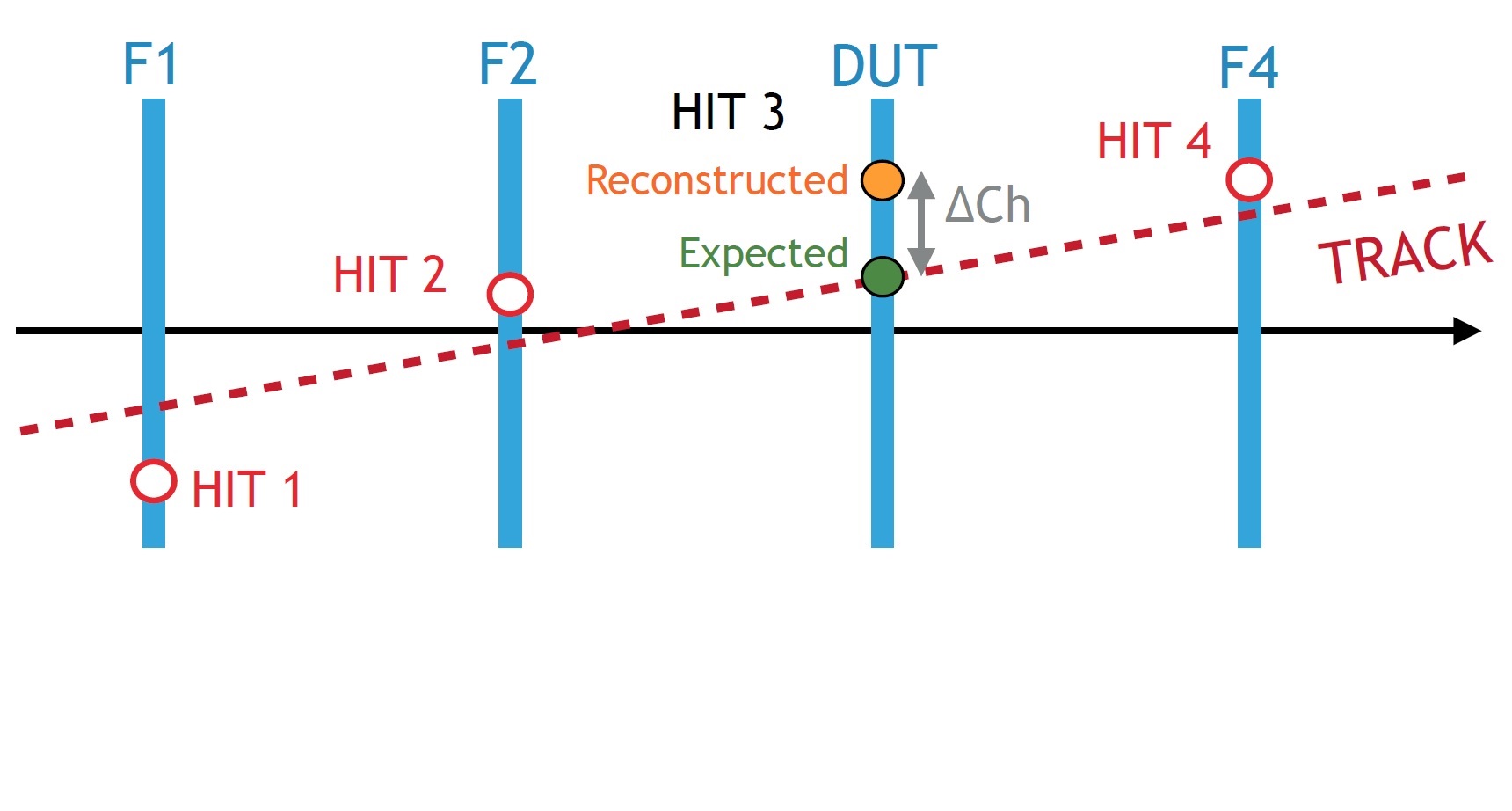}
   \includegraphics[width=0.49\textwidth]{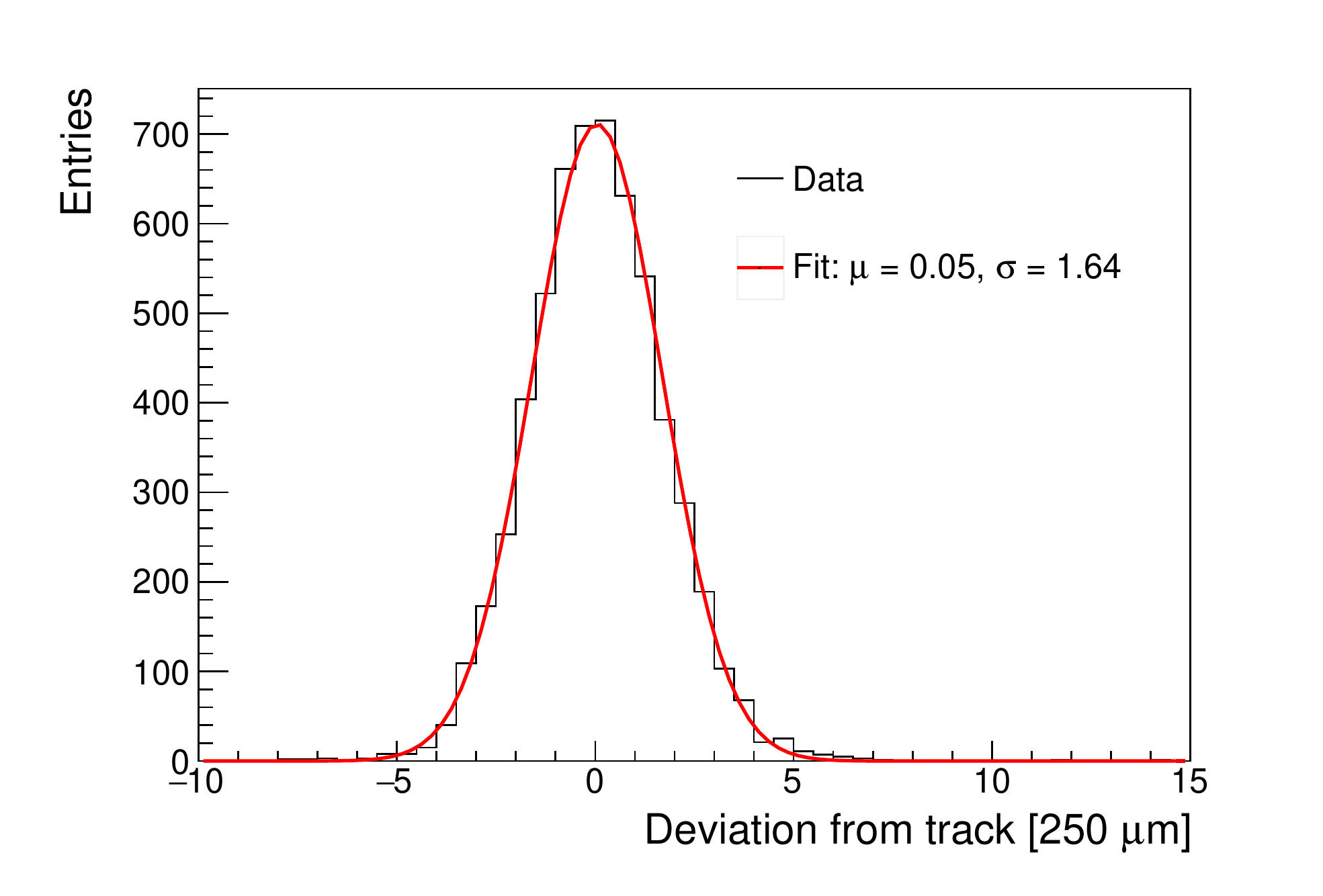}
   \caption{left) Illustration of the SciFi telescope alignment procedure.
right) Distribution of residuals between the clusters reconstructed in the DUT and the tracks reconstructed in the SciFi telescope.
The units are expressed in terms of the SiPM array pitch of $250~\mu{\rm m}$.}
   \label{fig:residuals}
\end{figure}

The SciFi detector of interest or Device Under Test (DUT) is placed placed between the second and fourth most downstream
Scifi detector (Figure~\ref{fig:setup}).
When evaluating the DUT performance, the DUT is excluded from the track reconstruction. 
Figure~\ref{fig:residuals} (right) shows the distribution of residuals for the DUT in units of the SiPM array pitch, i.e. $250~\mu{\rm m}$.
It should be noted that these residuals do not reflect the space resolution of the SciFi detector (Section~\ref{sec:muscatt}),
but the effects of multiple scattering of the low energy pion beam of 161~MeV/$c$ in the SciFi telescope.
The residual distribution is broader than what one would expect from the spatial resolution of $\sim 100~\mu{\rm m}$
of the SciFi detector discussed in Section~\ref{sec:space}.

\subsection{Cluster width}

\begin{figure}[!t]
   \centering
   \includegraphics[width=0.49\textwidth]{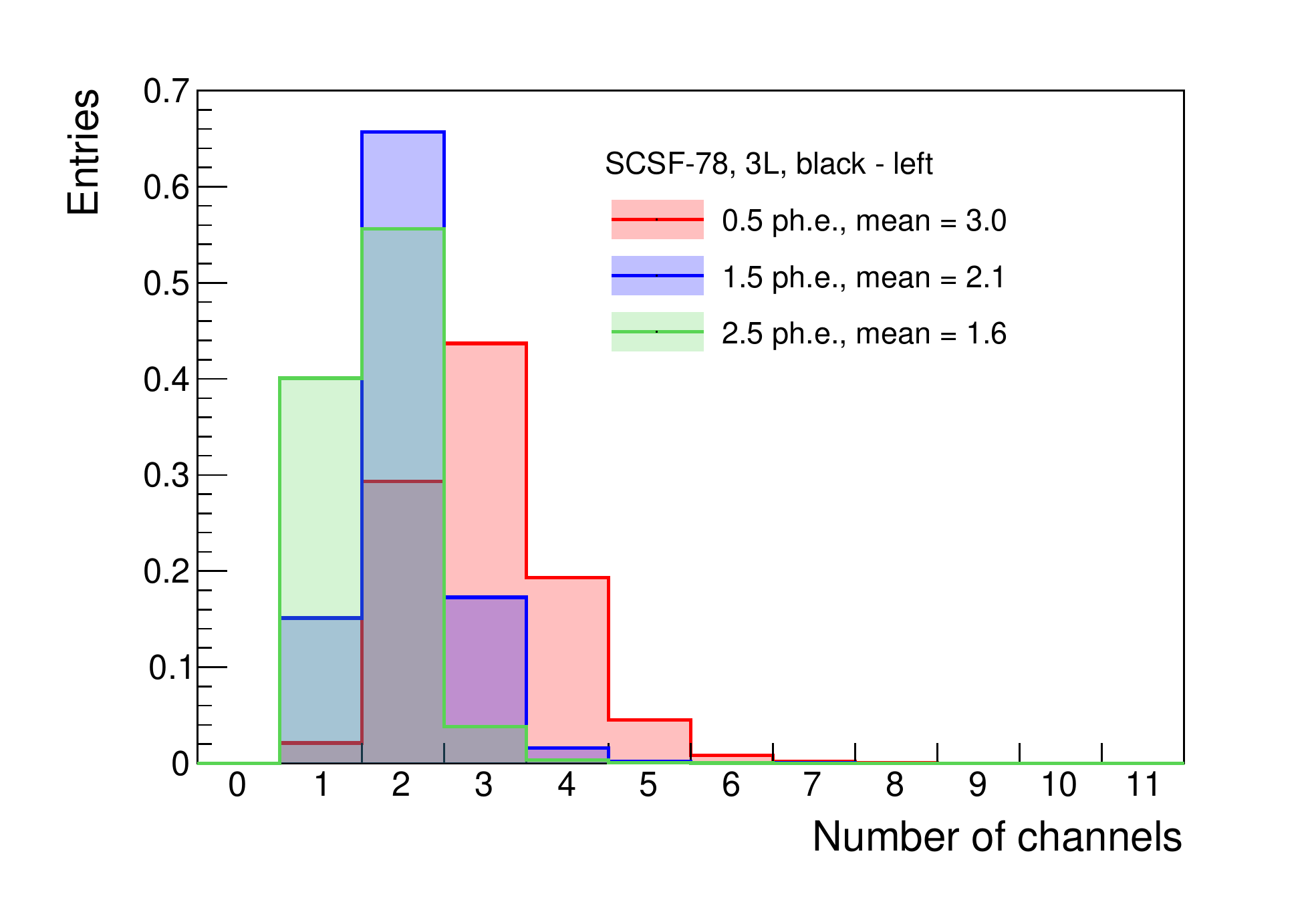}
   \includegraphics[width=0.49\textwidth]{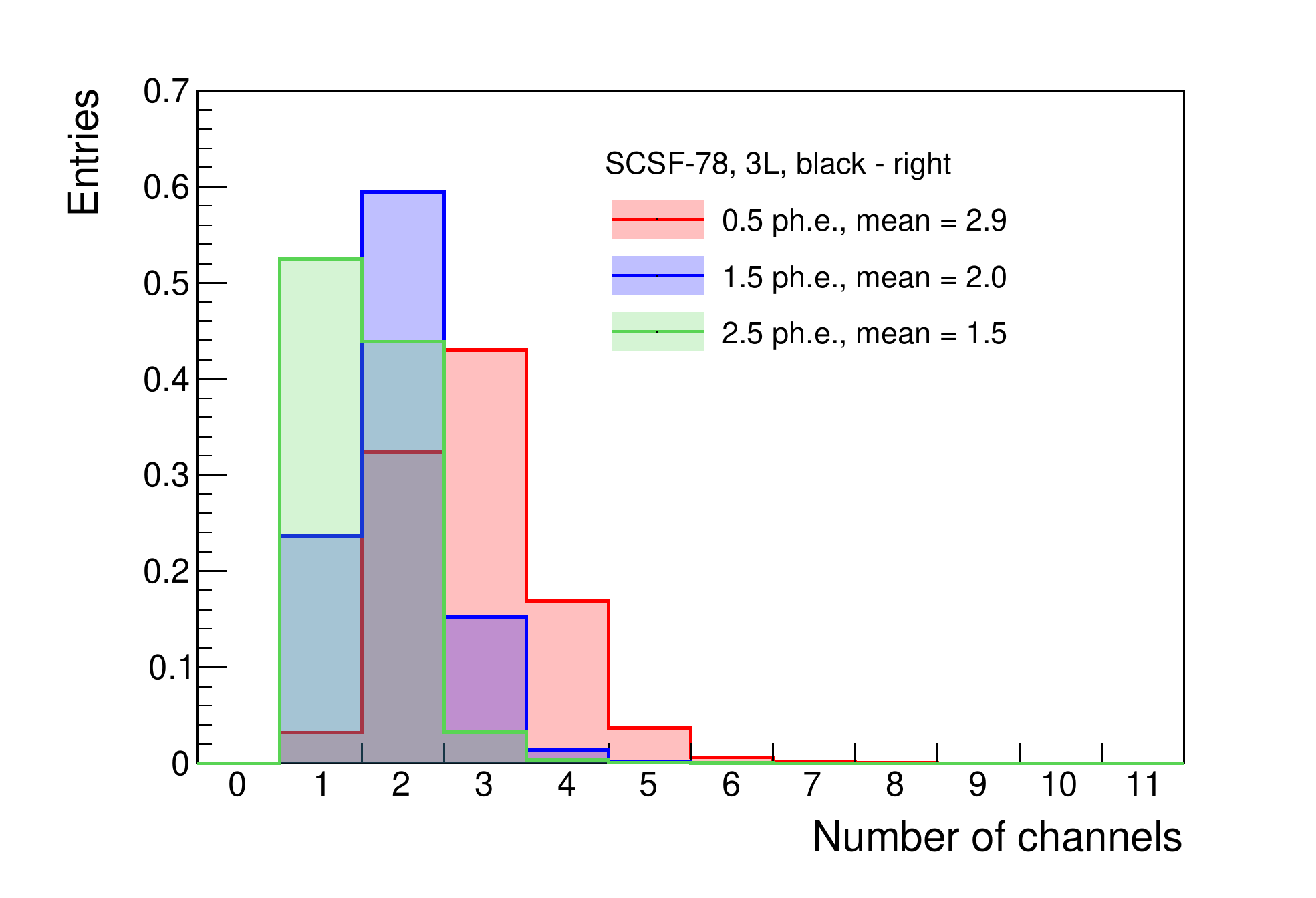}
   \caption{Cluster width for particles crossing a 3-layer SCSF-78 fiber ribbon prepared with black epoxy
for 0.5 ph.e., 1.5 ph.e., and 2.5 ph.e. thresholds 15~cm from the left ribbon's end (left) and 15~cm right ribbon's end (right).
The cluster width is reported w.r.t the mean of the distribution.
The distributions are normalized to the total number of events.}
   \label{fig:cl_size}
\end{figure}

\begin{figure}[!b]
   \centering
   \includegraphics[width=0.49\textwidth]{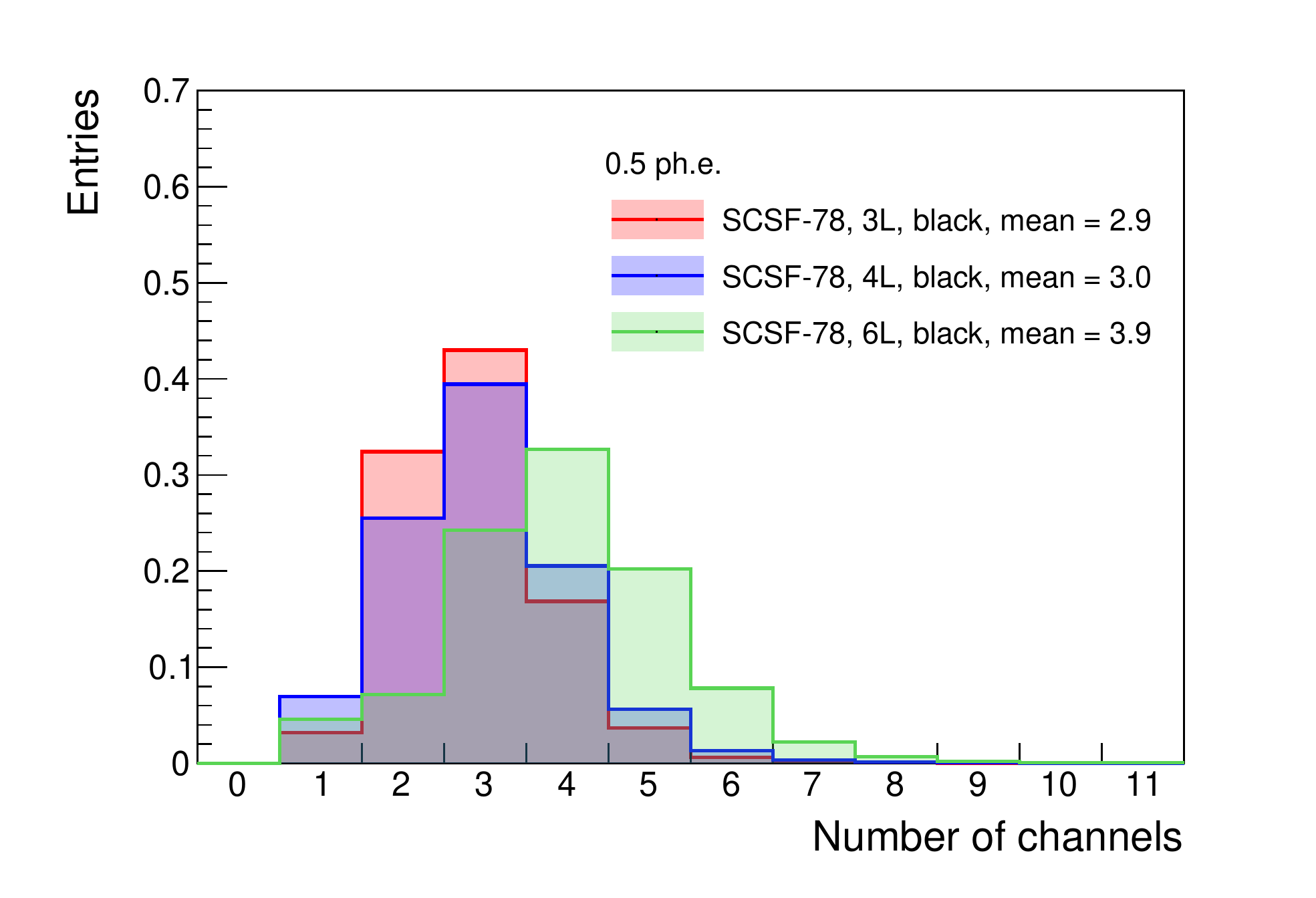}
   \includegraphics[width=0.49\textwidth]{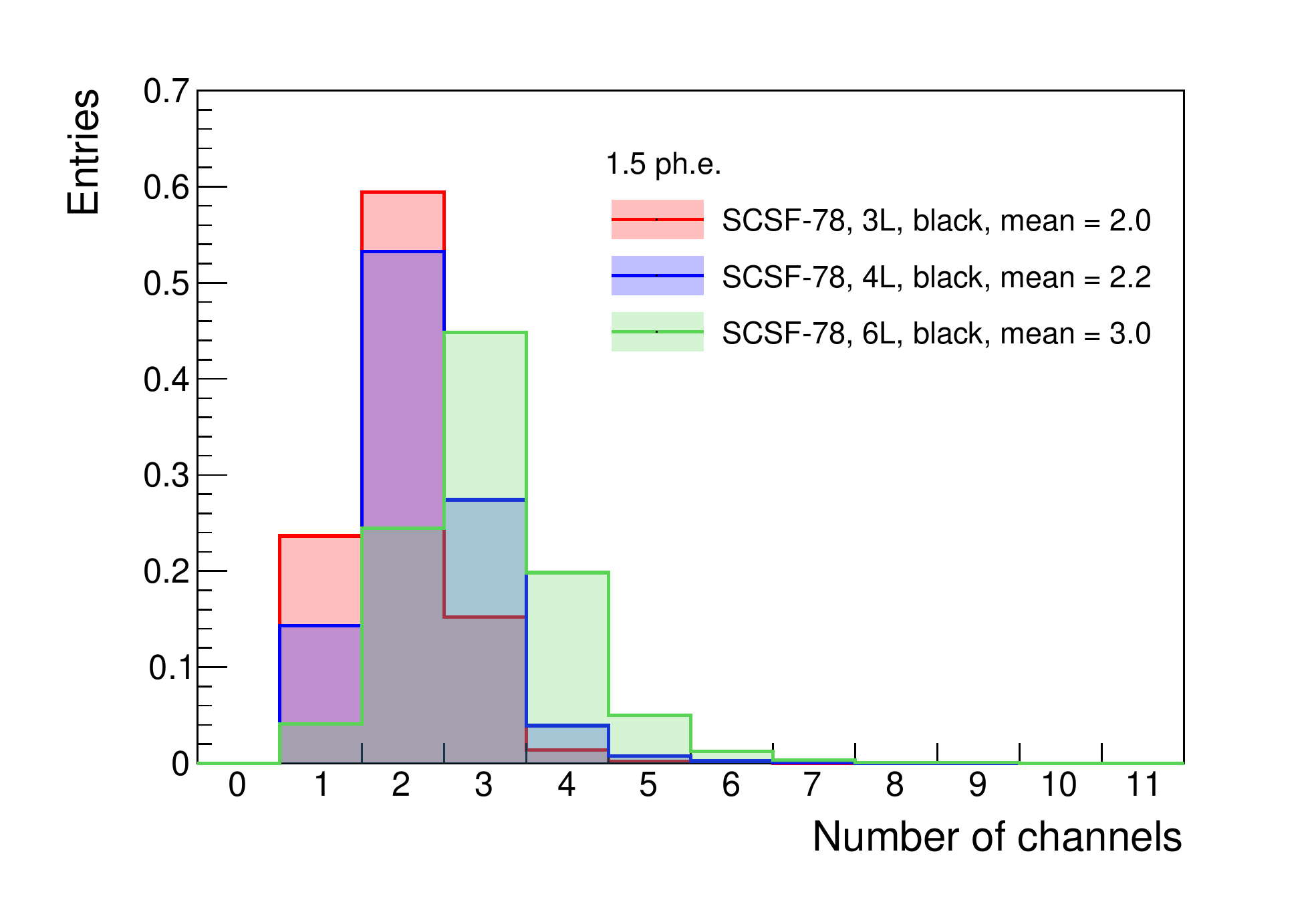}
   \includegraphics[width=0.49\textwidth]{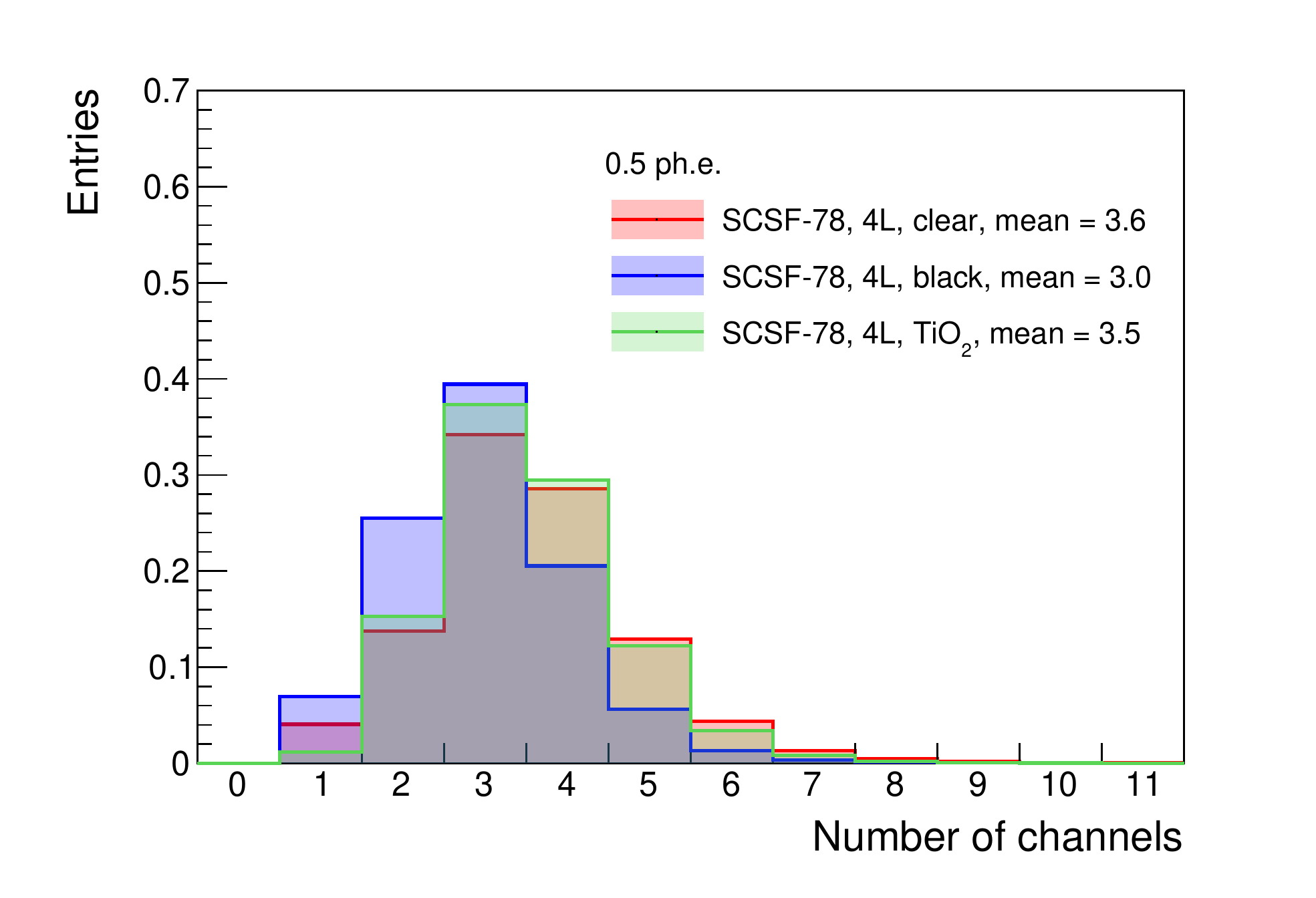}
   \includegraphics[width=0.49\textwidth]{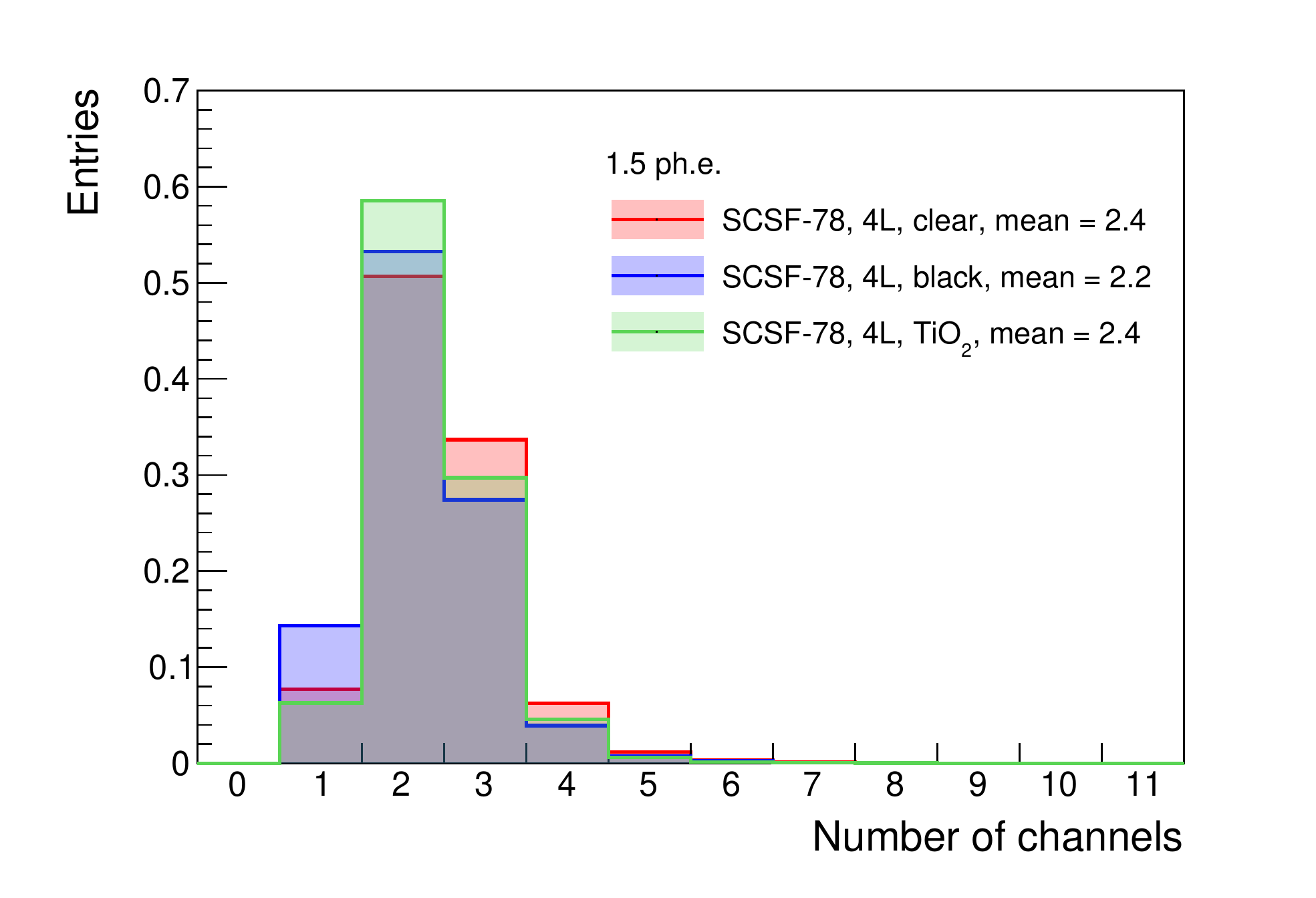}
   \caption{Comparison of cluster widths for 0.5 ph.e. (left) and 1.5 ph.e. (right) thresholds for
top) 3-, 4- and 6-layer SCSF-78 fiber ribbons prepared with black epoxy, and
bottom)  4-layer SCSF-78 ribbons prepared with different adhesives (clear, black, and clear with a 20\% \TiO admixture).}
   \label{fig:cl_compare}
\end{figure}

The clusters thus reconstructed are shown in Figure~\ref{fig:cl_size} for a 3-layer SCSF-78 fiber ribbon prepared with black epoxy
for 0.5 ph.e., 1.5 ph.e., and 2.5 ph.e. thresholds.
The same threshold is applied to all channels in the cluster.
Beam particles cross the SciFi ribbon in the middle, i.e. 15~cm from the edges and impinge vertically on the ribbon.
The cluster width is reported with respect to the mean of the cluster distribution.
The vertical scale in Figure~\ref{fig:cl_size} and in the following Figures is normalized to the total number of analyzed events.
As the threshold increases, the cluster shrinks, since less channels pass the cluster selection criteria.
Typically, these are channels on the outskirts of the cluster characterized by smaller signal amplitudes.
From geometry consideration, on average 2 consecutive SiPM columns are excited (Figure~\ref{fig:cluster}),
while broader clusters are observed.
This is driven by the optical cross-talk, which spreads the optical signal.
The cluster widths at both SciFi ribbon ends are also compared in Figure~\ref{fig:cl_size}.
It can be seen that they are comparable.
The precise calibration of the threshold levels assures a uniform response on both sides of the SciFi ribbon
and indirectly supports the reproducibility of the measurement.

\begin{figure}[!t]
   \centering
   \includegraphics[width=0.49\textwidth]{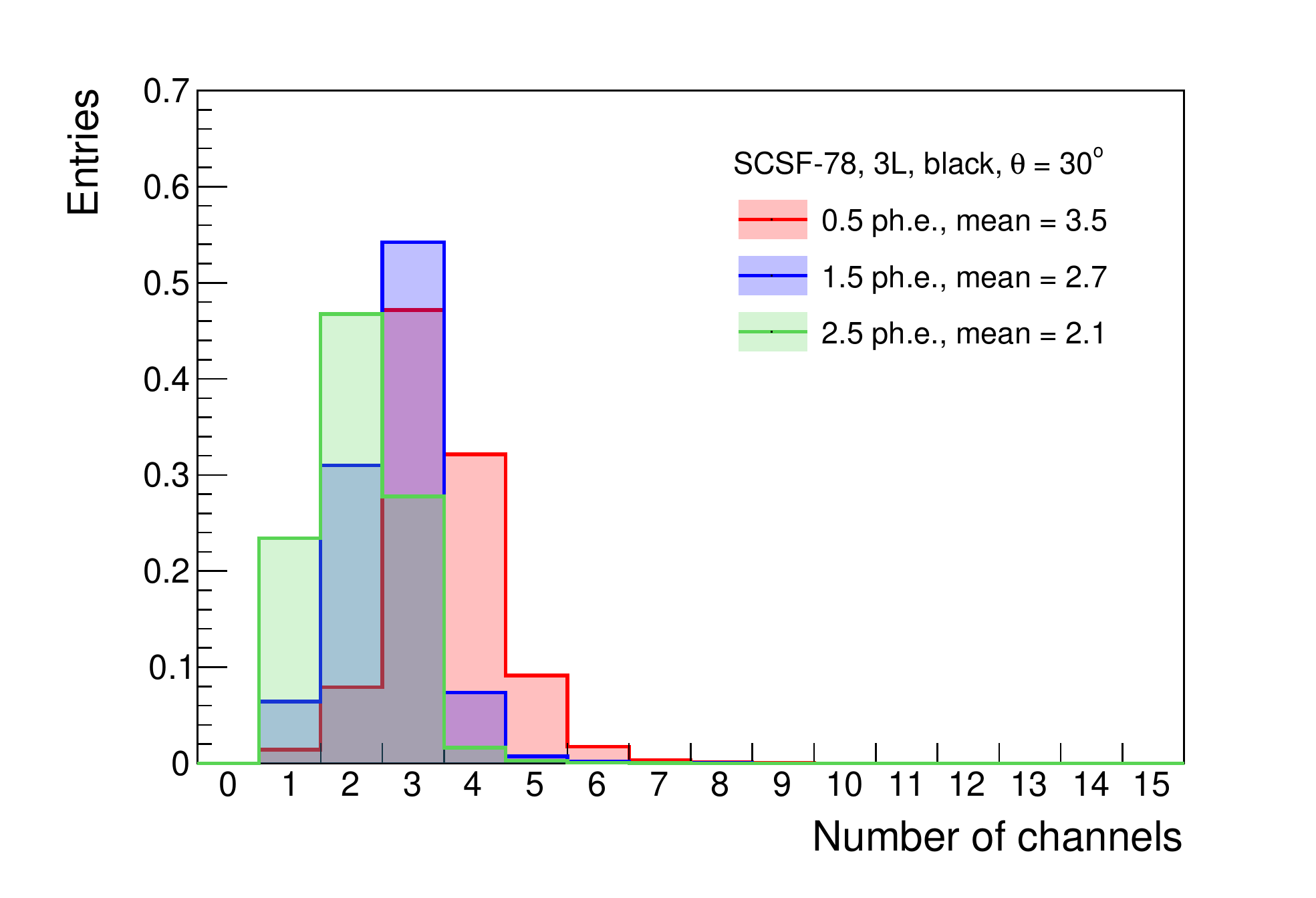}
   \includegraphics[width=0.49\textwidth]{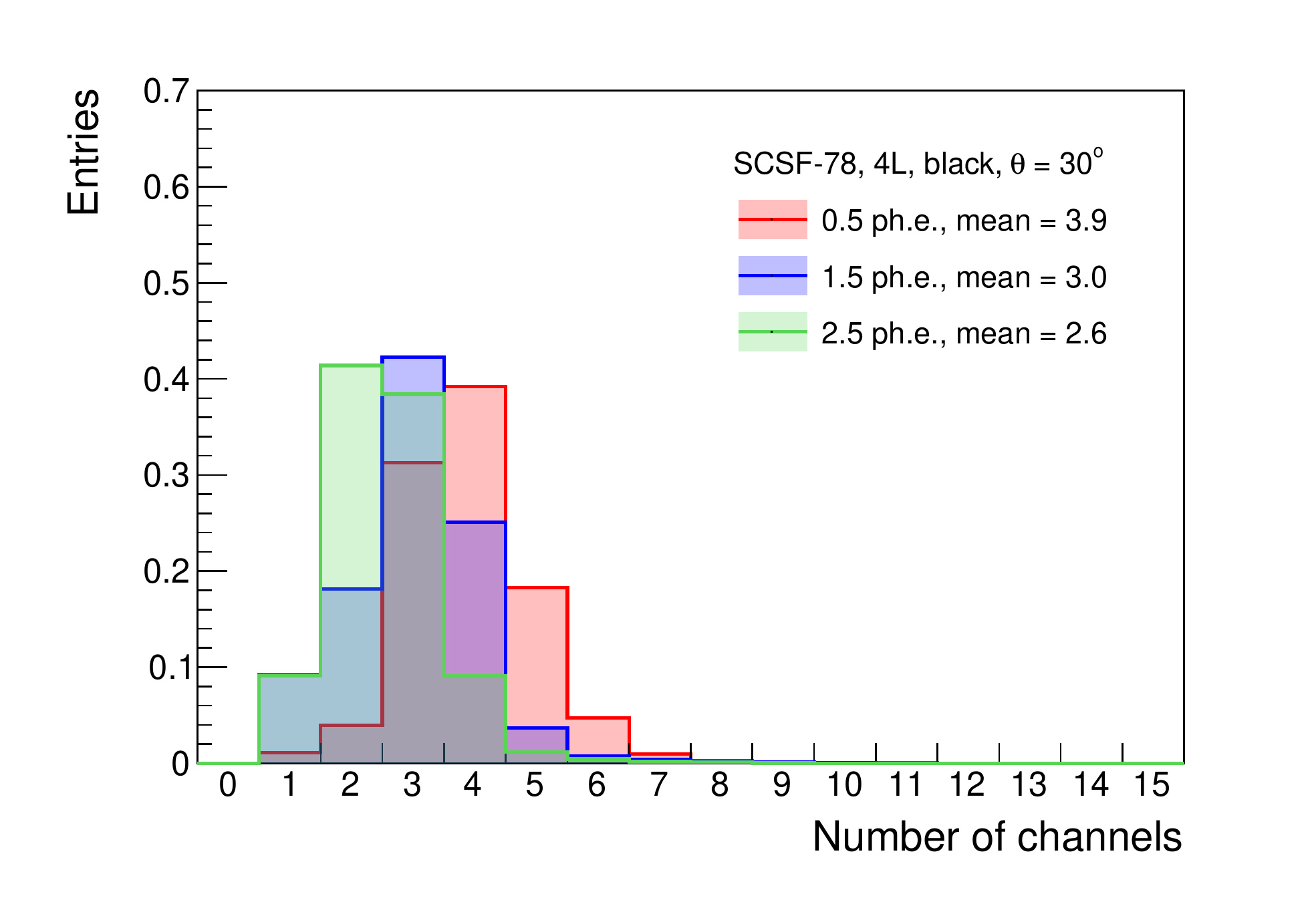}
   \includegraphics[width=0.49\textwidth]{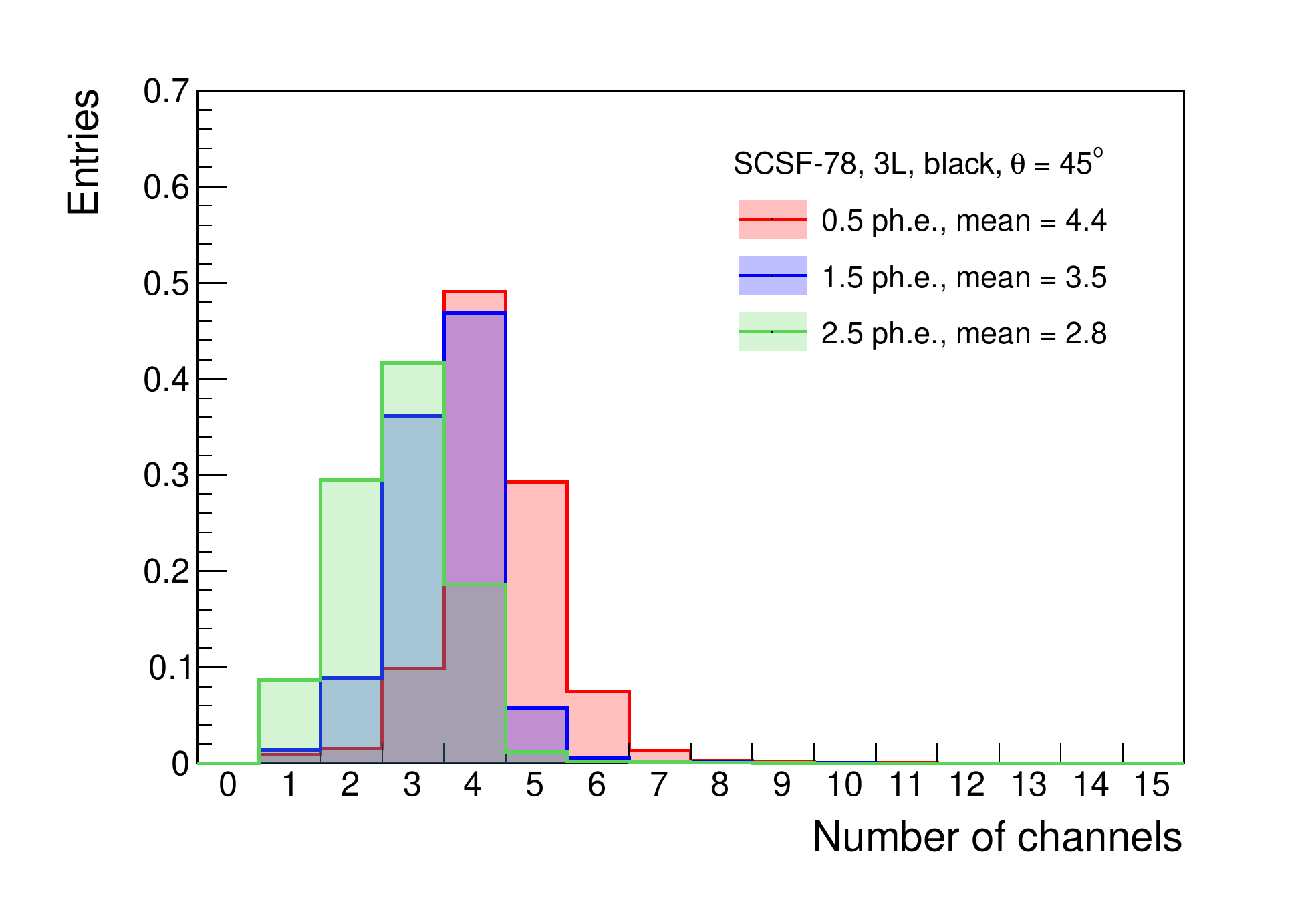}
   \includegraphics[width=0.49\textwidth]{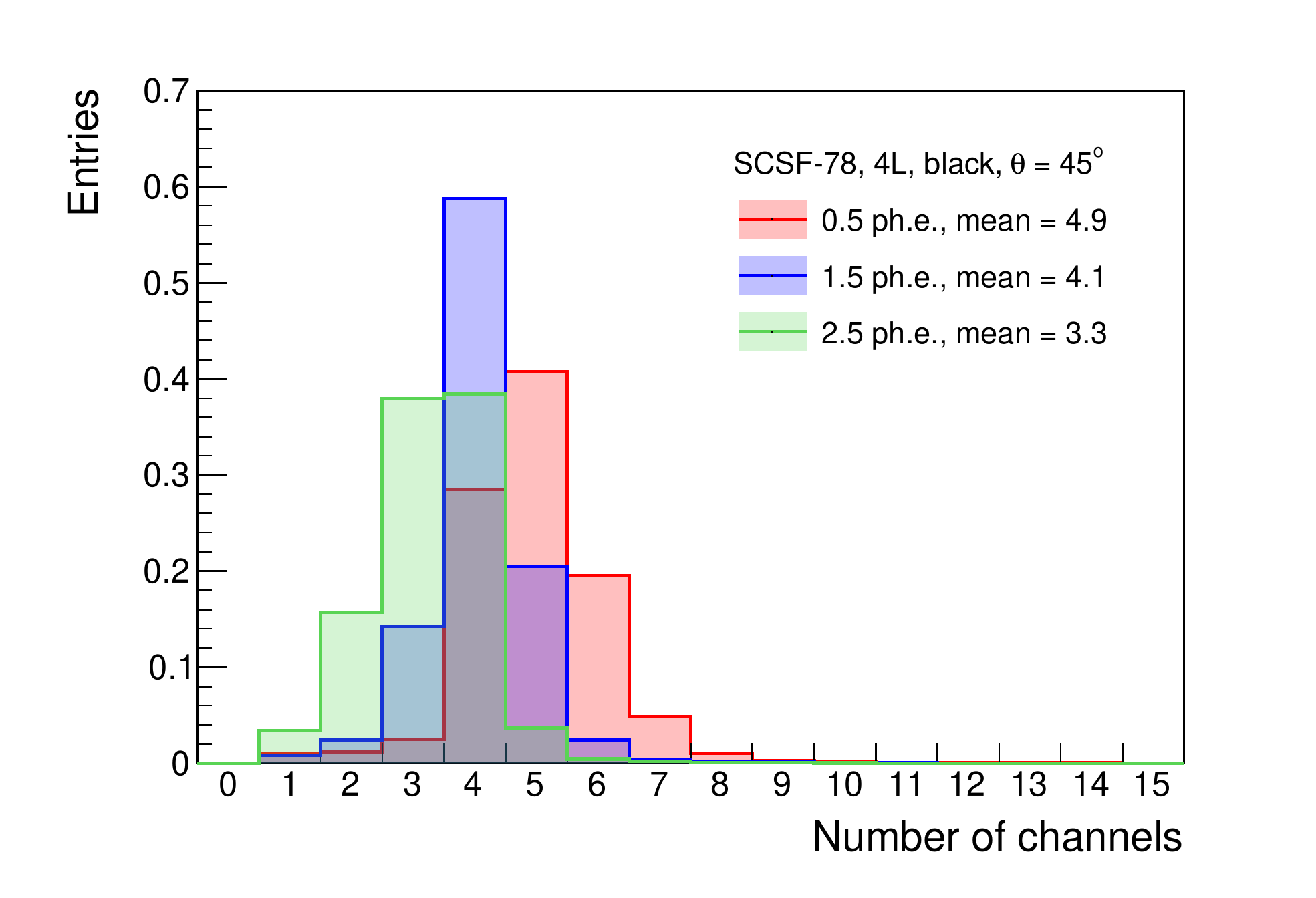}
   \includegraphics[width=0.49\textwidth]{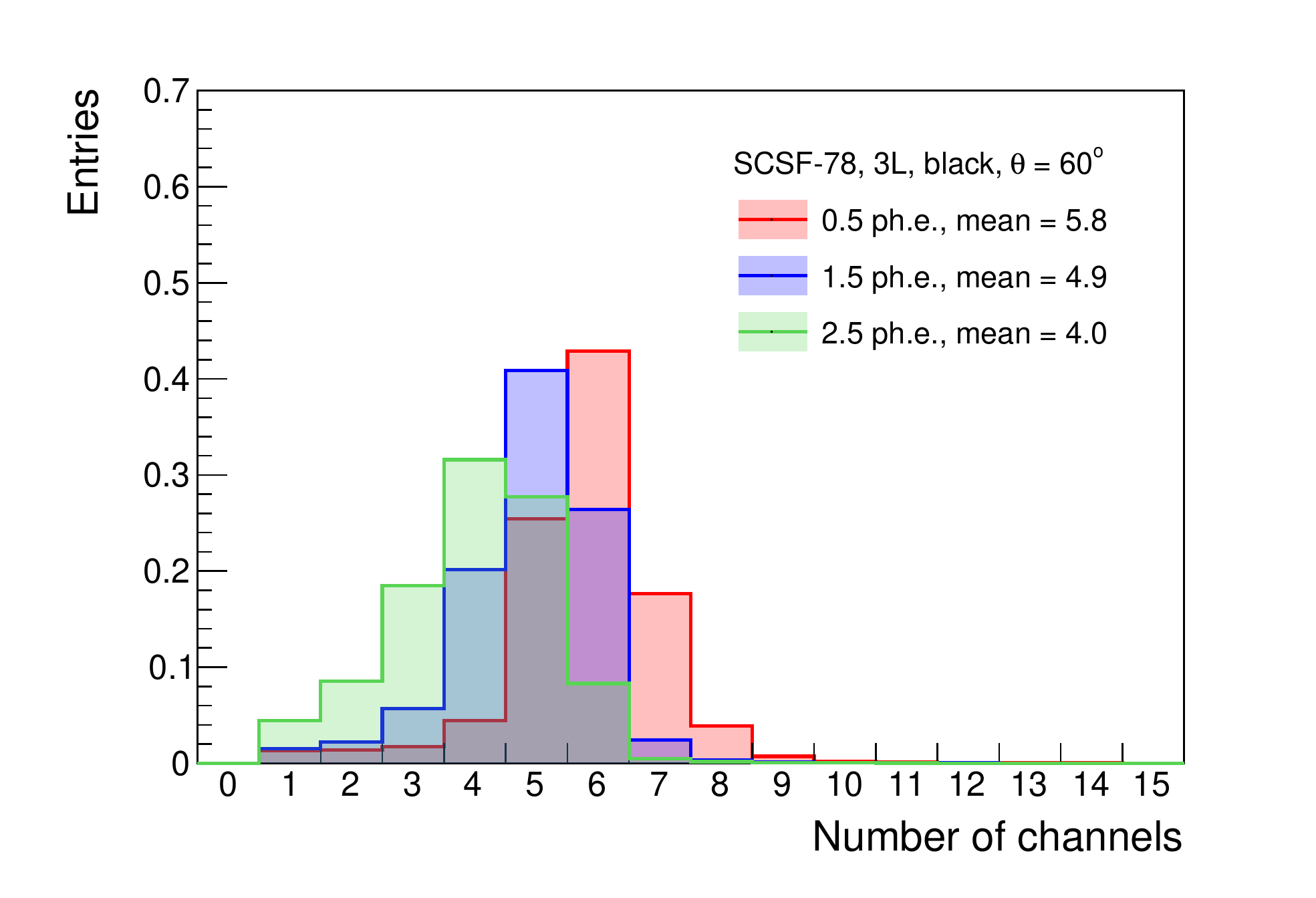}
   \includegraphics[width=0.49\textwidth]{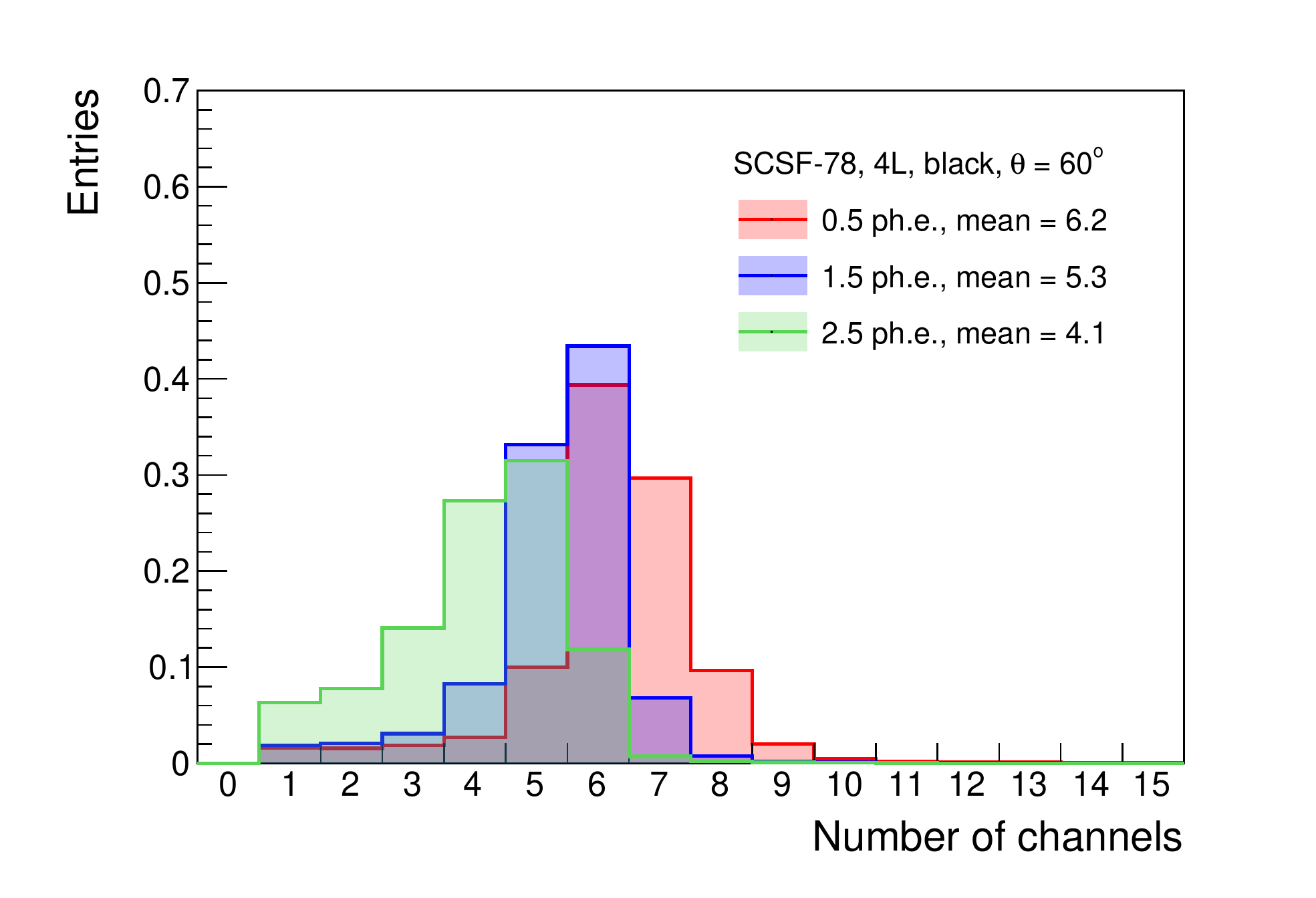}
   \caption{Cluster width for different crossing angles
for a 3- (left) and 4-layers (right) SCSF-78 fiber ribbon prepared with black epoxy.
The ribbon is rotated around a horizontal axis perpendicular to the beam.
The $0^\circ$ crossing angle is already shown in Figure~\ref{fig:cl_size}.}
   \label{fig:cl_angle}
\end{figure}

The cluster widths for 3-, 4-, and 6-layer SciFi ribbons are compared in Figure~\ref{fig:cl_compare} (top).
The cluster widths do not change much for particles impinging normally on the SciFi ribbons
confirming the good alignment of the fibers (Figure~\ref{fig:ribbonGeom})
and supporting the idea that the optical cross-talk is limited to adjacent channels. 
Figure~\ref{fig:cl_compare} (bottom) compares the cluster widths for ribbons prepared with different adhesives
(clear, black, and clear with a 20\% \TiO admixture).
The biggest reduction in the cluster width, although of $\sim 0.5$ channels only,
is observed for the SciFi ribbons prepared with the black epoxy.
This implies also the reduction of the optical cross-talk between fibers using the black epoxy.
As it will be discussed in Section~\ref{sec:LY}, the different types of adhesive do not affect much the light yield of the SciFi ribbons
nor its performance.

\subsection{Angular Scans}

Particles produced at the \mude target in muon decays will cross the SciFi ribbons at an average angle of $27^\circ$
w.r.t. to the normal to the ribbon in the transverse direction (i.e. across the ribbon),
ranging from $10^\circ$ up to virtually $90^\circ$,
because of the bending in the magnetic field.
This increases the spread of the light signal over several columns as illustrated in Figure~\ref{fig:cluster}.
Figure~\ref{fig:cl_angle} shows the cluster width for different incident angles (and thresholds, as well)
for a 3-layer SCSF-78 fiber ribbon prepared with black epoxy.
The cluster width scales as $1 / \cos \vartheta$.
To reduce the cluster width at large crossing angles, one has to reduce the number of staggered fiber layers.
This is one of the reasons for selecting 3-layer SciFi ribbons.

\subsection{Cross-talk}

\begin{figure}[!t]
   \centering
   \includegraphics[width=0.33\textwidth]{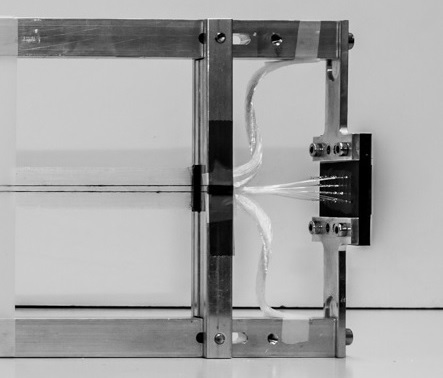}
   \hspace*{5mm}
   \includegraphics[width=0.28\textwidth]{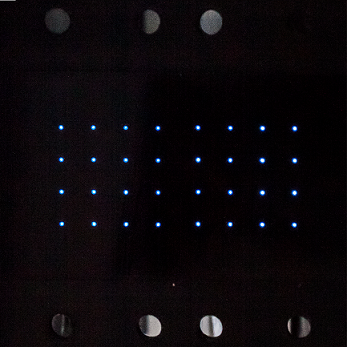}
   \hspace*{5mm}
   \includegraphics[width=0.30\textwidth]{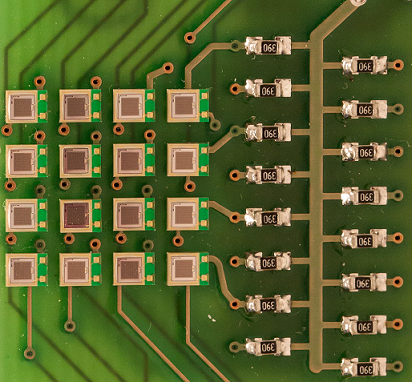}
   \caption{Individual fiber readout: each fiber in the Scifi ribbon (left) is placed in a socket (middle) and coupled to a single channel SiPM (right).}
   \label{fig:sf_setup}
\end{figure}

Optical cross-talk between fibers is observed when scintillation photons originating from one fiber
are trapped and propagate to the photo-sensor through another, usually neighboring fiber.
To investigate in detail the optical cross-talk,
specially prepared SciFi ribbons, in which each fiber is coupled individually to a single channel SiPM, have been produced
(Figure~\ref{fig:sf_setup}).
Two different SciFi ribbons have been prepared, one using clear epoxy, the second by admixing \TiO powder, 20\% by weight, to the clear epoxy.

\begin{figure}[!h]
   \centering
   \includegraphics[width=0.49\textwidth]{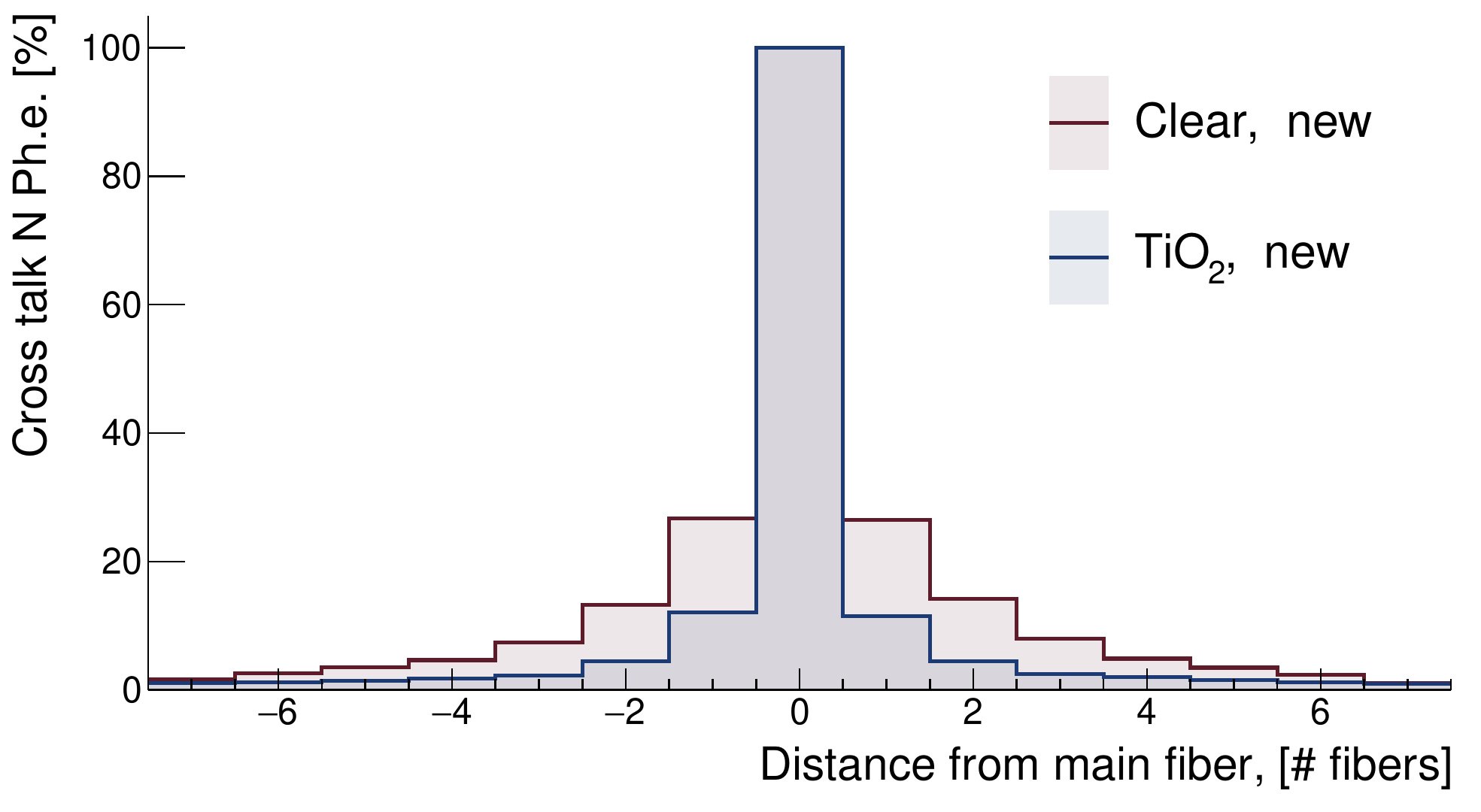}
   \caption{Fraction of events in which scintillation photons have been detected in neighboring fibers in the same layer in conjunction to the seed fiber.
The central bin (bin 0) corresponds to the seed fiber in each event.
A 0.5 ph.e. threshold has been applied to all SiPM channels.
The distribution is normalized to the total number of analyzed events.
right) Single layer cluster widths for a 0.5 ph.e. threshold to validate the hits in the SiPMs.}
   \label{fig:sf_xtalk}
\end{figure}

The crossing point, after cluster identification,
is set in the fiber with the largest signal amplitude (i.e. the {\it seed} fiber, the fiber with the highest number of detected scintillation photons)
in each fiber layer of the SciFi ribbon.
A 0.5 ph.e. threshold has been applied to all SiPM channels.
Figure~\ref{fig:sf_xtalk}
shows the fraction of events in which photons have been detected in fibers adjacent to the seed fiber
as a function of the distance from the seed fiber.
If photons are detected in neighboring fibers in the same layer around the {\it seed} fiber,
these fibers are added to a histogram bin depending on the distance from the seed fiber.
The seed fiber is assigned to the bin centered at 0.
The distribution is normalized to the total number of analyzed events and it is symmetric around the seed fiber. 
In absence of cross-talk, only one fiber would give a signal.
As it can be observed, admixing the \TiO powder to the clear epoxy reduces the cross-talk between the fibers by about 20\%.
This relatively low cross-talk reduction is due to the low amount of adhesive applied in preparing the SciFi ribbon
in order to minimize the amount of inactive materials.
Comparison with Figures~\ref{fig:cl_size} and~\ref{fig:cl_compare} suggests that most of the cross-talk is generated in the optical interface between
the SciFi ribbon and the SiPM array and not between the fibers.

\clearpage
\section{Light Yield}
\label{sec:LY}

To determine the light yield of a SciFi ribbon we have to sum over all active channels of the SiPM array,
since the light signal is spread over several channels, as discussed in Section~\ref{sec:cluster}.
The channel selection procedure is the same as for the cluster formation:
we require that the selected channels pass a threshold of 0.5 ph.e.,
and that they are matched to the trigger time and to a track reconstructed in the SciFi telescope.
However, there is no minimum multiplicity requirement, i.e. the cluster can consist of only one channel. 

\begin{figure}[b!]
   \centering
   \includegraphics[width=0.80\textwidth]{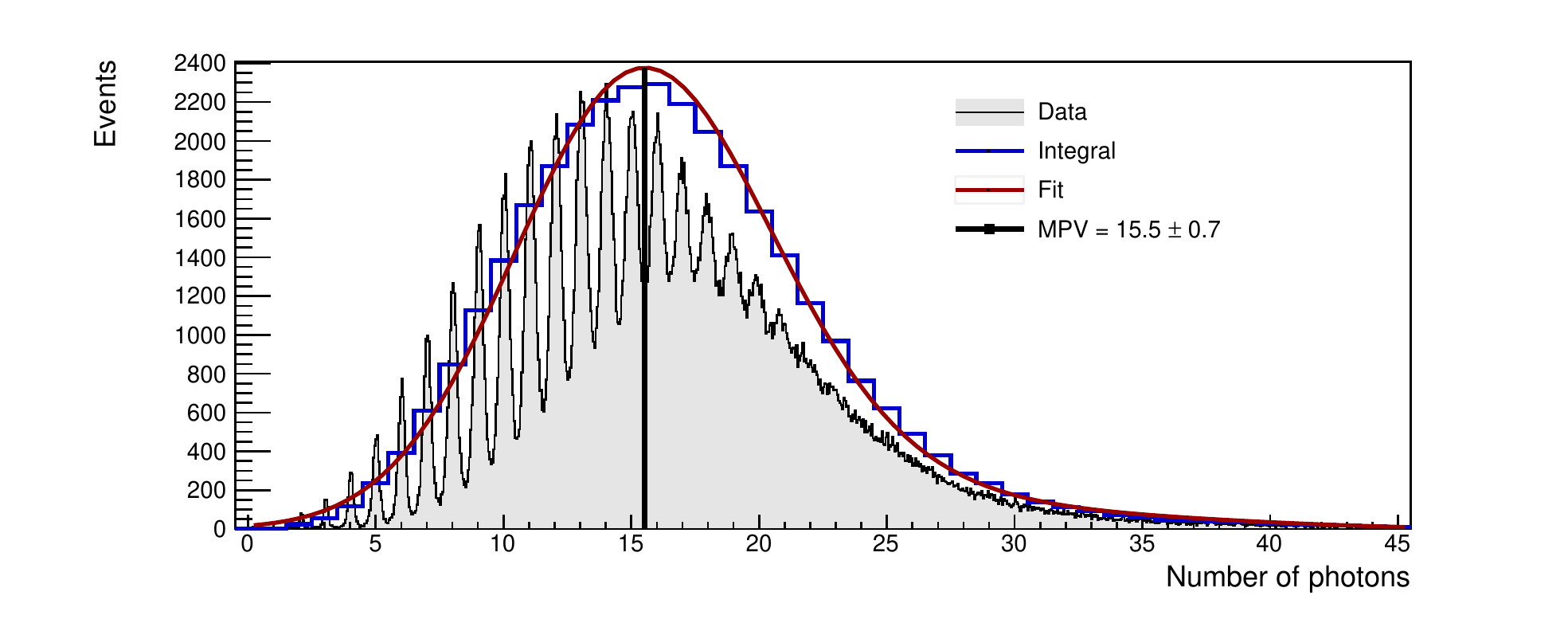}
   \caption{Charge spectrum summed over all active channels passing the 0.5 ph.e. threshold in a cluster
matched to the trigger time for a 3-layer SCSF-78 fiber ribbon prepared with black epoxy.
The spectrum is normalized to the charge generated by a single photo-electron.
Beam particles cross the SciFi ribbon in the center, i.e. 15~cm from both ribbon's ends.
The $n_{phe}$ distribution (blue) is obtained by integrating the charge in a region of $\pm 0.5$ ph.e. around each peak.
A convolution of a Gaussian with a Landau distribution is used to model the data (red) 
and the MPV of the convolution is extracted.}
   \label{fig:clCharge}
\end{figure}

\begin{figure}[t!]
   \centering
   \includegraphics[width=0.6\textwidth]{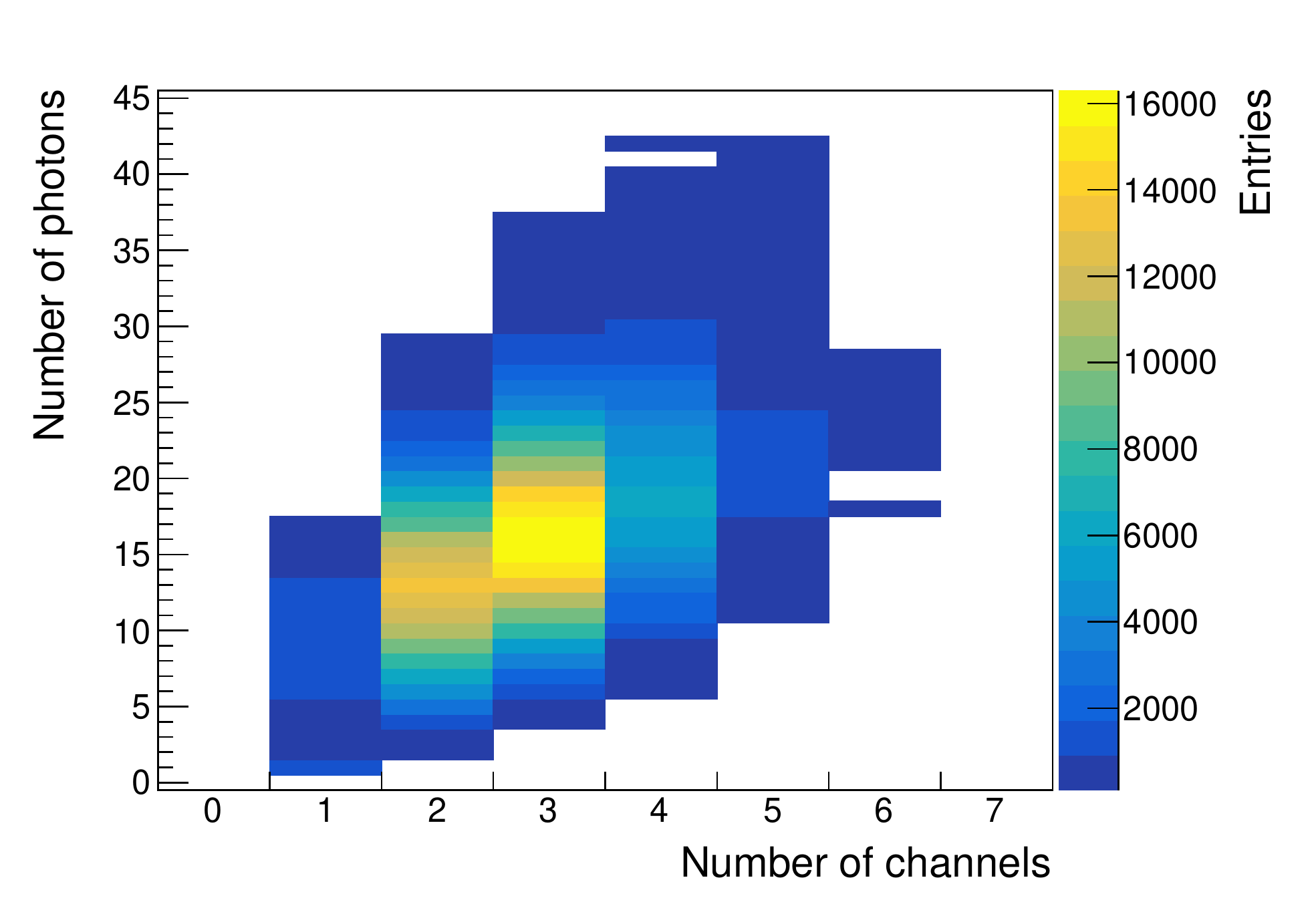}
   \caption{Cluster light yield vs cluster width for active channels passing the 0.5 ph.e. threshold
for a 3-layer SCSF-78 fiber ribbon prepared with black epoxy.}
   \label{fig:clCharge_width}
\end{figure}

Figure~\ref{fig:clCharge} shows the integrated charge spectrum summed over all active channels in the cluster
normalized to the charge generated by a single photo-electron (see Section~\ref{sec:chNorm})
for a 3-layer SCSF-78 fiber ribbon prepared with black epoxy.
Each peak corresponds to a specific number of detected photons $n_{phe}$.
The peaks are evenly spaced up to high multiplicities.
To extract $n_{phe}$ in an event,
the normalized charge distribution has been integrated in a region of $\pm 0.5$ around each peak.
The discrete distribution $n_{phe}$ has been fitted with a convolution of a Gaussian with a Landau distribution,
which interpolates well the observed spectrum,
and the Most Probable Value (MPV) of the convolution is extracted.
The Landau distribution describes the fluctuations in the energy deposit in a thin layer of material,
while the nature of the scintillation light generation results into a Poissonian distribution,
which approaches a Gaussian distribution.
Figure~\ref{fig:clCharge_width} shows the light yield for the same SciFi ribbon versus the number of active channels,
i.e. the cluster's width.
There is a clear correlation between the two quantities: the higher the light yield, the wider the cluster
because of the optical cross-talk.

\begin{figure}[t!]
   \centering
   \includegraphics[width=0.49\textwidth]{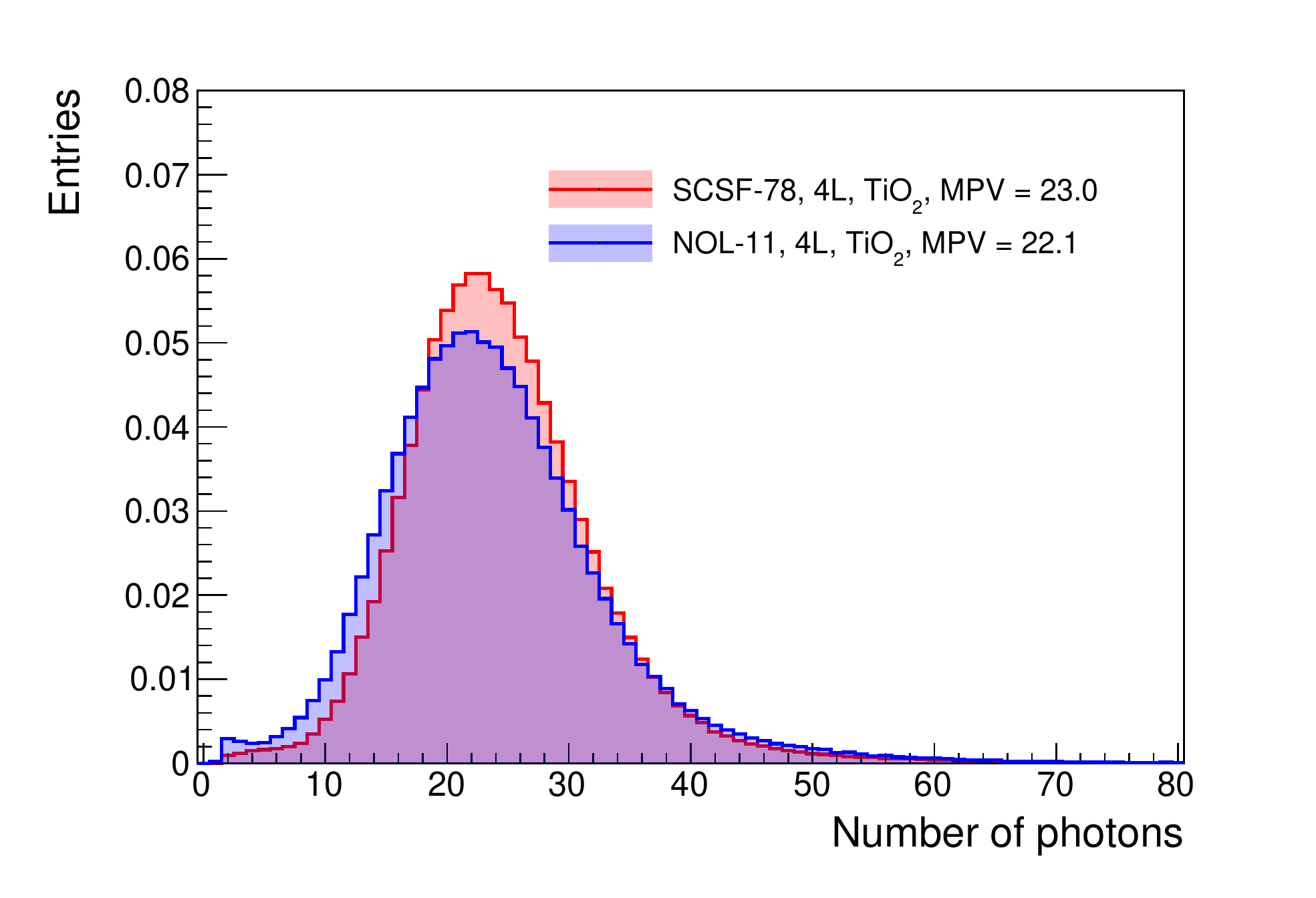}
   \includegraphics[width=0.49\textwidth]{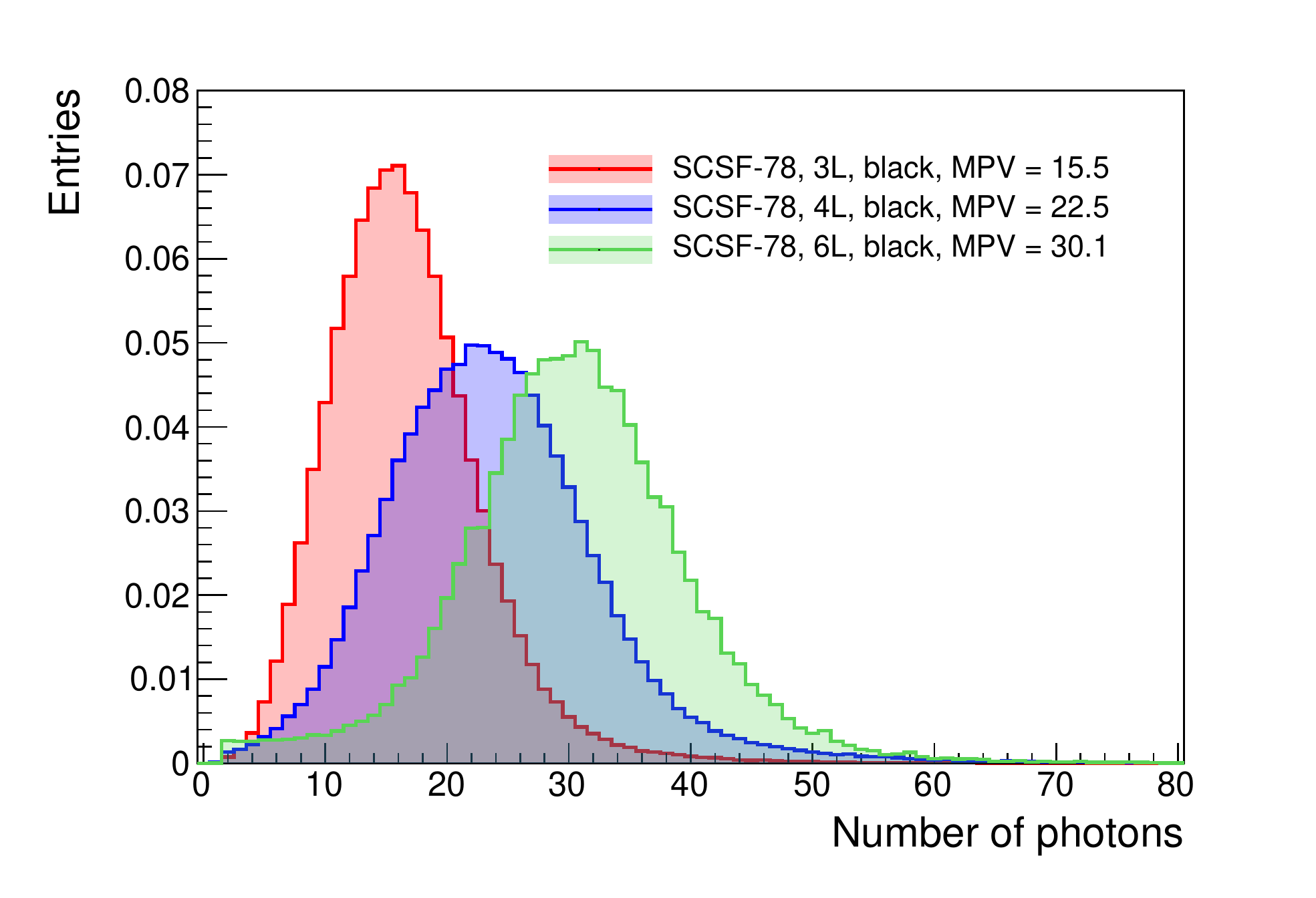}
   \includegraphics[width=0.49\textwidth]{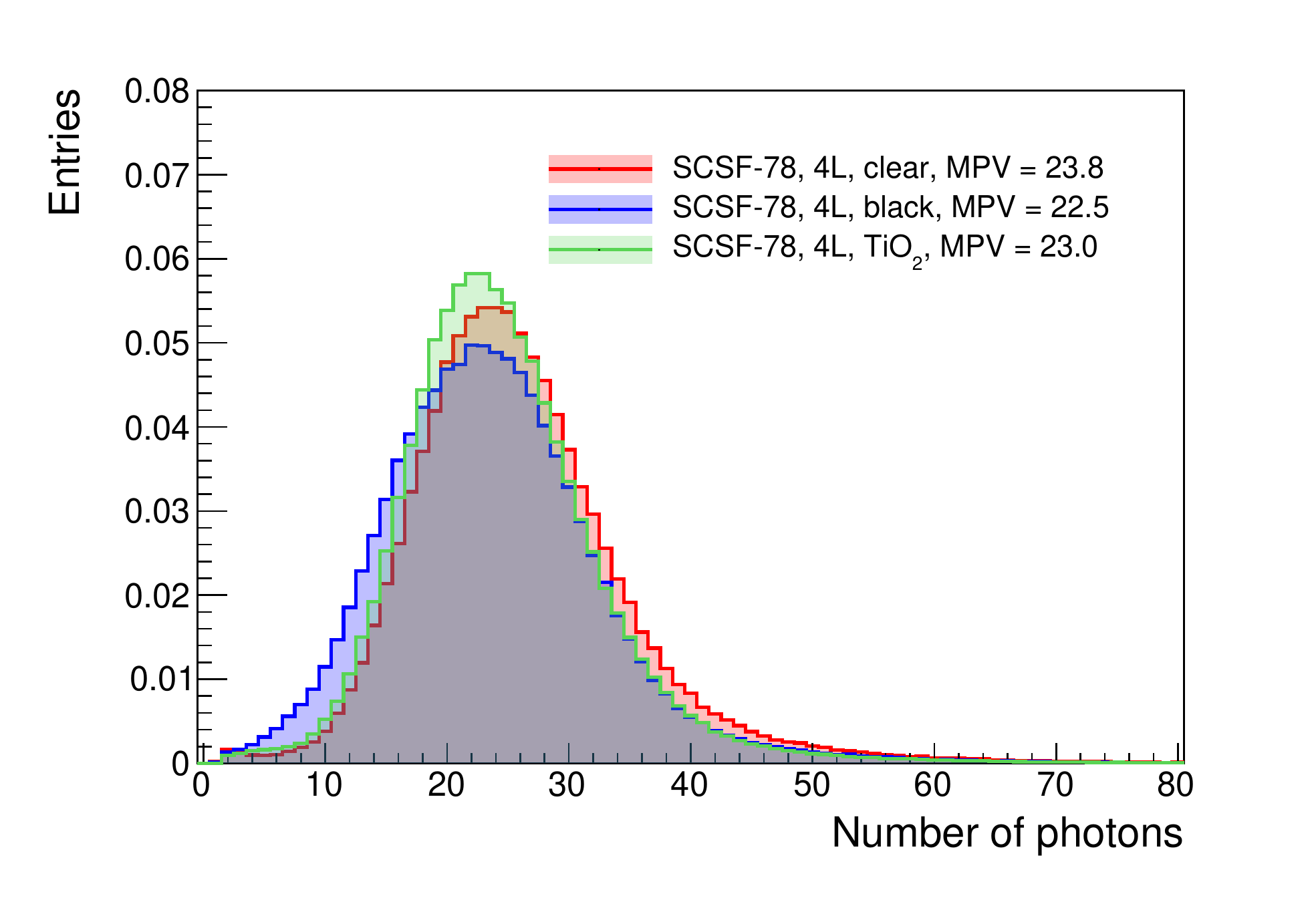}
   \includegraphics[width=0.49\textwidth]{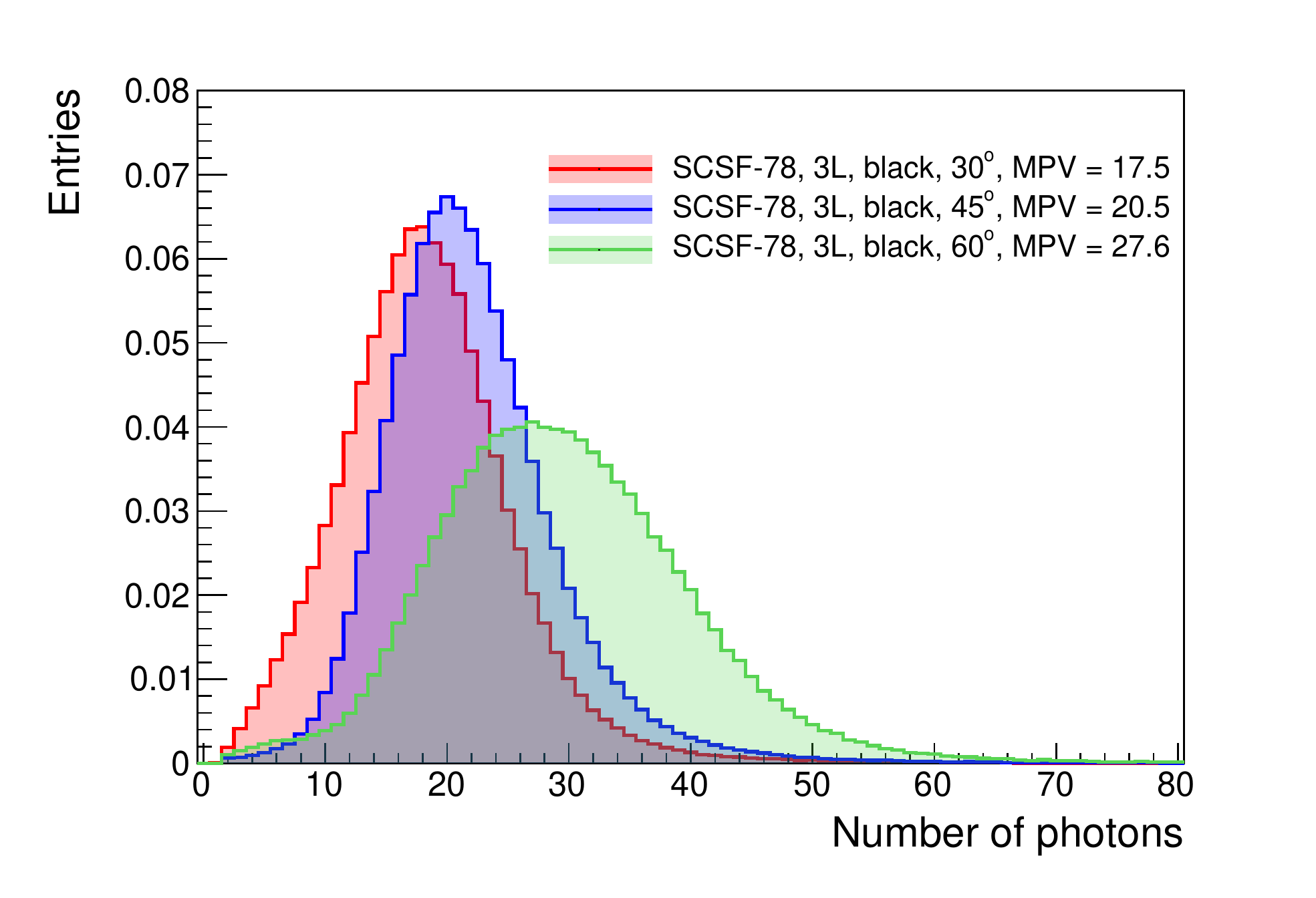}
   \caption{Comparison of the light yield of different SciFi ribbons:
top left) 4-layer SCSF-78, NOL-11, and SCSF-81 fiber ribbons (all are prepared with clear epoxy with a 20\% \TiO admixture),
top right) SCSF-78 fiber ribbons consisting of 3, 4, and 6 staggered fiber layers and assembled with black epoxy, 
bottom left) 4-layer SCSF-78 fiber ribbons assembled with different adhesives (clear epoxy, black epoxy, clear epoxy with a 20\% \TiO admixture),
bottom right) 3-layer SCSF-78 fiber ribbon prepared with black epoxy for different crossing angles.
The histograms are normalized to the number of analyzed events.
Beam particles cross the SciFi ribbon in the center, i.e. 15~cm from the ribbon's ends.}
   \label{fig:LY_comp}
\end{figure}

Figure~\ref{fig:LY_comp} compares the light yield of SCSF-78 and NOL-11 fiber ribbons
consisting of different numbers of staggered fiber layers, and assembled with different types of adhesives.
The MPV of the fits are reported in Table~\ref{tab:MPV}.
These values are not corrected for the PDE of the SiPM array, which is around 45\% (including the inefficiency from the fill factor $\varepsilon_{fill-factor}$).
A detailed study of the light yield of different blue-emitting scintillating fibers can be found in e.g.~\cite{fibers}.
The light yield for the SCSF-78 and the NOL-11 fiber ribbons is comparable (Figure~\ref{fig:LY_comp} top left),
while the light yield of a SCSF-81 fiber ribbon is $\sim 50\%$ lower.
The light yield increases with the ribbon's thickness, i.e. the number of staggered layers (Figure~\ref{fig:LY_comp} top right),
and it is twice as large for a 6-layer fiber ribbon compared to a 3-layer fiber ribbon.
Different adhesives do not influence strongly the light yield of the SciFi ribbons (Figure~\ref{fig:LY_comp} bottom left).
Within the errors, all light yields are comparable.
As the crossing angle $\vartheta$ increases, the thickness of the SciFi ribbon traversed by a particle increases as $1/\cos \vartheta$.
The measured light yield follows the same trend (Figure~\ref{fig:LY_comp} bottom right).

\begin{table}[b!]
\centering
\begin{tabular}{|l|c||l|c|}
\hline
ribbon type & MPV & ribbon type & MPV  \\
\hline
SCSF-78 3-layer black   &   $15.5 \pm 0.7$   &   NOL-11 2-layer \TiO   &   $11.7 \pm 1.0$     \\
SCSF-78 4-layer black   &   $22.5 \pm 0.9$   &   NOL-11 3-layer \TiO   &   $19.2 \pm 1.2$     \\
SCSF-78 6-layer black   &   $30.1 \pm 1.1$   &   NOL-11 4-layer \TiO   &   $22.1 \pm 1.3$     \\
\hline
SCSF-78 4-layer \TiO  &  $23.0 \pm 1.0$   &   SCSF-78 4-layer clear   &   $23.8 \pm 1.4$     \\
SCSF-81 4-layer \TiO  &  $11.9 \pm 1.0$   &   SCSF-78 4-layer black   &   $22.5 \pm 0.9$    \\
 NOL-11 4-layer \TiO  &  $22.1 \pm 1.3$   &   SCSF-78 4-layer \TiO    &   $23.0 \pm 1.0$     \\
\hline
\end{tabular}
\caption{MPV for SciFi ribbons made of different fiber types,
consisting of a different number of staggered fiber layers,
and prepared with different adhesives.
The reported errors are from the fits to the charge spectra
and are comparable to the differences in the number of detected photons at the two ribbon's ends.
These values are not corrected for the PDE of the SiPM array, which is around 45\%.}
\label{tab:MPV}
\end{table}

\begin{figure}[t!]
   \centering
   \includegraphics[width=0.6\textwidth]{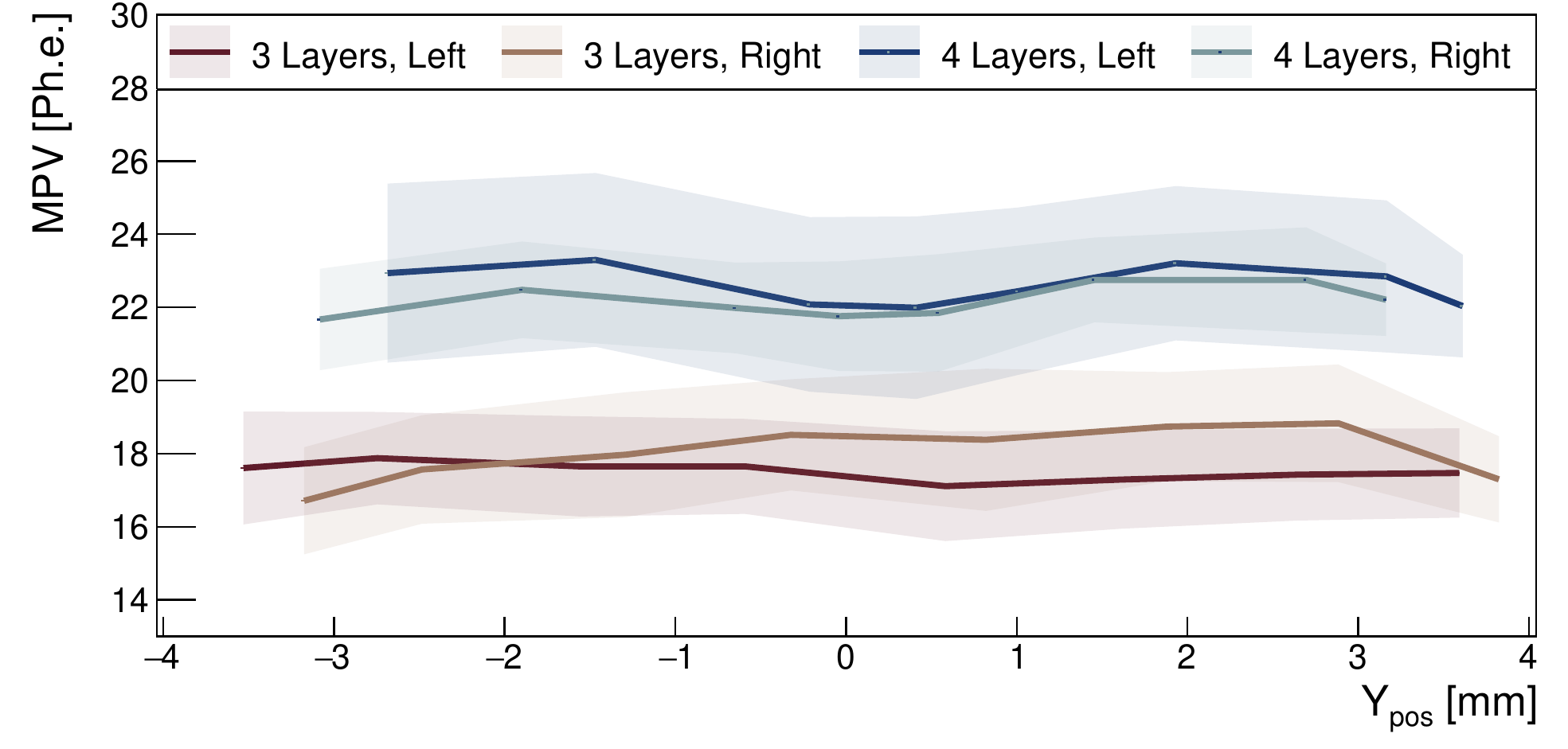}
   \caption{Light yield across the SciFi ribbon (i.e. vertical scan), always at the same distance from the photo-sensors.
The light yield at the left and right ends of a 3-layer and 4-layer SciFi ribbon are compared.
Both ribbons are made with the NOL-11 fiber using clear epoxy containing a 20\% \TiO admixture.}
   \label{fig:clVscan}
\end{figure}

To evaluate the uniformity of the response, various SciFi ribbons have been scanned vertically (i.e. across the ribbon)
and the light yields at the left and right sides have been compared.
From Figure~\ref{fig:clVscan}, which shows an example of a vertical scan performed in the middle of the ribbon,
it can be observed that indeed the response (i.e. the light yield) is quite uniform across the SciFi ribbon.
The fact that the light yield from the left and right ends are comparable also supports indirectly the reproducibility of the measurements. 

\subsection{Light Attenuation}

To study the light attenuation, the SciFi ribbons have been scanned horizontally (i.e. along the SciFi ribbon length)
at different distances from the ribbon's end from 5~cm to 25~cm.
The light intensity $I$ along the fiber as a function of the propagation distance $d$
is usually described in terms of a short and a long component as
(it depends also on the scintillation light wavelength $\lambda$)~\cite{fibers}:
\begin{equation}
    I(\lambda,d) = I_0^\mathrm{short}(\lambda) \cdot  \exp \left(-d / \Lambda^\mathrm{short}(\lambda)\right) +
                          I_0^\mathrm{long}(\lambda) \cdot  \exp \left(-d / \Lambda^\mathrm{long}(\lambda)\right) \; , 
\label{eq:att}
\end{equation}
\noindent with $\Lambda^\mathrm{short}(\lambda)$ and $\Lambda^\mathrm{long}(\lambda)$
the attenuation lengths for the short and long components, respectively.
At these distances (i.e. $\leq 30~{\rm cm}$), the absorption of light is controlled by the short component $\Lambda^\mathrm{short}$,
which dies off rather quickly after 10 -- 20 cm~\cite{fibers},
while the long component can be assumed constant over the length of the SciFi ribbon.

\begin{figure}[t!]
   \centering
   \includegraphics[width=0.49\textwidth]{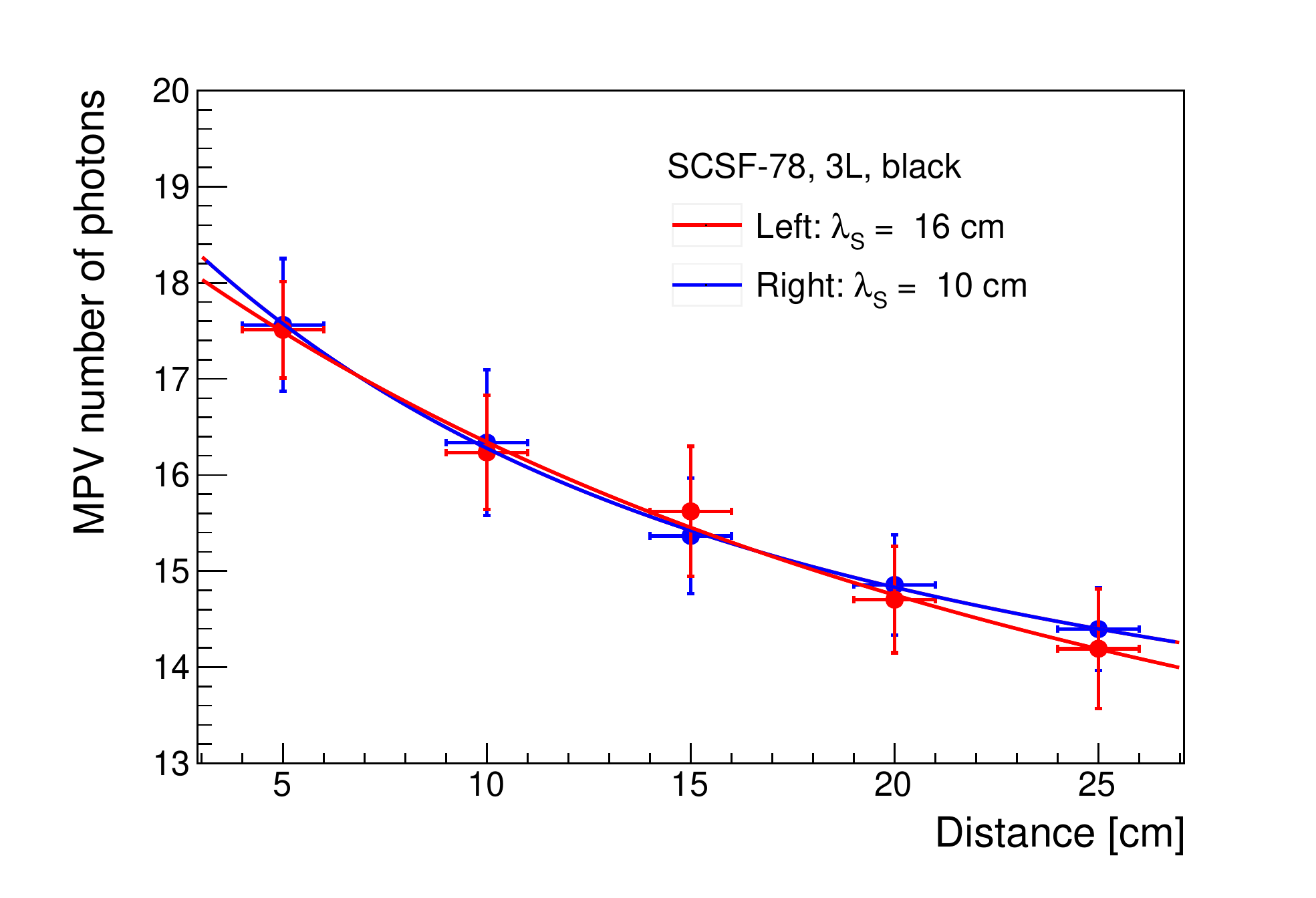}
   \includegraphics[width=0.49\textwidth]{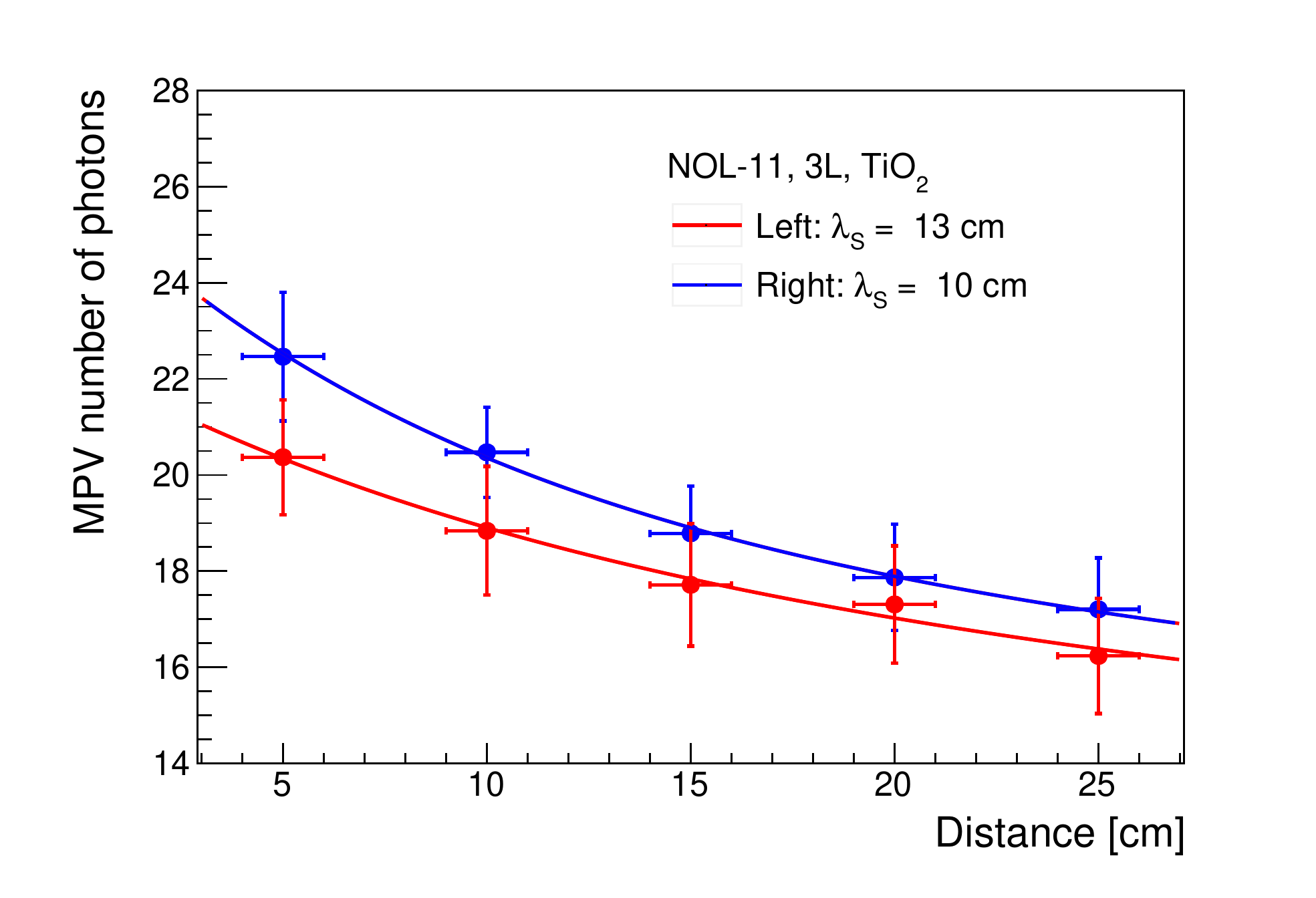}
   \caption{Number of detected photons $n_{phe}$ as a function of the distance from both SciFi ribbon ends for 
a 3-layers SCSF-78 fiber ribbon prepared with black epoxy (left)
and a 3-layers NOL-11 fiber ribbon prepared with clear epoxy with a 20\% \TiO admixture (right).
The blue circles are for measurements from the left side and the red circles for the right side of the ribbons.
The data points are fitted with a single exponential function.
The extracted attenuation lengths are also shown in the plots.}
   \label{fig:clHscan}
\end{figure}

Figure~\ref{fig:clHscan} shows the number of detected photons $n_{phe}$ as a function of the distance from the ribbon's end
for two different SciFi ribbons.
Measurements from both sided are compared.
The small differences are due to the optical couplings of the fibers to the SiPM arrays, the PDE of each SiPM array,
non-uniformities in the fibers, misalignments in the assembly of the fiber ribbons, etc.
From this comparison one can also asses the reproducibility of the measurements.
The measurements are fitted with a single exponential function describing the short component $\Lambda^\mathrm{short}$,
while keeping the long component constant (Equation~\ref{eq:att}),
since this component does not vary significantly over the length of the SciFi ribbons
The extracted attenuation lengths $\Lambda^\mathrm{short}$ are shown in the plots.

\clearpage
\section{Timing Analysis}
\label{sec:timing}

The timing performance of the SciFi detector is the most important feature for the background rejection in \mude. 
The intrinsic limit on the time resolution of low mass (i.e. thin) scintillation detectors is driven by the statistical processes
involved in the generation of the light signal and fluctuations in the light detection system.
The time resolution of a relatively low light yield SciFi detector ($\mathcal{O}(15)$ detected photons)
depends on the decay time $\tau$ of the {\it spectral shifter}
(the shorter the decay time, the better the achievable time resolution),
but also on the signal amplitude, i.e. the number of detected photons $n_{phe}$.
For a detailed discussion, see e.g.~\cite{fibers}.

\begin{figure}[!b]
   \centering
   \includegraphics[width=0.6\textwidth]{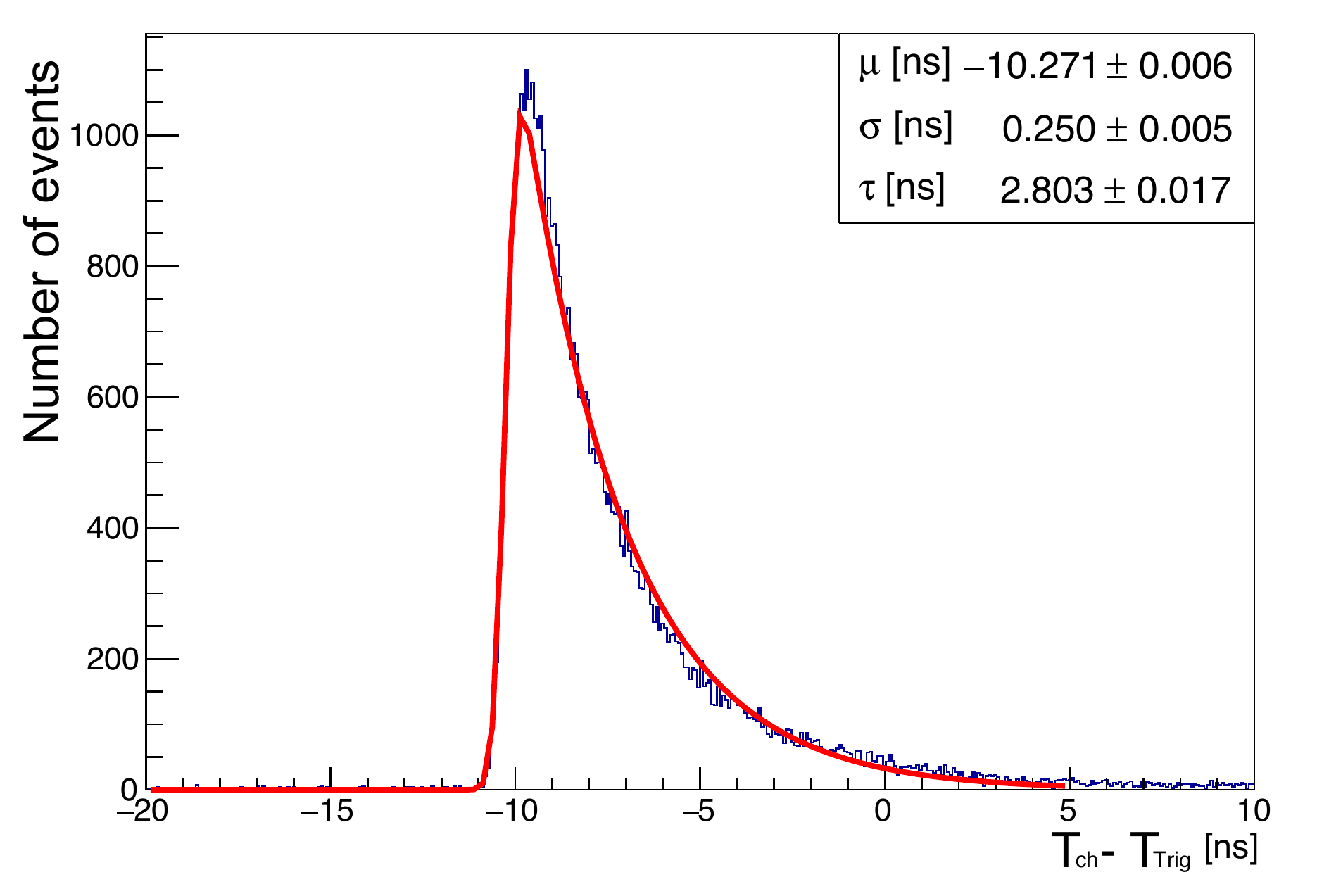}
   \caption{Time of arrival of the first detected photon $T_{\rm fiber} - T_{\rm trigger}$.
The data is described by the EMG distribution with parameters $\sigma$ and $\tau$.}
   \label{fig:TDecay}
\end{figure}

Figure~\ref{fig:TDecay} shows the arrival time distribution of the first detected photon
in a selected channel of the SiPM array coupled to a 3-layer SCSF-78 fiber ribbon prepared with black epoxy
(i.e. the time difference $T_{\rm fiber} - T_{\rm trigger}$ between the first detected photon and the external time reference - the trigger).
The shape of the arrival time distribution is best described by a convolution of a Gaussian,
which describes the time spread of the light pulse generation process
(and accounts also for the fluctuations in the light detection),
with an exponential decay function describing the de-excitation of the {\it spectral shifter},
the so called exponentially modified Gaussian distribution (EMG) or exGaussian distribution: 
\begin{equation}
F(t, \mu, \sigma, \tau) = A \frac{1}{2\tau} \exp \left( \frac{\mu - t}{\tau} + \frac{\sigma^2}{2\tau^2} \right)
\left[ 1 - {\rm erf} \left( \frac{1}{\sqrt 2} \left( \frac{\mu - t}{\sigma} + \frac{\sigma}{\tau} \right) \right)  \right] \; ,
\label{eq:EMG}
\end{equation}
where $\tau$ is the scintillation light decay time and $\sigma$ accounts for the time spread of the light collection
(which includes also the time jitter of the external time reference,
the transit time jitter of a single photon in the SiPMs of $\sim 220~{\rm ps}$,
the response of the electronics, etc.).
erf is the error function, $t = T_{\rm fiber} - T_{\rm trigger}$, and $A$ is a normalization factor.

The arrival time of the first scintillation photon is determined from the analysis of the recorded waveforms.
It is extracted by interpolating the rising edge of the signal with a straight line
extrapolated to the baseline of the waveform,
after correcting for baseline fluctuations.
The interpolation is performed on four samples on the rising edge of the waveform,
the first sample is below the single photon half amplitude, the next three samples above.
At 5~GHz sampling, the samples are spaced by $\sim 200~{\rm ps}$ for a time base of $\sim 600~{\rm ps}$.
The time walk of this algorithm has been estimated to be below 10~ps~\cite{fibers}.
The time resolution of the trigger scintillator $\sigma_{\rm trigger}$ of 80~ps
does not affect significantly the shape of the light pulse in this analysis.

Incidentally, Figure~\ref{fig:TDecay} gives also an appreciation of the time resolution achievable
with single ended readout of the fiber detector,
which is dominated by the decay time $\tau$ for low photon statistics.
The variance $\sigma_t$ of the EMG distribution is given by $\sigma_t = \sqrt{\sigma^2 + \tau^2}$,
where $\sigma$ is the intrinsic resolution of the system.
$\sigma_t$ is not the best choice for quantifying the time resolution of the SciFi detector
given the asymmetric nature of the timing distribution.
Instead, one could quote the parameters $\sigma$ and $\tau$ of the EMG fit
or the FWHM for the left and the right side of the distribution.

\subsection{Time Alignment}
\label{sec:Talign}

\begin{figure}[!b]
   \centering
   \includegraphics[width=0.49\textwidth]{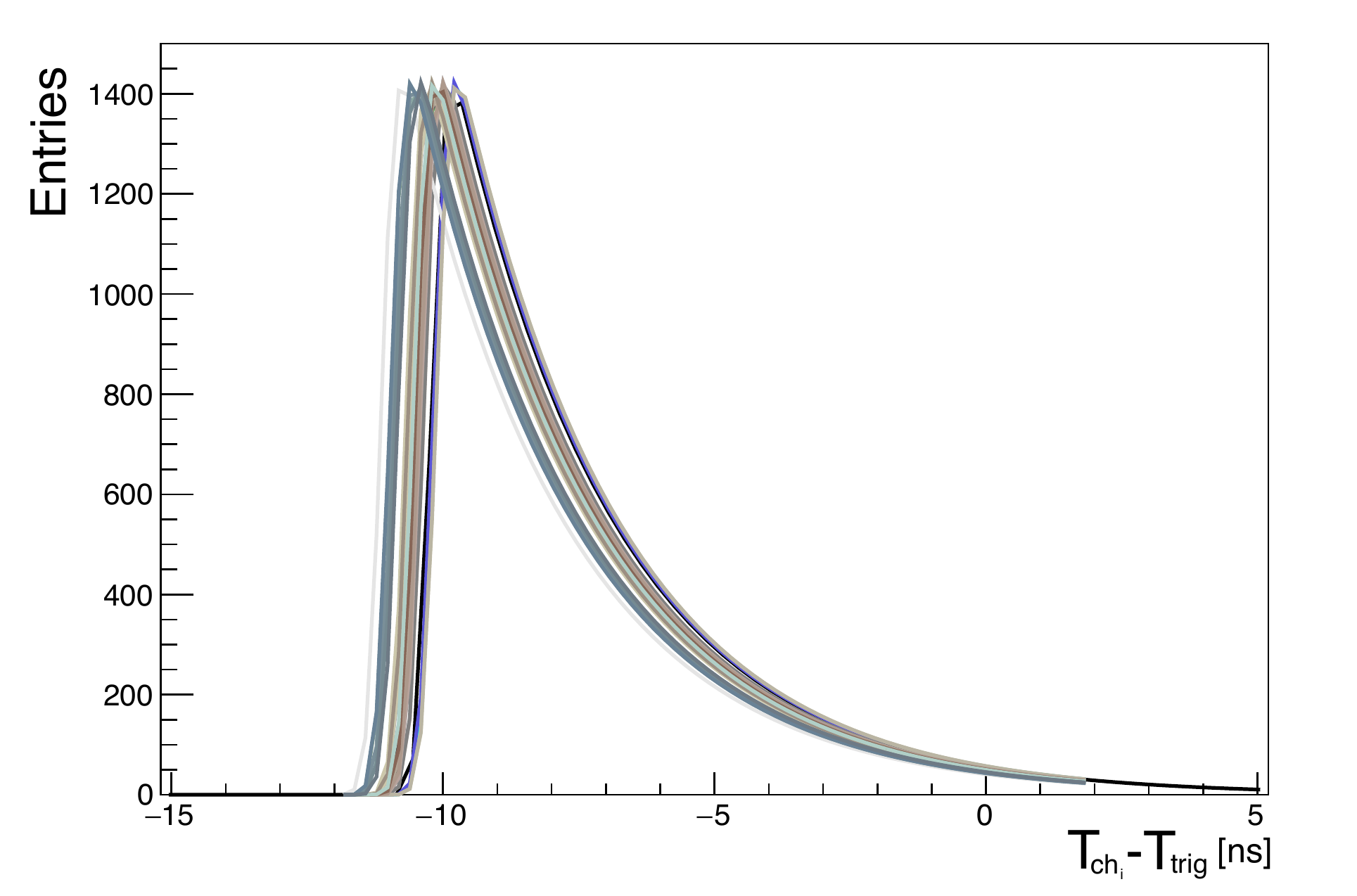}
   \includegraphics[width=0.49\textwidth]{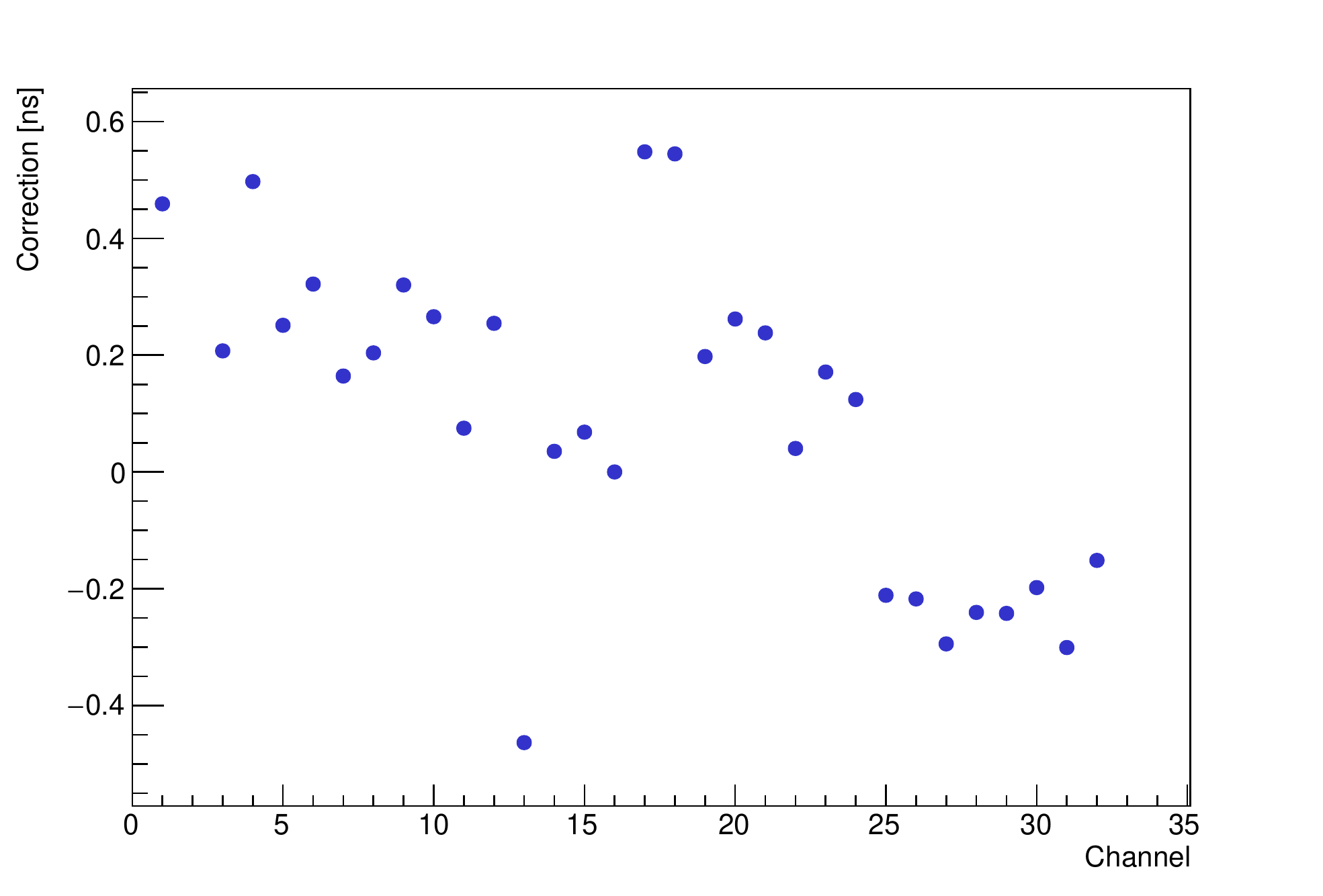}
   \caption{left) Paremetrized $T_{\rm fiber} - T_{\rm trigger}$ distribution, based on fits to the same distribution (Figure~\ref{fig:TDecay},
for 32 consecutive channels of the SiPM array.
right) Time shifts for each channel.}
   \label{fig:TCalib}
\end{figure}

To extract the timing information from the SciFi detector, all SiPM channels must be timed in.
Figure~\ref{fig:TCalib} shows the time response of several SiPM channels from the same SiPM array.
As it can be noticed, the signals are spread in time.
This is due to different trace lengths on the amplifier boards, different cable lengths, SiPM response, electronics response, etc. 
Starting from the $T_{\rm fiber} - T_{\rm trigger}$ distributions in Figure~\ref{fig:TCalib}
we have determined the time shift of all channels w.r.t. the the external time reference $T_{\rm trigger}$ (Figure~\ref{fig:TCalib} right).
Once the time shifts have been determined for one SciFi detector configuration,
they can be applied to different SciFi ribbons,
since we have always used the same detection chain.

\subsection{Time Difference $\Delta T$}
\label{sec:timeres}

\begin{figure}[t!]
   \centering
   \includegraphics[width=0.6\textwidth]{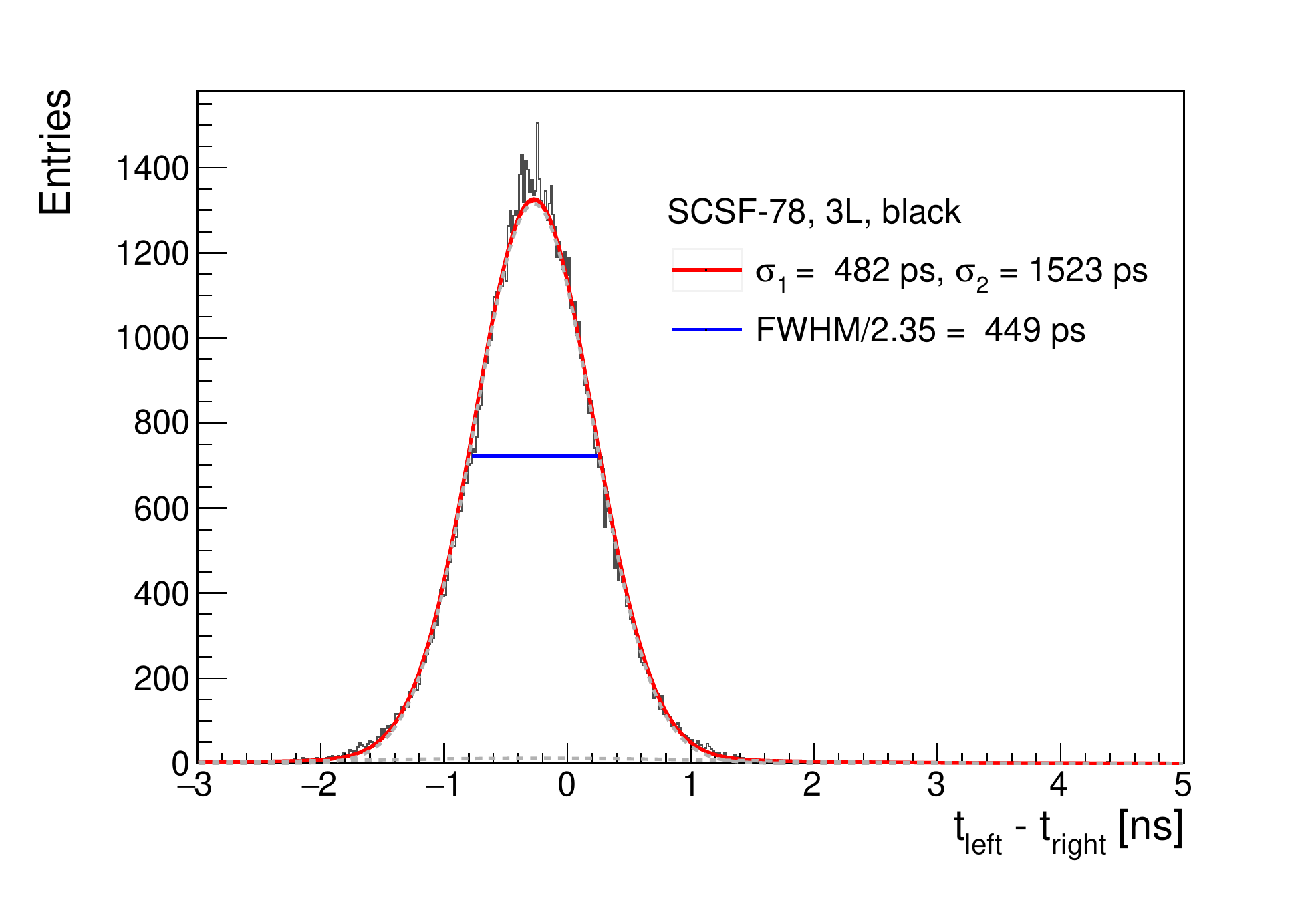}
   \caption{$\Delta T = T_{\rm left} - T_{\rm right}$ distribution for a 3-layer SCSF-78 SciFi ribbon prepared with black epoxy.
The distribution is fitted with different models:
a double Gaussian distribution with a common mean (red and dashed gray),
a single Gaussian distribution (green),
and an exponential distribution convoluted with a Gaussian (blue).
The latter two are fitted symmetrically around the mean but are shown only on one side to enhance clarity.
Also shown is the FWHM of the distribution.}
   \label{fig:DeltaT}
\end{figure}

The time resolution of the SciFi detector can be significantly improved by reading out the detector at both ends and combining the two time measurements.
That, however, requires time matched clusters at both SciFi ribbon's ends, which reduces slightly the detection efficiency of the detector
(see~Section~\ref{sec:efficiency}).
If not stated otherwise, the cluster time is determined by taking the first detected photon in the cluster,
which corresponds to the earliest time within a cluster.
Firstly, we have studied the time difference $\Delta T = T_{\rm left} - T_{\rm right}$ distribution,
where $T_{\rm left}$ and $T_{\rm right}$ are the time measurements for the {\it left} and {\it right} ribbon ends.
$\Delta T$ is self-contained in the sense that it can be formed without an external time reference.
$\Delta T$, however, cannot be used to determine the crossing time of a particle.

Figure~\ref{fig:DeltaT} shows the $\Delta T$ distribution for a 3-layer SCSF-78 fiber ribbon prepared with black epoxy
for beam particles crossing the SciFi ribbon in the center (i.e. 15~cm from both ribbon's ends).
The $\Delta T$ distribution is symmetric around the peak,
since the fluctuations in the time measurements add/subtract symmetrically for $T_{\rm left}$ and $T_{\rm right}$.
The $\Delta T$ distribution is not necessarily centered around 0, for instance because of different cable lengths. 
The tails, which extend symmetrically around the peak, are driven by the fiber's decay time.
Several models have been used to describe the $\Delta T$ distribution and extract  the time-difference resolution $\sigma_{\Delta T}$.
The $\Delta T$ distribution is fitted
with the sum of two Gaussian distributions centered around a common mean value:
\begin{equation}
f(t) = N_1 \cdot \frac{1}{\sqrt{2 \pi \sigma_1^2}} \exp \left( (t-\mu)^2 / 2 \sigma_1^2 \right) +
         N_2 \cdot \frac{1}{\sqrt{2 \pi \sigma_2^2}} \exp \left( (t-\mu)^2 / 2 \sigma_2^2 \right) \; .
\label{eq:2G}
\end{equation}
The first Gaussian, which describes the core of the distribution,
can be interpreted as due to fluctuations in the excitation of the primary die in the fibers (i.e. the {\it activator}) and in the light collection,
while the second Gaussian describes the tails, which are driven by the fiber's decay time.
More than 95\% of events fall under the first Gaussian, while less than 5\% under the second.
As an indication of the time-difference resolution $\sigma_{\Delta T}$ we quote the FWHM/2.355 of the $\Delta T$ distribution,
which is close to the width of the first Gaussian.
Alternatively, the $\Delta T$ distribution can be modeled with a symmetric exponential function such as the Laplace distribution,
which is a direct consequence of the Poissonian nature of the scintillation process,
smeared with a Gaussian distribution to take into account the time spread of the light collection:
\begin{equation}
f(t) = A \cdot \exp \left( -|t-t_0| / \tau \right) \ast \frac{1}{\sqrt{2 \pi \sigma^2}} \exp \left( (t-t_0)^2 / 2 \sigma^2 \right) \; .
\end{equation}
This convolution models rather well the $\Delta T$ distribution, in particular the tails (blue line in Figure~\ref{fig:DeltaT}).
Finally, the $\Delta T$ distribution is fitted with a single Gaussian within $\pm 4 \sigma$ around the peak position
(green line in Figure~\ref{fig:DeltaT}).
Also a single Gaussian describes well the $\Delta T$ distribution, since the fraction of events in the tails of the distribution is rather small (i.e. $< 5\%$).
All fits are stable with respect to the fitting ranges.

\begin{figure}[t!]
   \centering
    \includegraphics[width=0.49\textwidth]{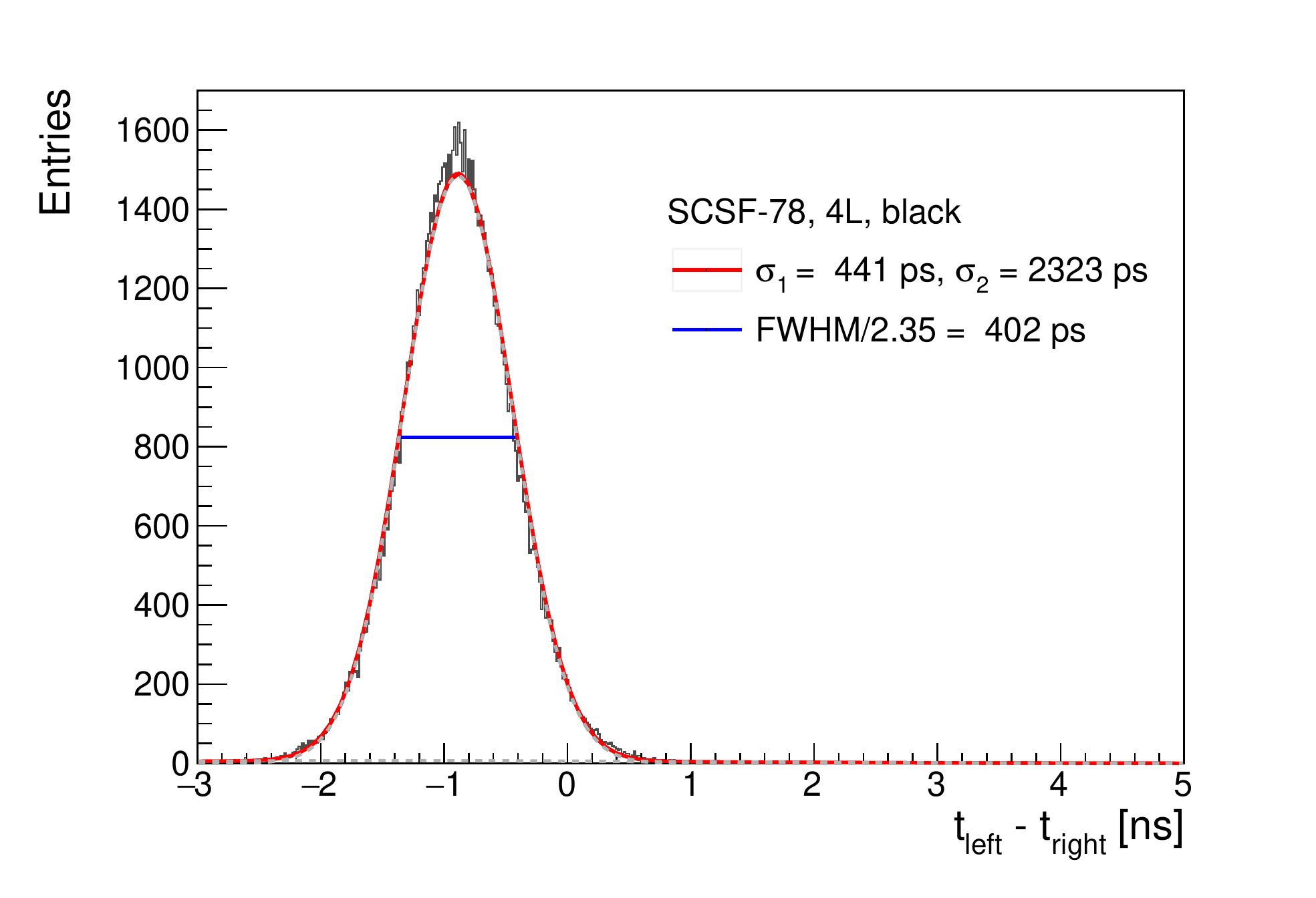}
   \includegraphics[width=0.49\textwidth]{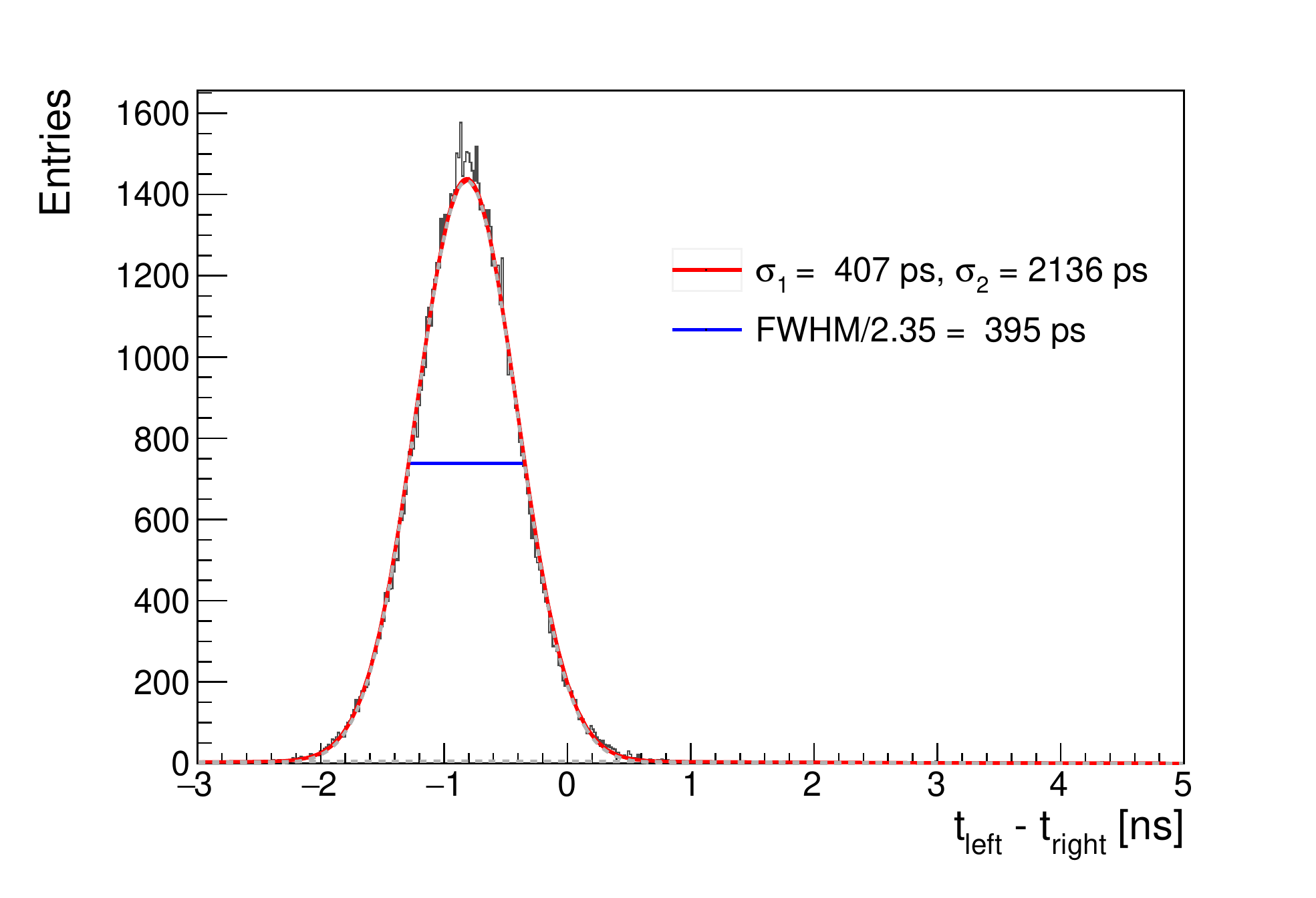}  
   \caption{$\Delta T = T_{\rm left} - T_{\rm right}$ distribution for a 4-layer SCSF-78 fiber ribbon prepared with black epoxy (left)
and a 4-layer NOL-11 fiber ribbon prepared with clear epoxy with 20\% \TiO admixture (right).}
   \label{fig:DeltaTcomp}
\end{figure}

Figure~\ref{fig:DeltaTcomp} compares the $\Delta T$ distribution for 4-layer SCSF-78 and NOL-11 fiber ribbons.
The $\Delta T$ distributions are fitted with the sum of two Gaussians centered around a common mean value (Equation~\ref{eq:2G}).
As an indication of the time-difference resolution $\sigma_{\Delta T}$ we quote the FWHM/2.355 of the $\Delta T$ distributions,
which is close to the width of the first Gaussian.
The measured values range between 400~ps and 500~ps depending on the fiber type used to make the SciFi ribbons
and the number of staggered layers.
(in Figure~\ref{fig:DeltaTcomp} only two SciFi ribbons are shown).
The best $\sigma_{\Delta T}$ resolution of 400~ps has been achieved with a 4-layer SciFi ribbon made of NOL-11 fibers
assembled with clear epoxy with 20\% \TiO admixture.
Despite the significantly smaller decay time of 1.1~ns of the NOL-11 fiber,
SciFi prototypes made with this fiber show a small improvement w.r.t. the SCSF-78 fiber,
which has a 2.8~ns decay time.
That suggest that $\sigma_{\Delta T}$ is dominated by the light yield,
which is comparable for the two SciFi ribbon types (Figure~\ref{fig:LY_comp}),
and not by the decay time of the scintillation light in the fibers.

\subsection{Light Propagation In the Fibers}

\begin{figure}[t!]
   \centering
   \includegraphics[width=0.49\textwidth]{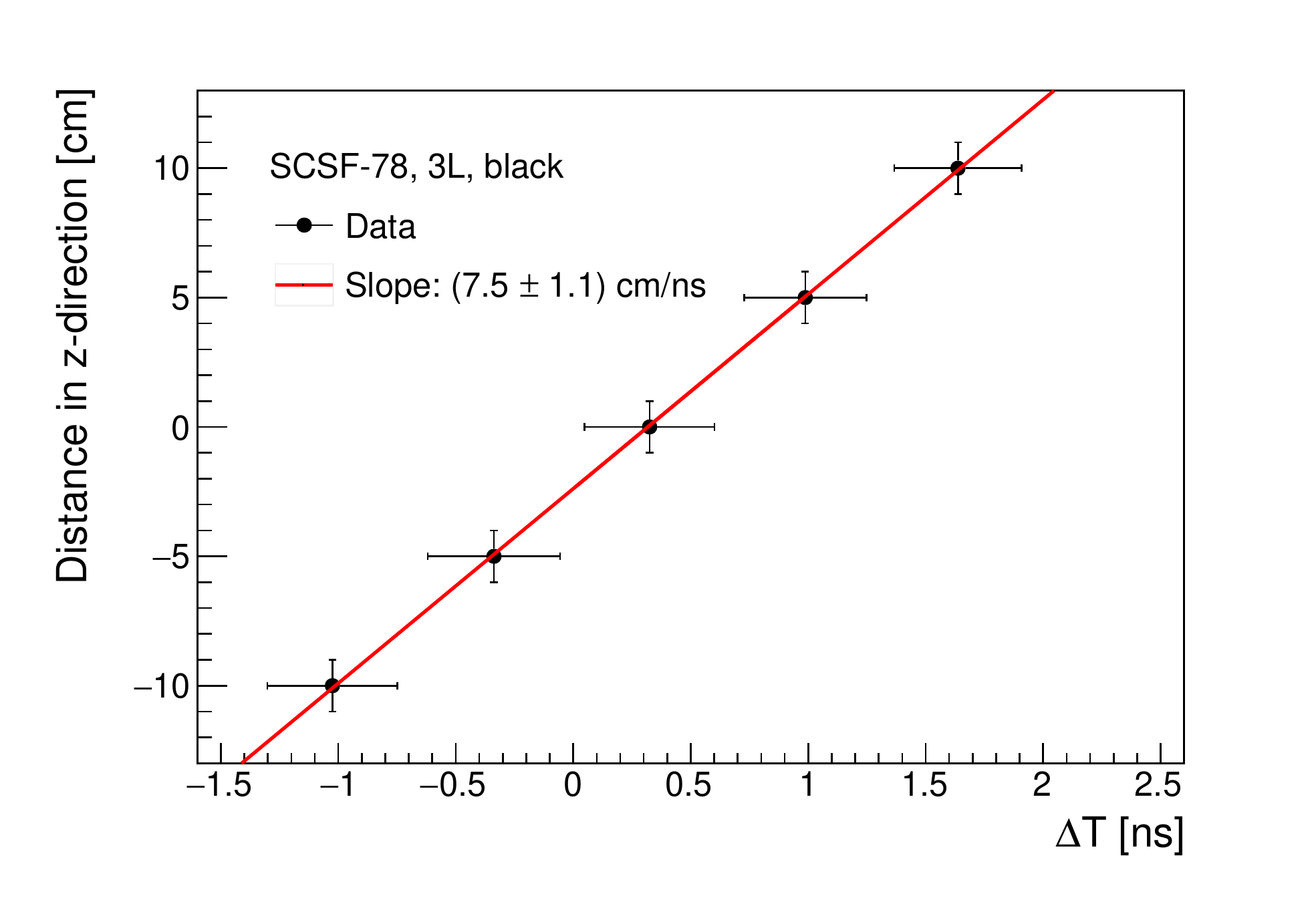}
   \includegraphics[width=0.49\textwidth]{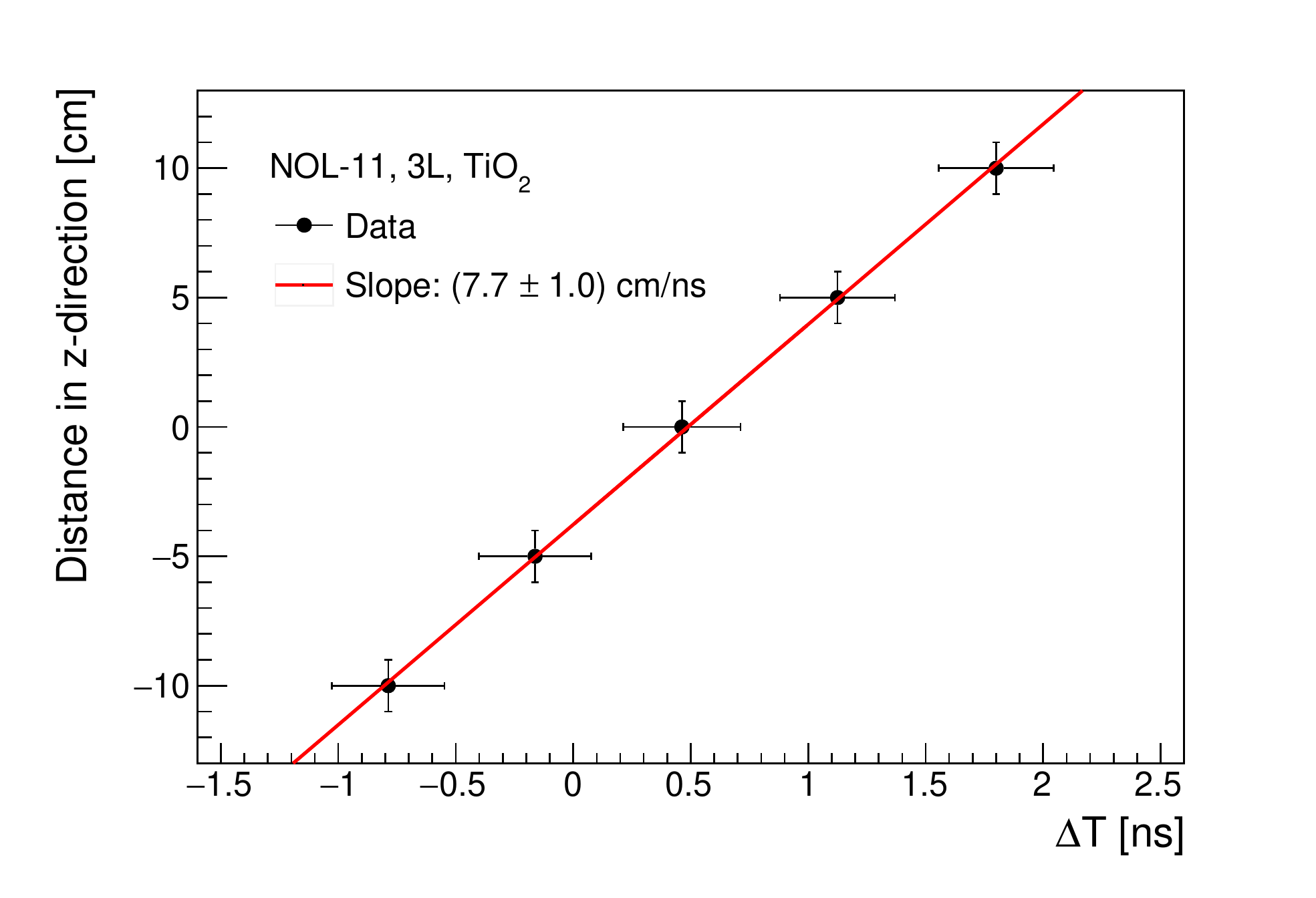}
   \caption{Peak position of the $\Delta T$ distribution vs hit position along the SciFi ribbon
for a 3-layer SCSF-78 fiber ribbon prepared with black epoxy (left)
and for a 3-layer NOL-11 fiber ribbon prepared with clear epoxy with a 20\% \TiO admixture (right).
Taking into account that the difference in the traveled distance by the scintillation photons is twice the displacement,
from the slope one can estimate the speed of light propagation in the fibers $v_{\rm fiber}$ which is $\sim \frac{1}{2} c$.}
   \label{fig:DeltaTz}
\end{figure}

Figure~\ref{fig:DeltaTz} shows the peak position of the $\Delta T$ distribution for different positions along the SciFi ribbon.
Since the distance to the photo-sensor changes,
the peak of the $\Delta T$ distribution drifts accordingly.
The width of the $\Delta T$ distributions along the ribbon remains almost constant,
indicating that $\sigma_{\Delta T}$ does not depend appreciably on the impact position.
The data points can be interpolated with a straight line
supporting the idea that the light propagates uniformly in both directions along the fibers.
Within our resolution, we do not observe edge effects as we approach the end of the fiber ribbon. 
Taking into account that the difference in the traveled distance by the scintillation photons is twice the displacement,
from the slope one can derive the speed of light propagation in the fibers $v_{\rm fiber} = 2 \times \Delta z / \Delta T$.
The scintillation photons propagate at the speed,
which is roughly half of the speed of light in vacuum, i.e. $v_{\rm fiber} \sim 0.5 \times c$.
$v_{\rm fiber}$ is significantly slower
compared to the speed that one would derive from the refractive index $n$ of the fiber's core, i.e. $c/n$.
This can be understood by the fact that during propagation the photons are reflected internally from the cladding many times and 
therefore travel over a longer distance. 
From $\Delta T$ one can also determine the impact position $z$ along the fiber 
$z = \frac{1}{2} L + \Delta T \times \frac{1}{2} v_{\rm fiber}$ with
a position resolution of $\sigma_z = \sigma_{\Delta T} \times \frac{1}{2} v_{\rm fiber} \sim 3~{\rm cm}$
(for $\sigma_{\Delta T} = 400~{\rm ps}$),
where $L$ is the length of the fiber ribbon.

\begin{figure}[b!]
   \centering
   \includegraphics[width=0.8\textwidth]{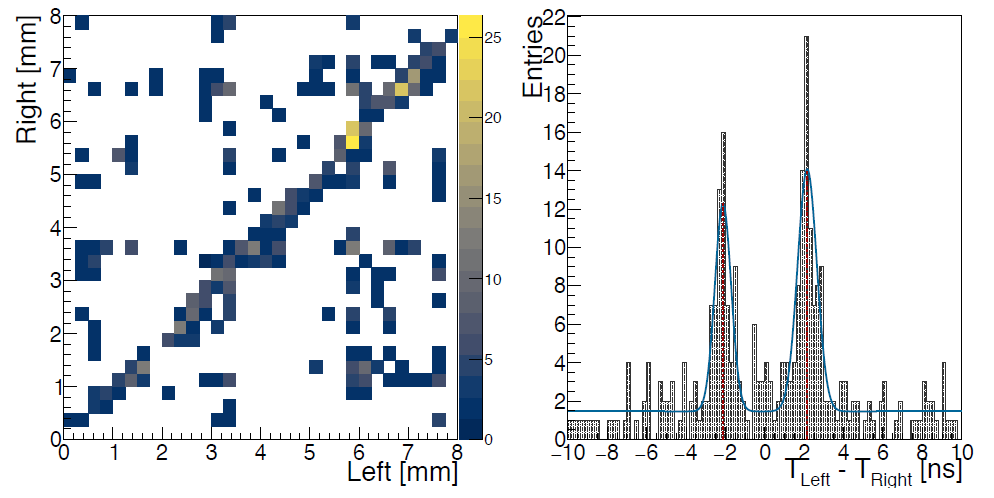}
   \caption{left) Correlation between {\it hits} at both ribbon's ends generated by random discharges in one SiPM array (dark counts).
Some IR photons are trapped and propagate in the fibers.
When they reach at the opposite end of the fiber ribbon, they can trigger discharges in the SiPM array at the opposite side.
right) Time difference between {\it hits} in the SiPM arrays placed on both sides of the SciFi ribbon.
The time difference between the two peaks is indicative of the speed of propagation of the photons in the SciFi ribbon.}
   \label{fig:speed2}
\end{figure}

A dataset has also been collected in absence of ionizing particles
to study the efficiency of the clustering algorithms in rejecting dark counts (see Sections~\ref{sec:cluster} and~\ref{sec:efficiency}).
When a SiPM cells discharges (dark count) it generates also infrared photons.
Some IR photons escape from the SiPM and can be trapped and transported by the scintillating fibers to the opposite end,
where they can trigger a discharge in the SiPM array coupled at the opposite side of the SciFi ribbon.
Figure~\ref{fig:speed2} (left) shows the correlation between {\it hits} at both ribbon's ends.
A clear correlation between the two sides emerges proving that such events indeed occur.
Figure~\ref{fig:speed2} (right) shows the time difference $T_{\rm left} - T_{\rm right}$ between {\it hits} at opposite ribbon's ends from such events.
The two peaks, which in this Figure are separated by 4.2~ns, are due to the fact that we show $\Delta T = T_{\rm left} - T_{\rm right}$
and not the absolute value $\Delta T$ (i.e. the time between the two events).
The speed of light propagation in the fibers is $v_{\rm fiber} = L / |\Delta T| = 30~{\rm cm}~/~2.1~{\rm ns} \sim 0.5 c$.
$v_{\rm fiber}$ thus estimated is in agreement with the above measurement.

\subsection{Mean-Time}

\begin{figure}[b!]
   \centering
   \includegraphics[width=0.49\textwidth]{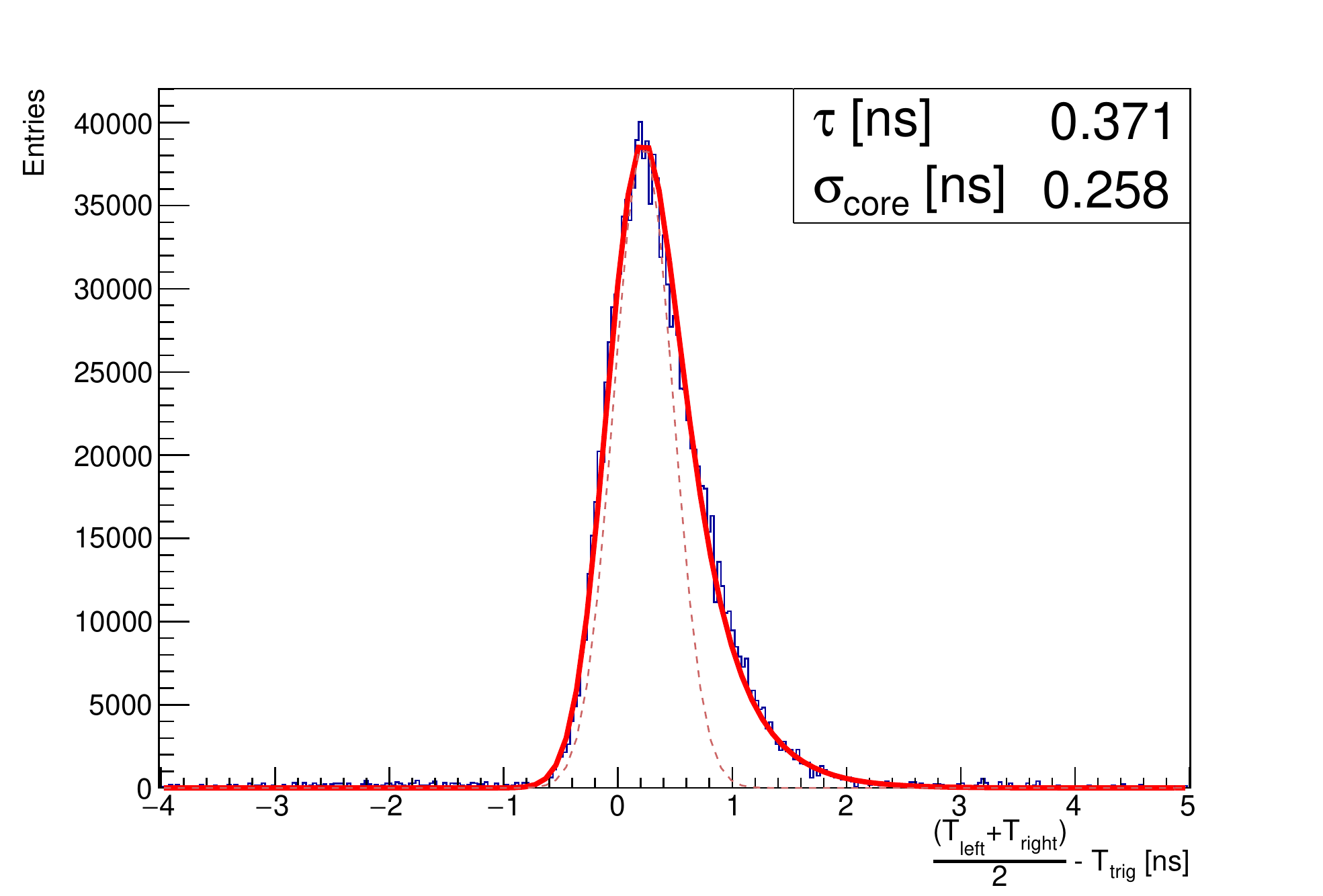}
   \includegraphics[width=0.49\textwidth]{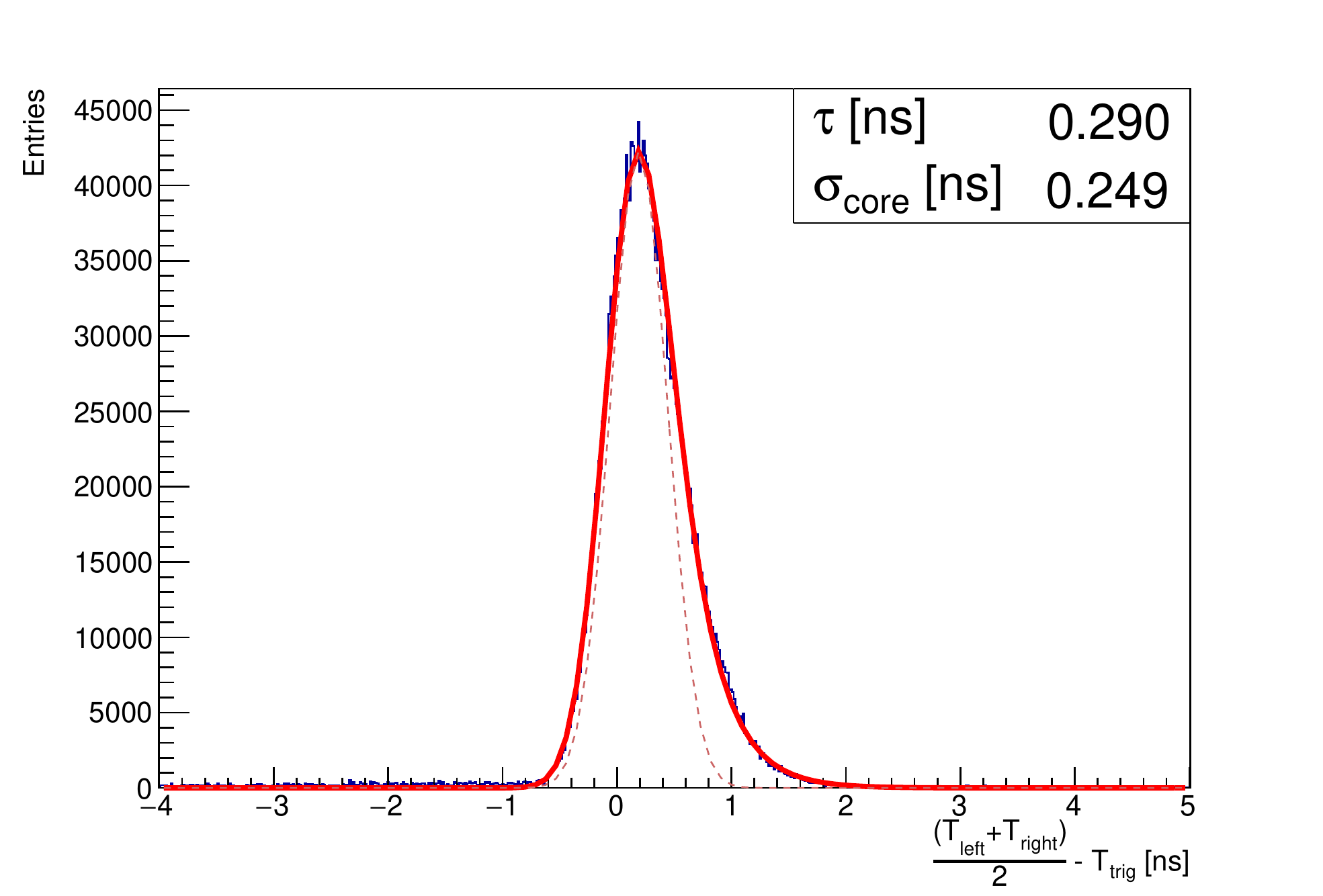}
   \includegraphics[width=0.49\textwidth]{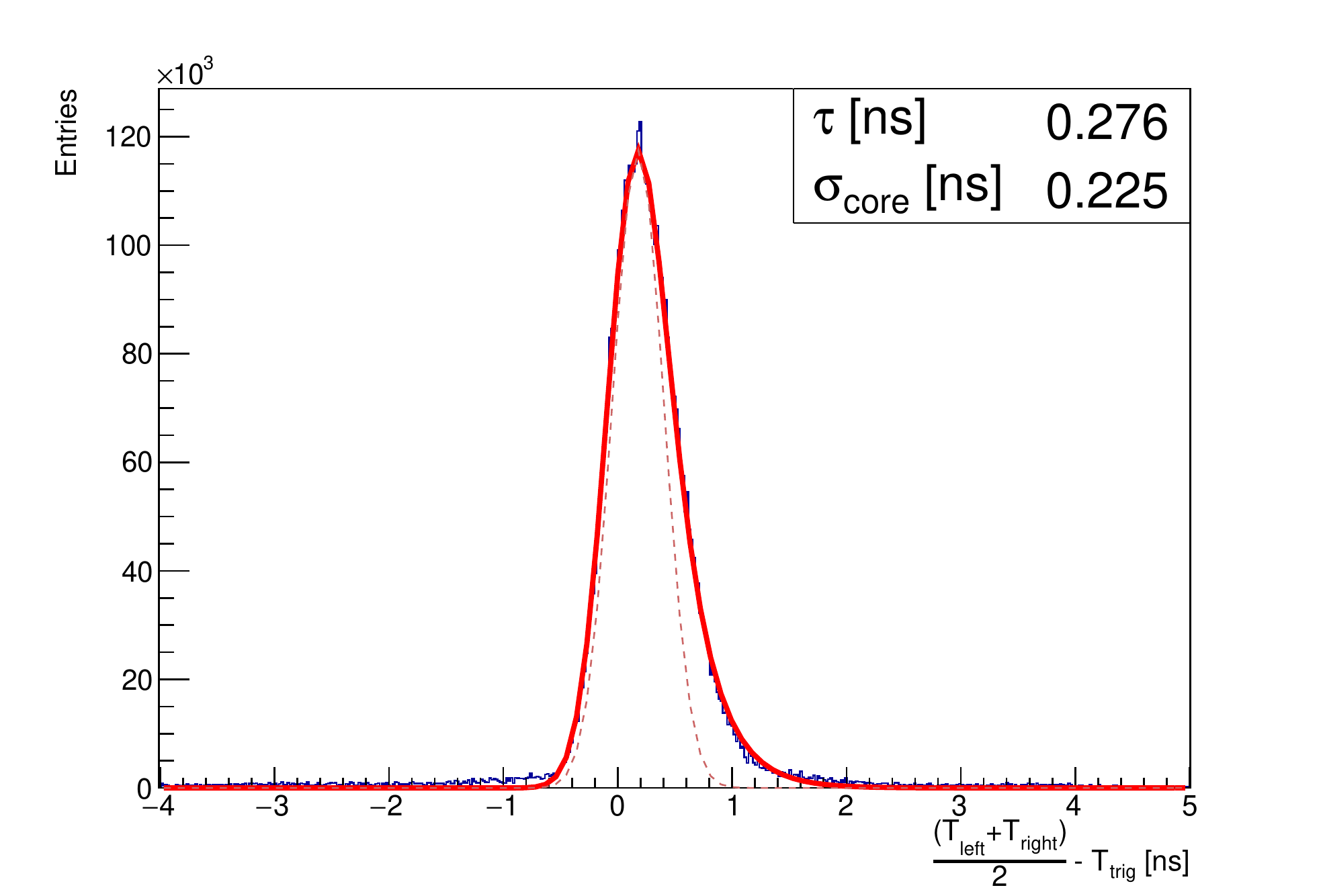}
   \includegraphics[width=0.49\textwidth]{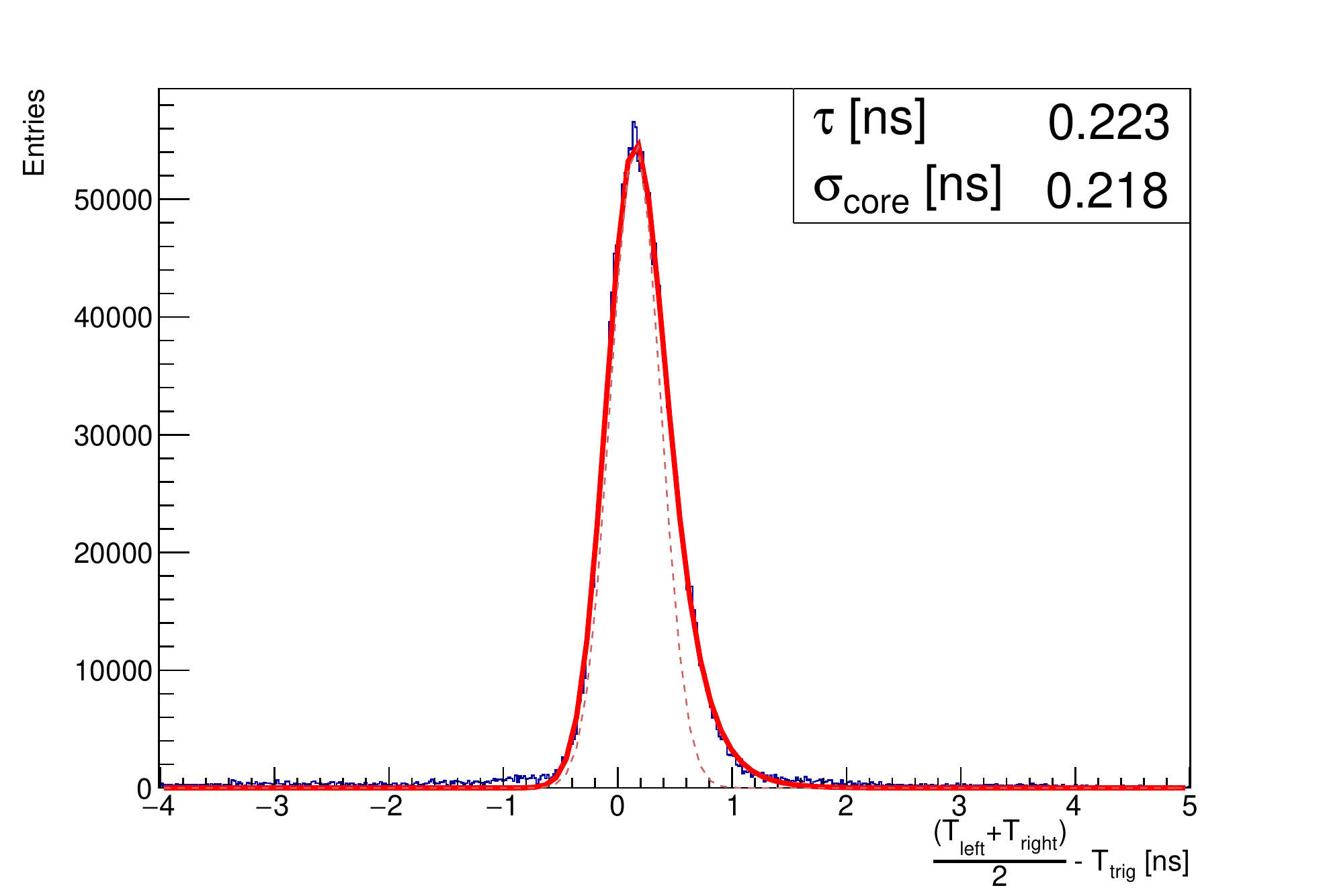}
   \caption{top) Mean-Time distribution $MT = \frac{1}{2} (T_{\rm left} + T_{\rm right})$ for a 3-layer SCSF-78 fiber ribbon
prepared with black epoxy (left) and for a 3-layer NOL-11 ribbon prepared with clear epoxy with a 20\% \TiO admixture (right).
bottom) Same, but for 4-layer SciFi ribbons.
The distributions are well described by the EMG distribution, i.e. a convolution of a Gaussian and an exponential function (Equation~\ref{eq:EMG}). 
The tails on the right of the peaks are driven by the decay time of the scintillation light.
The jitter of the external time reference of 80~ps has not been subtracted in these plots.}
   \label{fig:MTcomp}
\end{figure}

\begin{figure}[t!]
   \centering
   \includegraphics[width=0.49\textwidth]{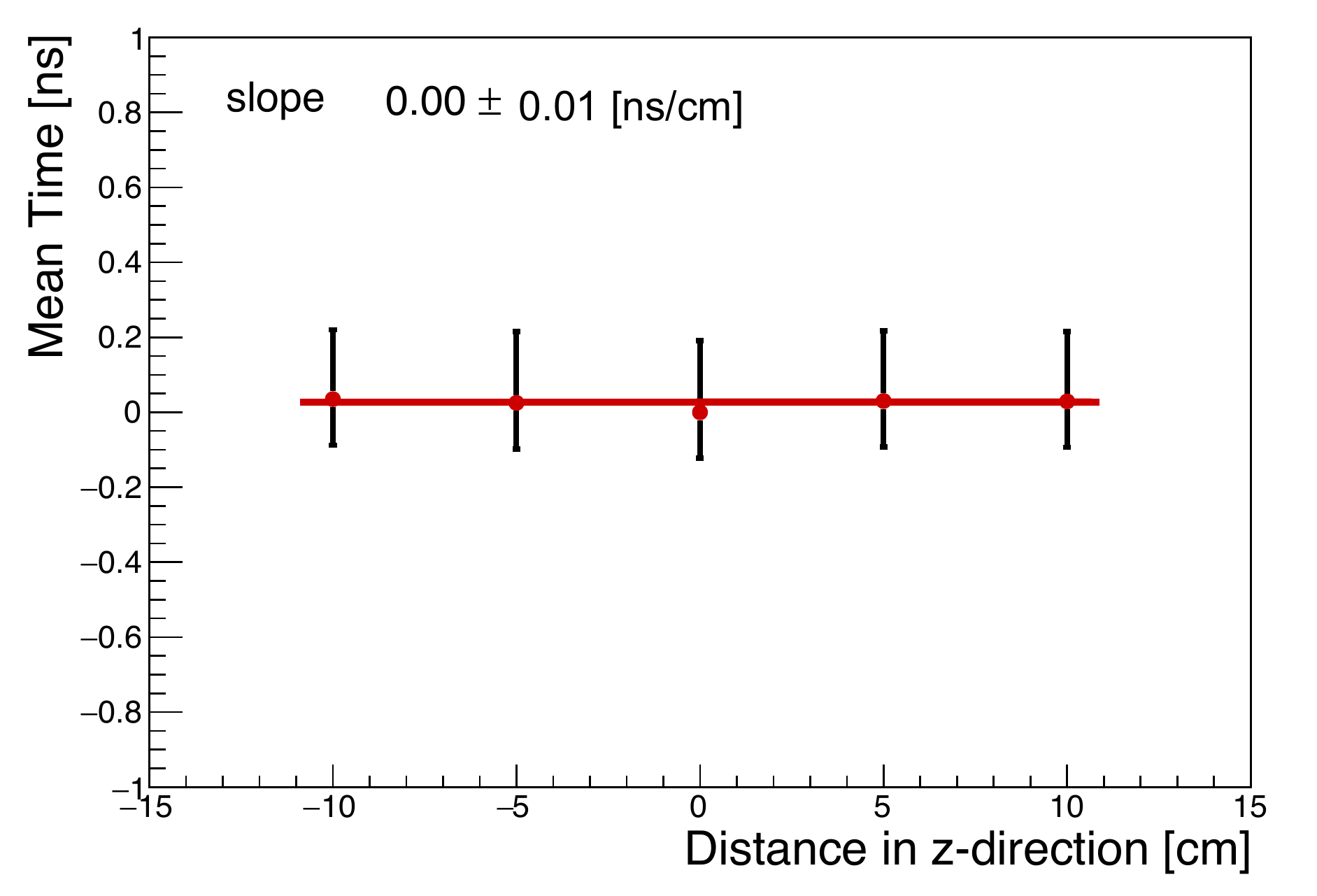}
   \includegraphics[width=0.49\textwidth]{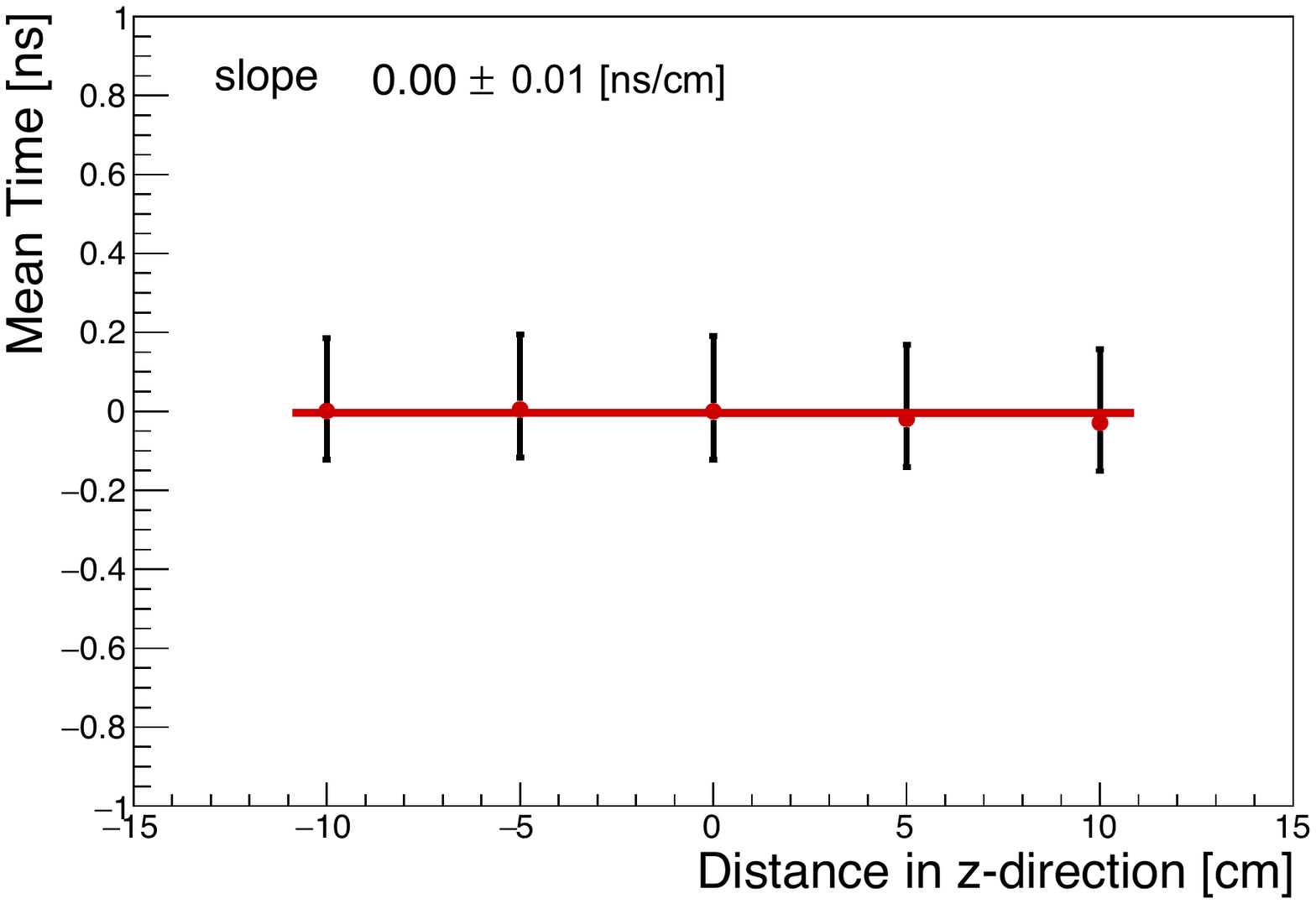}
   \caption{Mean-time $MT$ as function of the impact point along the SciFi ribbon
for a 3-layer SCSF-78 fiber ribbon prepared with black epoxy (left)
and for a 3-layer NOL-11 fiber ribbon prepared with clear epoxy with a 20\% \TiO admixture (right).
The vertical error bars are asymmetric to reflect the shape of the $MT$ distribution.
The data points are fitted with a straight line.}
   \label{fig:MTz}
\end{figure}

The {\it common} approach to measure time, when reading out a scintillator at both ends,
is to form the so called Mean-Time $MT$,
defined as 
\begin{equation}
MT = \frac{1}{2}(T_{\rm left} + T_{\rm right}) - T_0 \; ,
\end{equation}
where $T_0$ represents an external time reference, which could come from a low jitter trigger,
or from a second timing measurement, like in a time of flight measurement,
or from the system clock.
In the study of the $MT$ distribution, $T_0$ can be considered as an additive constant,
provided that $\sigma_{T_0}$ is small compared to the width of the $MT$ distribution.
In these studies, $T_0$ is provided by the trigger time $T_{\rm trigger}$
with $\sigma_{\rm trigger} = 80~{\rm ps}$ (Section~\ref{sec:setup})
and is subtracted from the  $MT$ distribution.
The principal reason for forming the mean time is that $MT$ does not depend, to a good degree of accuracy,
on the hit position in the scintillator (see Figure~\ref{fig:MTz}).
Hence one does not need to take into account the hit position in the scintillator and to correct for the light propagation in the scintillator.
By construction one would expect that the width of the $MT$ distribution is half the width of the $\Delta T$ distribution, i.e. 
\begin{equation}
\sigma_{MT} = \frac{1}{2} \sigma(T_{\rm left} + T_{\rm right}) = \frac{1}{2} \sigma(T_{\rm left} - T_{\rm right}) = \frac{1}{2} \sigma_{\Delta T} \; .
\end{equation}
In general this assumption holds, provided that the time measurements are normally distributed,
as would be the case with a thick scintillator ($\mathcal{O}(100)$ detected photons).
The $MT$ distributions for 3-layer and 4-layer SciFi ribbons made of SCSF-78 and NOL-11 fibers are shown in Figure~\ref{fig:MTcomp}.
These measurements have been taken with the beam crossing the SciFi ribbons in the middle (i.e. 15 cm from the ribbon's ends).
As it can be observed, the $MT$ distribution is not symmetric w.r.t. the peak
and exhibits a tail extending to the right of the peak,
which is driven by fiber's decay time because of the relatively low photon statistics
With a thicker scintillator the tail would be completely absorbed below the Gaussian.
The $MT$ distributions in Figure~\ref{fig:MTcomp} are well described by the EMG distribution (Equation~\ref{eq:EMG}).
More than 85\% of events (it depends also on the SciFi ribbon type) are described by the Gaussian. 
Since the $MT$ distributions are not too skewed,
we take $\sigma_{\rm core}$ of the $MT$ distribution as indication for the time resolution $\sigma_{MT}$.
Otherwise one could report separate errors for the left and right sides of the peak,
for instance the FWHM for the left and right side of the distribution.
After subtraction of the $T_{\rm trigger}$ jitter of 80~ps,
the time resolution $\sigma_{MT}$ achievable with the 3-layer SCSF-78 SciFi ribbon prepared with black epoxy is $\sim 250~{\rm ps}$,
and it is $\sim 210~{\rm ps}$ for a 4-layer SCSF-78 SciFi ribbon.
Similarly it is $\sim 240~{\rm ps}$ for the 3-layer NOL-11 SciFi ribbon prepared with clear epoxy with a 20\% \TiO admixture,
and it is $\sim 200~{\rm ps}$ for a 4-layer NOL-11 SciFi ribbon.
The effect of the shorter decay time of the NOL-11 fibers of 1.1~ns results in a reduced tail and is visible when comparing the different $MT$ distributions.
However, the effect of going from a 3-layer to a 4-layer SciFi ribbon is much more marked,
resulting also in an improved time resolution,
which proves the importance of the light yield.
The best timing performance in these studies has been achieved with a 4-layer NOL-11 SciFi ribbon.
For the \mude experiment, however, the SCSF-78 fibers have been selected,
since they were available at the time of the SciFi detector development and construction.

Figure~\ref{fig:MTz} shows the peak values of the mean time $MT$ distribution for different positions along the SciFi ribbon.
Since the total distance traveled by light emitted in opposite hemispheres is constant (i.e. the length of the SciFi ribbon)
one can expect that the mean time $MT$ will remain constant along the ribbon and independent of the hit position.
Figure~\ref{fig:MTz} shows that indeed this is the case within our resolution,
and that the width of the $MT$ distribution does not change appreciably along the SciFi ribbon, as well.
The errors shown in Figure~\ref{fig:MTz} are taken at the FWHM of the $MT$ distribution separately for the left and right sides of the distribution
to reflect the shape of the $MT$ distribution.
Therefore, the mean-time $MT$ is a good observable for timing measurements and $MT$ does not depend on the hit position
(i.e. it does not require to take into account the hit position and correct for the light propagation in the fiber ribbon).

Table~\ref{tab:time} summarizes the timing performance of different fiber ribbons.
Both, the time-difference resolution and mean-time resolution are reported.

\begin{table}[h!]
   \centering
   \begin{tabular}{|l c |c | c c |}
    \hline
fiber type & \# of layers & $\sigma_{\Delta T}~[{\rm ps}]$ & $\sigma_{MT}~[{\rm ps}]$ & $\tau_{EMG}~[{\rm ps}]$ \\
   \hline
SCSF-78 & 3 & 449 & 245 & 371 \\
SCSF-78 & 4 & 402 & 210 & 276 \\
 NOL-11 & 3 & 409 & 236 & 290 \\
 NOL-11 & 4 & 395 & 203 & 223 \\
SCSF-81 & 4 & 544 & n/a & n/a \\
   \hline   
   \end{tabular}
   \caption{Timing performance of 3- and 4-layers Scifi ribbons made of different fibers.
Both, the time difference resolution $\sigma_{\Delta T}$ taken as the FWHM/2.355 of the $\Delta T$ distribution,
and the mean-time distribution resolution $\sigma_{MT}$, taken as the core of the $MT$ distribution after subtracting $\sigma_{\rm trigger}$,
and the parameter $\tau_{EMG}$ of the EMG fit, 
are reported.}
   \label{tab:time}
\end{table}

\clearpage
\section{Spatial Resolution}
\label{sec:space}

\begin{figure}[!h]
   \centering
   \includegraphics[width=0.6\textwidth]{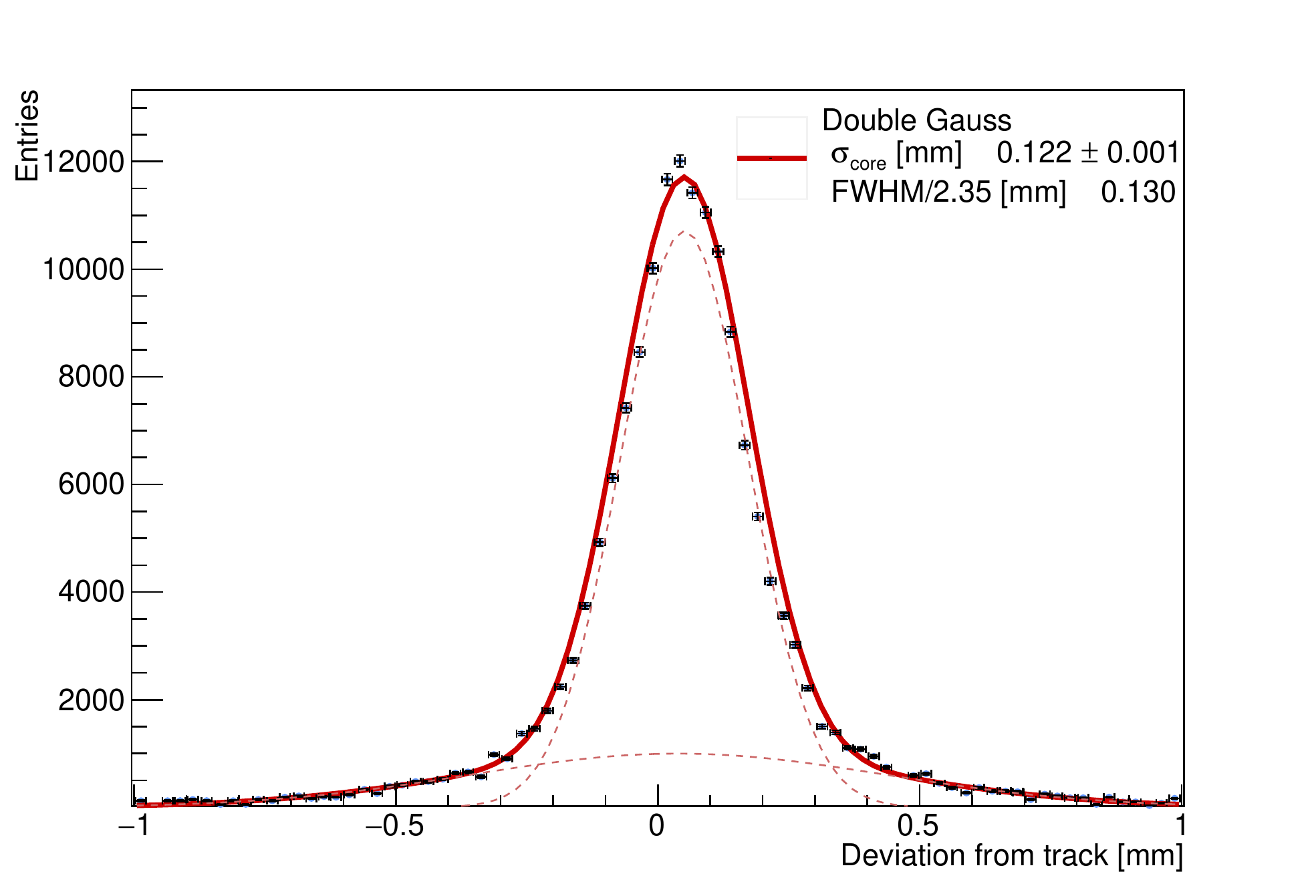}
   \caption{Distribution of residuals between the clusters reconstructed in the DUT
and the tracks reconstructed in the SciFi telescope. 
A spatial resolution of about $100~\mu{\rm m}$ is obtained after subtracting the track reconstruction uncertainty of $80~\mu{\rm m}$
with a 150~GeV proton beam.}
   \label{fig:spatial}
\end{figure}

The spatial resolution of the SciFi detector has been studied 
with a setup similar to the one illustrated in Figure~\ref{fig:sf_setup}
at the CERN SPS using a 150~GeV proton beam~\cite{DamyanovaNIM}.
At this energy the multiple scattering in the SciFi telescope is negligible
compared to the tracking resolution of the SciFi telescope.
Figure~\ref{fig:spatial} shows the distribution of residuals
between the hit position measured in the third most downstream 4-layer SCSF-78 fiber detector (i.e. the DUT)
and the hit position predicted by the tracks measured with the three other SciFi detectors.
The residual distribution is fitted with a sum of two Gaussians with the same central value to describe better the tails of the distribution.
The FWHM/2.355 of the residual distribution is $130~\mu{\rm m}$.
After subtracting the uncertainty on the track reconstruction of $80~\mu{\rm m}$
a spatial resolution of about $100~\mu{\rm m}$ has been extracted.
While this resolution is several times worse than the spatial resolution of the \mude tracking Si-pixel detectors,
it will be essential for associating correctly a hit in the fiber detector to the tracks reconstructed with the Si-pixels.

\clearpage
\section{Efficiency}
\label{sec:efficiency}

The second most important feature of the SciFi detector for \mude is the detection efficiency. 
To associate a hit in the SciFi detector to a track reconstructed in the \mude Si-pixel tracking detectors (Figure~\ref{fig:experiment})
and to extract the timing information (i.e. the mean-time),
clusters are required to be identified at both sides of the SciFi ribbons
and to be matched in space and time.
The cluster identification and selection are discussed in Section~\ref{sec:cluster}.
The {\it left} and {\it right} clusters are matched together by requiring that they have in common at least one scintillating fiber,
i.e. at least one matching channel in the two SiPM arrays at opposite ends.
From the uncorrelated hit rate in Figure~\ref{fig:speed2} (left) one can asses that the totality (i.e. $> 99.9\%$) of uncorrelated clusters
formed from dark counts are rejected by this requirement.
The clusters are matched in time by requiring that the time difference $| \Delta T = T_{\rm left} - T_{\rm right} |$ is smaller than a predefined value, 
typically $3 \times$ or $5 \times$ the width of the $\Delta T$ distribution $\sigma_{\Delta T}$.
In applying this selection, care has to be taken to account for the drift of the $\Delta T$ induced by the hit position
along the SciFi ribbon (Figure~\ref{fig:DeltaTz}).
Moreover, the time matching requirement rejects accidental combinations of clusters from track pileup at high rates.

Following the method described in~\cite{PDG},
the efficiency estimator is taken as the mode of the probability distribution $P(\epsilon | k, N, I)$,
where $k$ are the observed events (i.e. the number of matched clusters)
out of $N$ trials (i.e. the total number of tracks reconstructed in the SciFi telescope crossing the SciFi ribbon)
with the term $I$ encoding any knowledge available prior to the measurement.
The maximum of the efficiency distribution is always at
\begin{equation}
{\hat \epsilon} = \frac{k}{N} = \frac{\rm \#~of~matched~clusters}{\rm \#~of~reconstructed~tracks}\; .
\end{equation}
The uncertainties of the efficiency estimator are taken at the boundaries of a 68.27\% central confidence interval
exploiting the Fisher-Snedecor distribution~\cite{James}.
For a low number of events (trial and observed events) of $\mathcal{O}{(1000)}$,
the boundaries of the confidence interval are asymmetric around the mode of the distribution,
however they approach the $\mu \pm 1 \sigma$ region of a Normal distribution at much larger numbers.

The cluster detection and matching efficiency is evaluated for the SciFi ribbon (i.e. the DUT) 
placed between the second and fourth most downstream Scifi ribbons (Figure~\ref{fig:setup}).
The efficiency is evaluated for different values of the space and time intervals, hit selection threshold, cluster multiplicity,
and SiPM bias voltage.
The efficiency is evaluated in horizontal and vertical scans over the fiber ribbons, as well.
The beam is impinging normally in the middle of the SciFi ribbon, if not stated otherwise.
The efficiencies are presented in the following Figures.
The impact of different cluster selections on the cluster finding and matching efficiency is summarized in Table~\ref{tab:eff_cl} for different timing intervals.

\begin{table}[t!]
   \centering
   \begin{tabular}{| c | c| c | c | c | c |}
    \hline
      threshold     &    cluster   & single sided & \multicolumn{3}{c|}{matched clusters}  \\
$[{\rm ph.e.}]$  & multiplicity &    cluster     & no $\Delta T$ cut & $|\Delta T| < 5 \times \sigma_{\Delta T}$ & $|\Delta T| < 3 \times \sigma_{\Delta T}$ \\
    \hline
    0.5 & $\geq 2$ & $97.5^{+0.3}_{-0.4}$ & $97.5^{+0.3}_{-0.4}$ & $96.4^{+0.4}_{-0.5}$ & $93.6^{+0.5}_{-0.6}$ \\
    1.5 & $\geq 1$ & $99.4^{+0.1}_{-0.2}$ & $99.4^{+0.1}_{-0.2}$ & $98.3^{+0.2}_{-0.3}$ & $95.5^{+0.4}_{-0.5}$ \\
    1.5 & $\geq 2$ & $84.3^{+0.6}_{-0.7}$ & $84.3^{+0.6}_{-0.7}$ & $83.9^{+0.7}_{-0.8}$ & $80.3^{+0.7}_{-0.8}$ \\
    2.5 & $\geq 1$ & $96.6^{+0.4}_{-0.5}$ & $96.5^{+0.4}_{-0.5}$ & $95.5^{+0.4}_{-0.5}$ & $93.0^{+0.5}_{-0.6}$ \\
    2.0 & $\geq 2$ & $72.7^{+1.1}_{-1.2}$ & $72.6^{+1.1}_{-1.2}$ & $72.3^{+1.1}_{-1.2}$ & $71.1^{+0.2}_{-1.2}$ \\  \hline   
  \end{tabular}
  \caption{Cluster matching efficiency in \% for different thresholds, cluster multiplicities, and different timing cuts
for a 3-layer SCSF-78 fiber ribbon prepared with black epoxy.}
  \label{tab:eff_cl}
\end{table}

\subsection{Threshold Scans}

\begin{figure}[b!]
   \centering
   \includegraphics[width=0.49\textwidth]{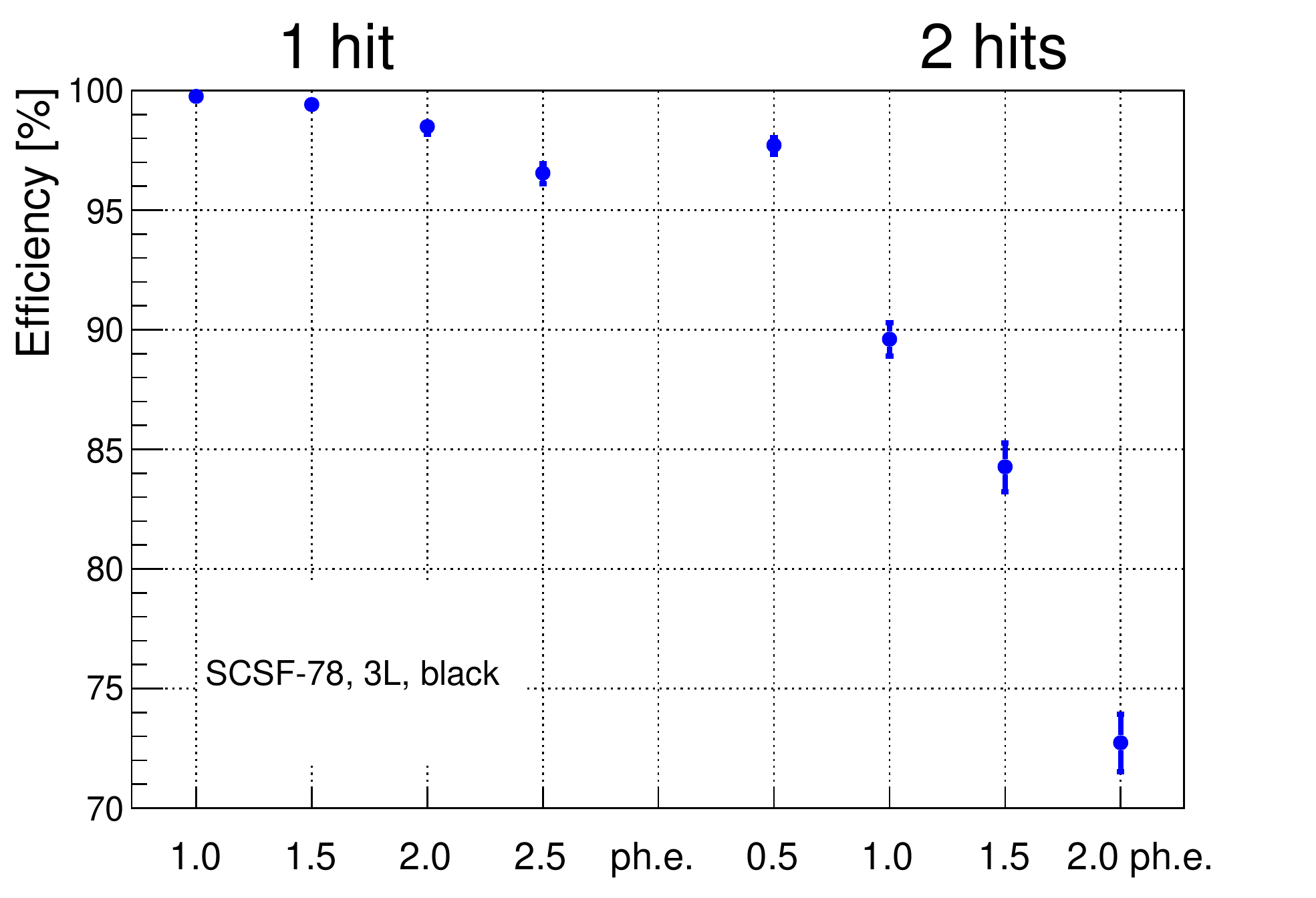}
   \includegraphics[width=0.49\textwidth]{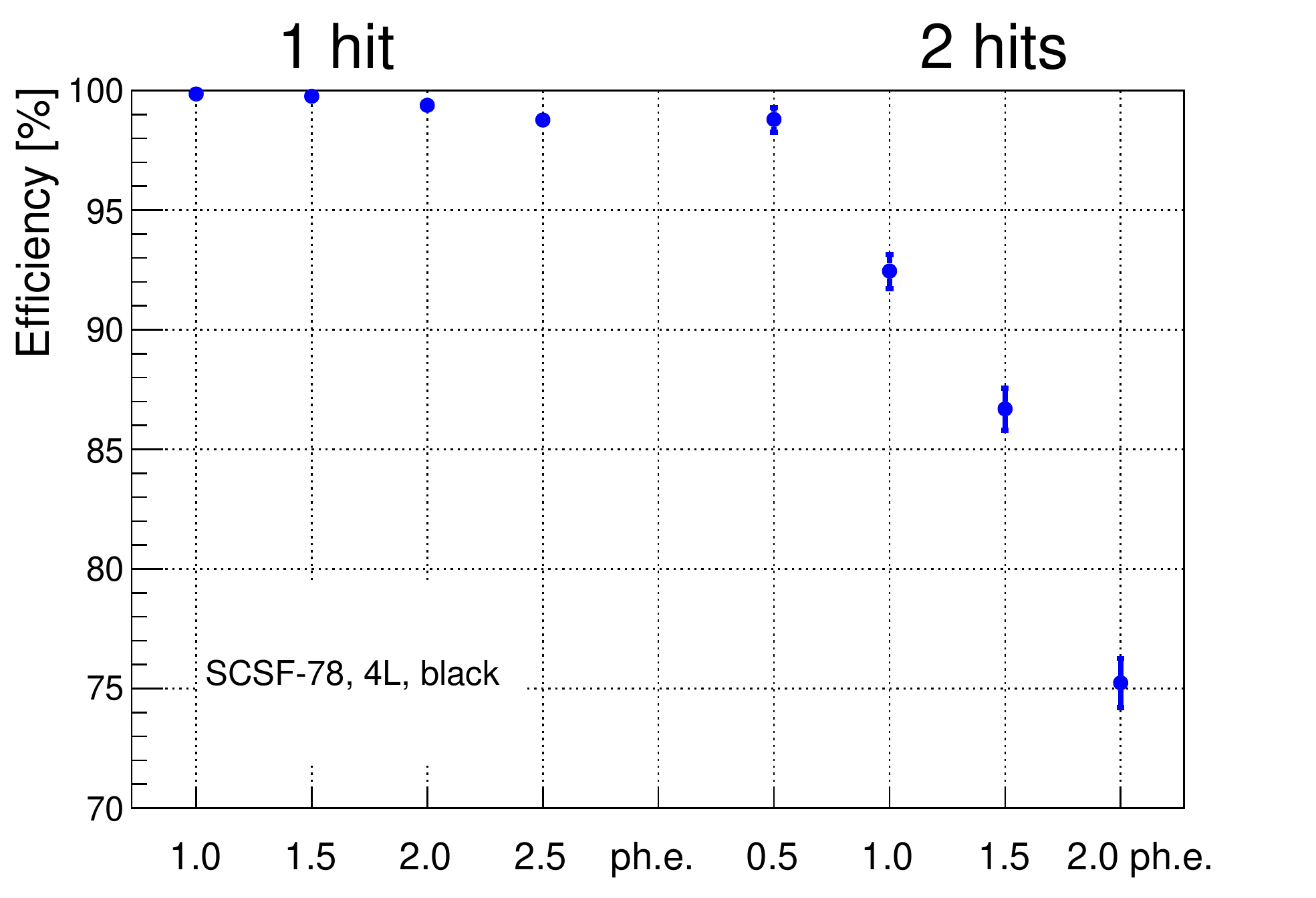}
   \caption{Single sided cluster detection efficiency vs amplitude thresholds
for a 3-layer (left) and a 4-layer (right) SCSF-78 fiber ribbon prepared with black epoxy.
To validate the cluster at least one or at least two channels forming the cluster are required to pass the preselected thresholds.}
   \label{fig:eff_thrSE}
\end{figure}

\begin{figure}[t!]
   \centering
   \includegraphics[width=0.49\textwidth]{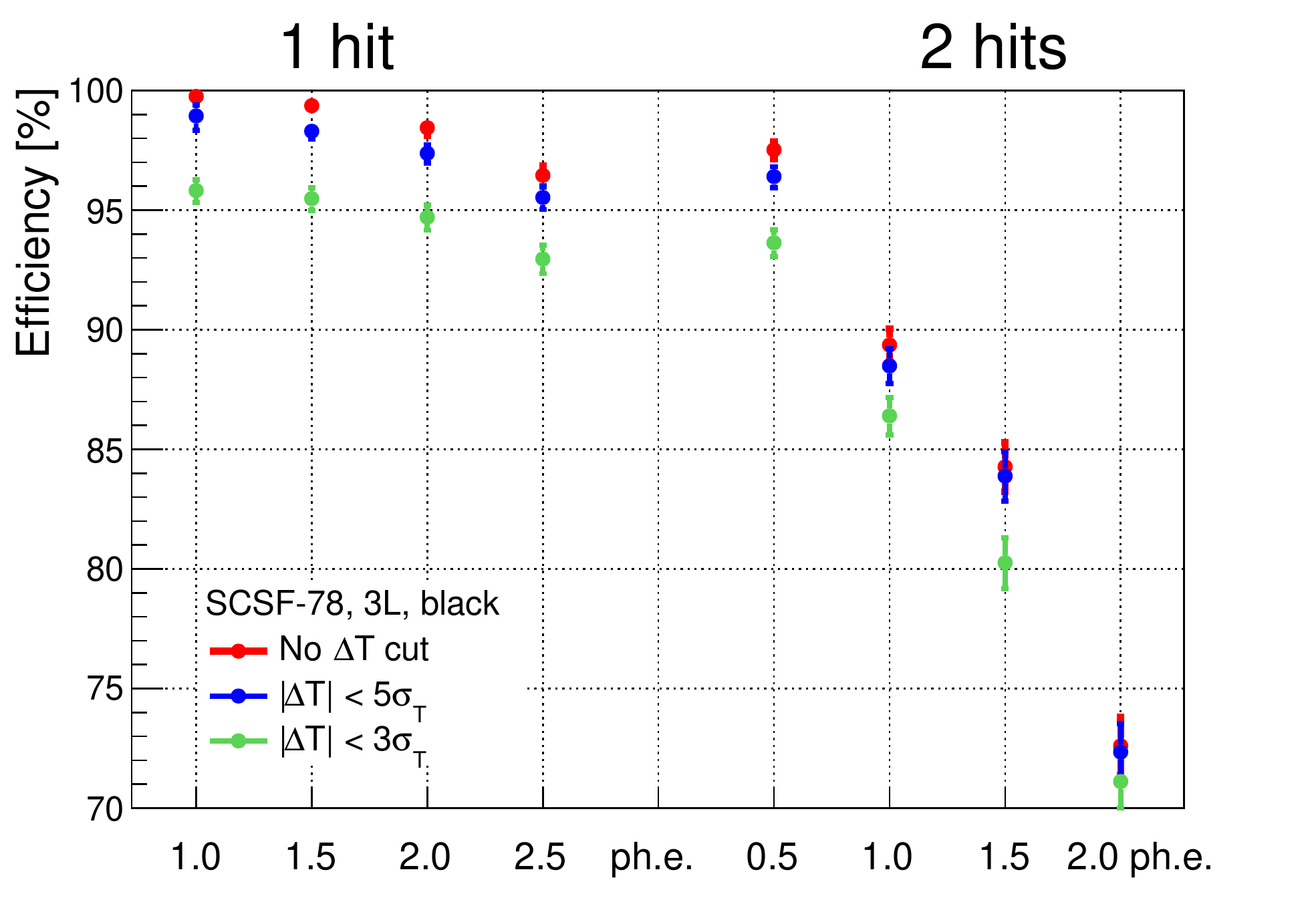}
   \includegraphics[width=0.49\textwidth]{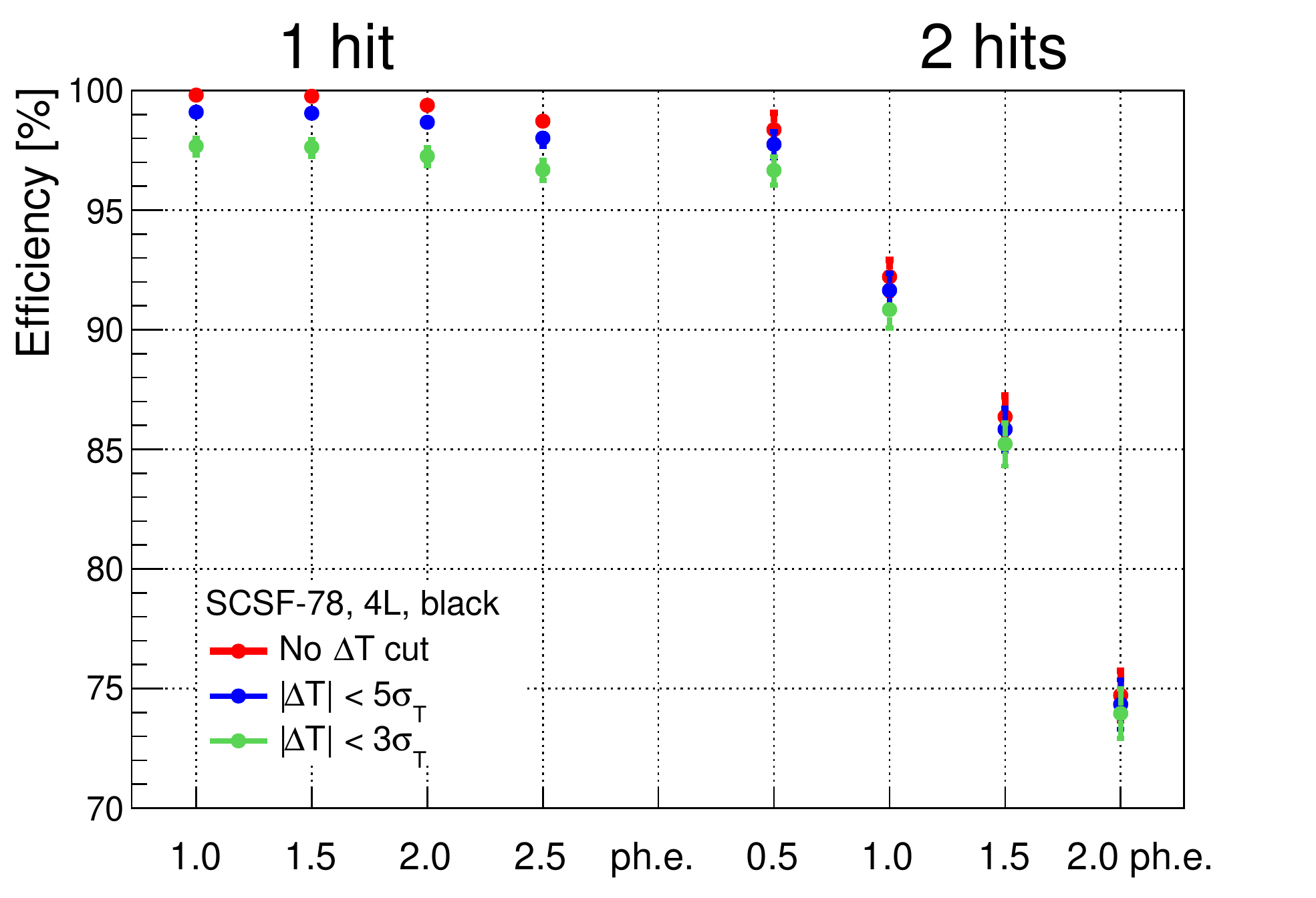}
   \caption{Cluster matching efficiency vs amplitude thresholds
for different $\Delta T$ timing selections (no $\Delta T$ cut, $5 \times \sigma_{\Delta T}$, and $3 \times \sigma_{\Delta T}$)
for a 3-layer (left) and a 4-layer (right) SCSF-78 fiber ribbon prepared with black epoxy.}
   \label{fig:eff_thr}
\end{figure}

Figure~\ref{fig:eff_thrSE} shows the single sided cluster detection efficiency as a function of the hit validation threshold for
a 3- and a 4-layer SCSF-78 fiber ribbon,
while the {\it left} / {\it right} cluster matching efficiency is shown in Figure~\ref{fig:eff_thr} for different timing selections, as well.
A glimpse of the detection efficiency can already be obtained from the cluster widths shown in Figures~\ref{fig:cl_size} and~\ref{fig:cl_compare}.
When only one channel is required to pass the preset amplitude threshold, the detection efficiency does not drop much with increasing threshold.
However, if two channels are required to pass the same threshold, the efficiency drops rapidly.
This can be explained by the observation that most of the light signal is concentrated in one channel, the {\it seed} channel identified in Section~\ref{sec:cluster}. 
This efficiency drop is less pronounced for a 4-layer SciFi ribbon and it can be understood by the higher light yield of the 4-layer SciFi ribbon.

For the {\it left} -- {\it right} matched clusters the efficiency shows a similar behavior:
when only one channel is required to pass the preset amplitude threshold, the detection efficiency drops slowly with increasing threshold,
while the efficiency drops rapidly, when a cluster width of at least two contiguous channels is required.
To validate the {\it left} -- {\it right} cluster, a timing cut on $|\Delta T|$ is imposed, which induces an additional drop in the efficiency.
This originates from the fact that this selection removes the tails of the $\Delta T$ distribution (Figure~\ref{fig:DeltaT}).

Based on this study, two different working points for the SciFi detector have been identified:

\begin{itemize}
\itemsep0em
   \item require at least two contiguous channels to pass a threshold of 0.5 ph.e.  \\
(this removes most of the dark counts, since it is highly unlikely that two neighboring channels will discharge at the same time);
   \item require at least one channel to pass a threshold of 1.5 ph.e. (higher thresholds can also be applied) \\ 
(this suppresses the dark counts, since it is unlikely that the dark counts pass a high threshold, and also reduces the DCR)
\end{itemize}

\noindent with a $|\Delta T|$ timing selection of $5 \times \sigma_{\Delta T}$.
Different clustering algorithms are also possible, however in this study we tried to reproduce the operation mode of the MuTRiG ASIC,
which allows one to set only one threshold for the hit validation during data taking.
The final choice on how to validate the {\it hits} in the SciFi detector will be based on the behavior of the SciFi detector in the \mude experiment
and the DCR of the irradiated SiPM arrays (Section~\ref{sec:rad}).

\subsection{HV Bias Dependence}

\begin{figure}[b!]
   \centering
   \includegraphics[width=0.49\textwidth]{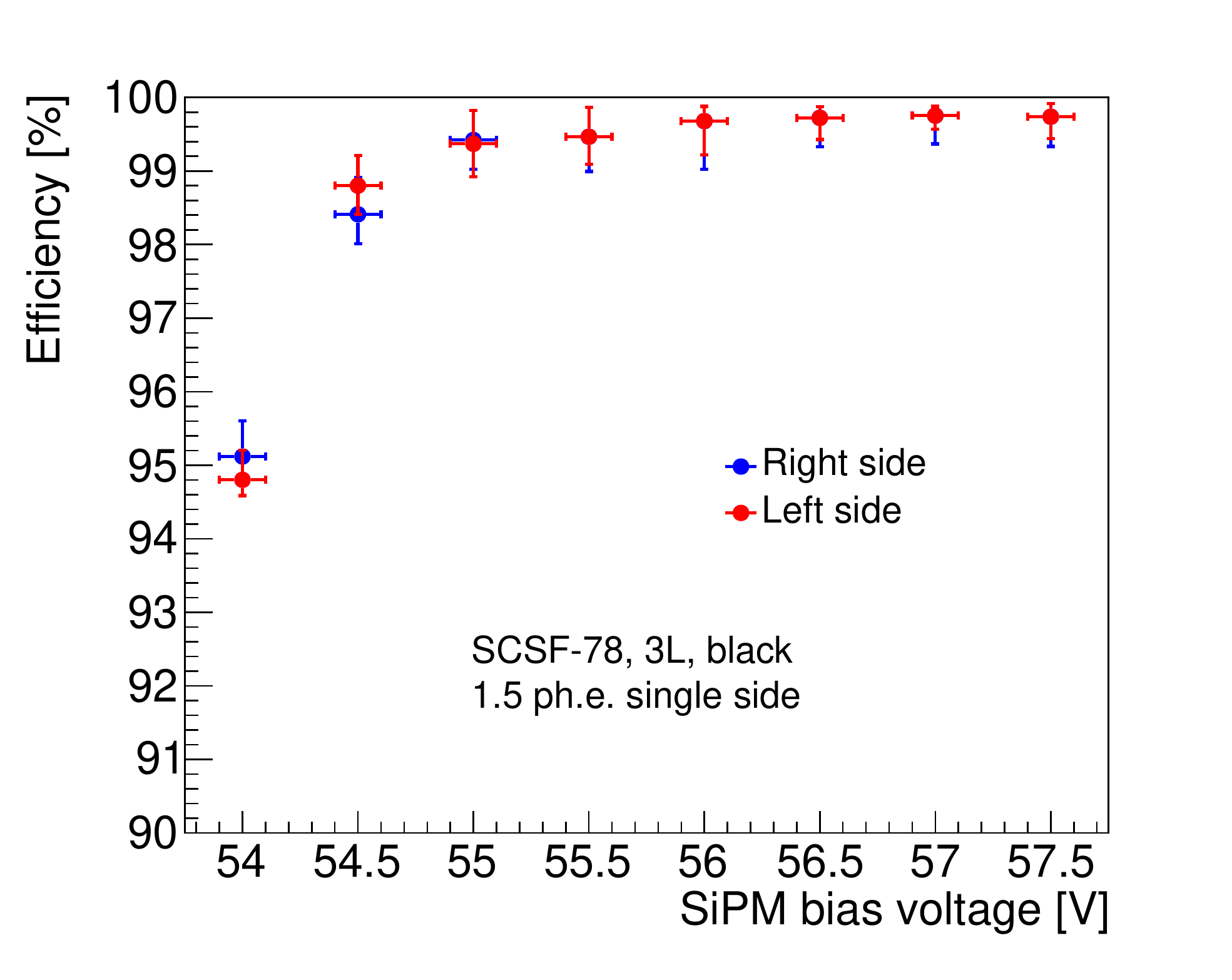}
   \includegraphics[width=0.49\textwidth]{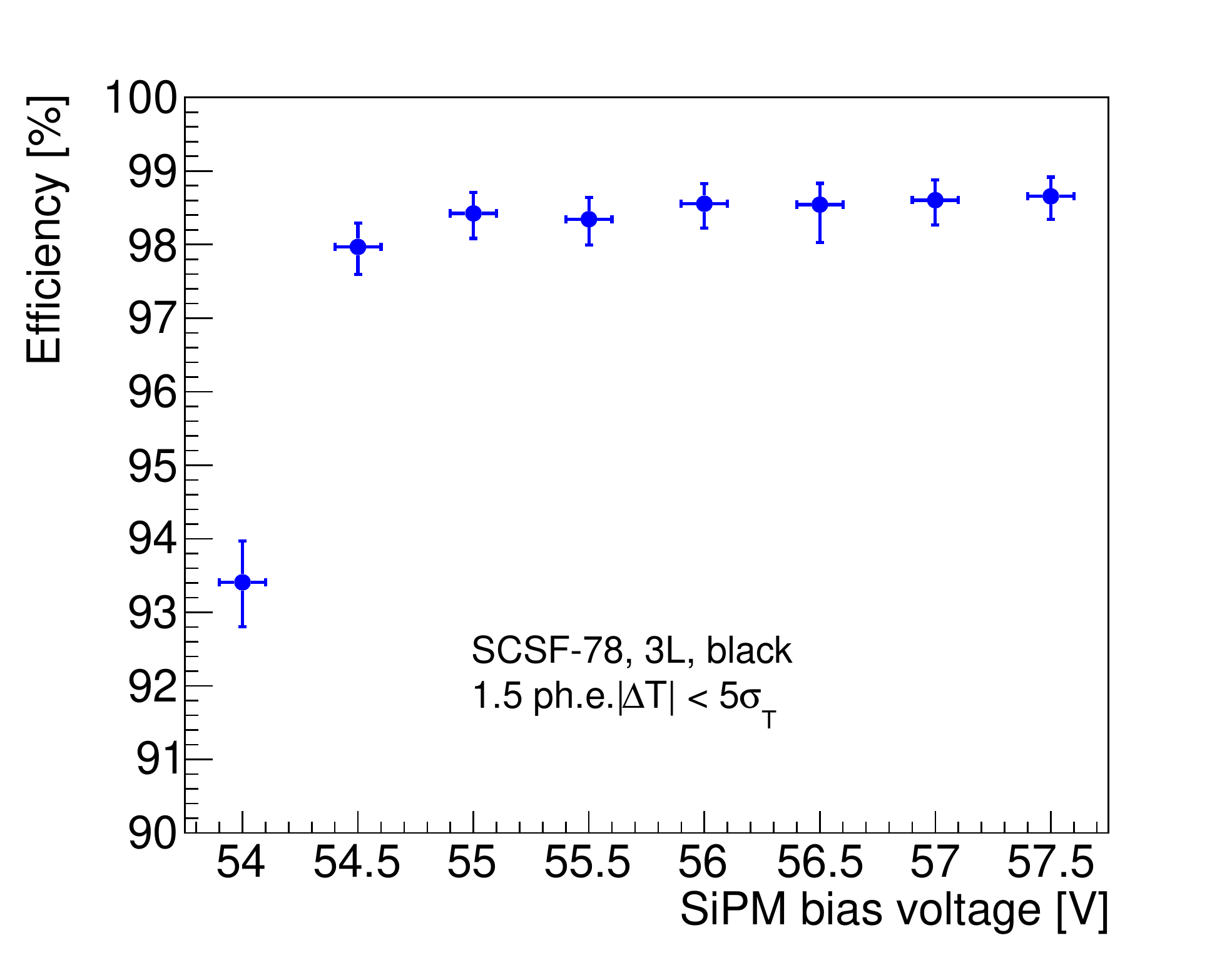}
   \caption{Single sided cluster detection efficiency (left) and cluster matching efficiency (right) vs $V_{\rm bias}$
for a 3-layer SCSF-78 fiber ribbon prepared with black epoxy.
A threshold of 1.5 ph.e. is applied to the hit validation and a timing cut of $5 \times \sigma_{\Delta T}$ is used for matched clusters.}
   \label{fig:eff_HV}
\end{figure} 

As mentioned in Section~\ref{sec:SiPM}, the photo-detection efficiency of a SiPM increases with the applied bias voltage $V_{\rm bias}$.
Since all channels of the SiPM array are operated at the same $V_{\rm bias}$, one can expect some inefficiencies at low over-bias voltages $V_{ob}$.
Figure~\ref{fig:eff_HV} shows the single sided cluster detection (left) and matching (right) efficiency dependence on the applied bias $V_{\rm bias}$.
The cluster selection requires at least one channel passing the 1.5 ph.e. threshold wit a $5 \times \sigma_{\Delta T}$ timing cut for matched clusters.
At a low over-bias ($V_{bd} \sim 52.4~{\rm V}$) some inefficiencies are indeed observed,
as well as differences between the two sensors at the opposite ends of the SciFi ribbon,
which were operated at the same $V_{\rm bias}$ (Figure~\ref{fig:eff_HV} left).
The efficiency reaches a plateau around $V_{ob} \approx 3~{\rm V}$ and
it is quite constant for higher values of $V_{\rm bias}$.
The efficiency for matched clusters behave similarly.
Therefore, the operation voltage of the SiPM arrays has been set at $V_{bias} = 55.5~{\rm V}$ (i.e. $V_{ob} \approx 3~{\rm V}$),
and it has been used for all efficiency studies discussed here.

\subsection{Position Scans}

\begin{figure} [t!]
   \centering
   \includegraphics[width=0.49\textwidth]{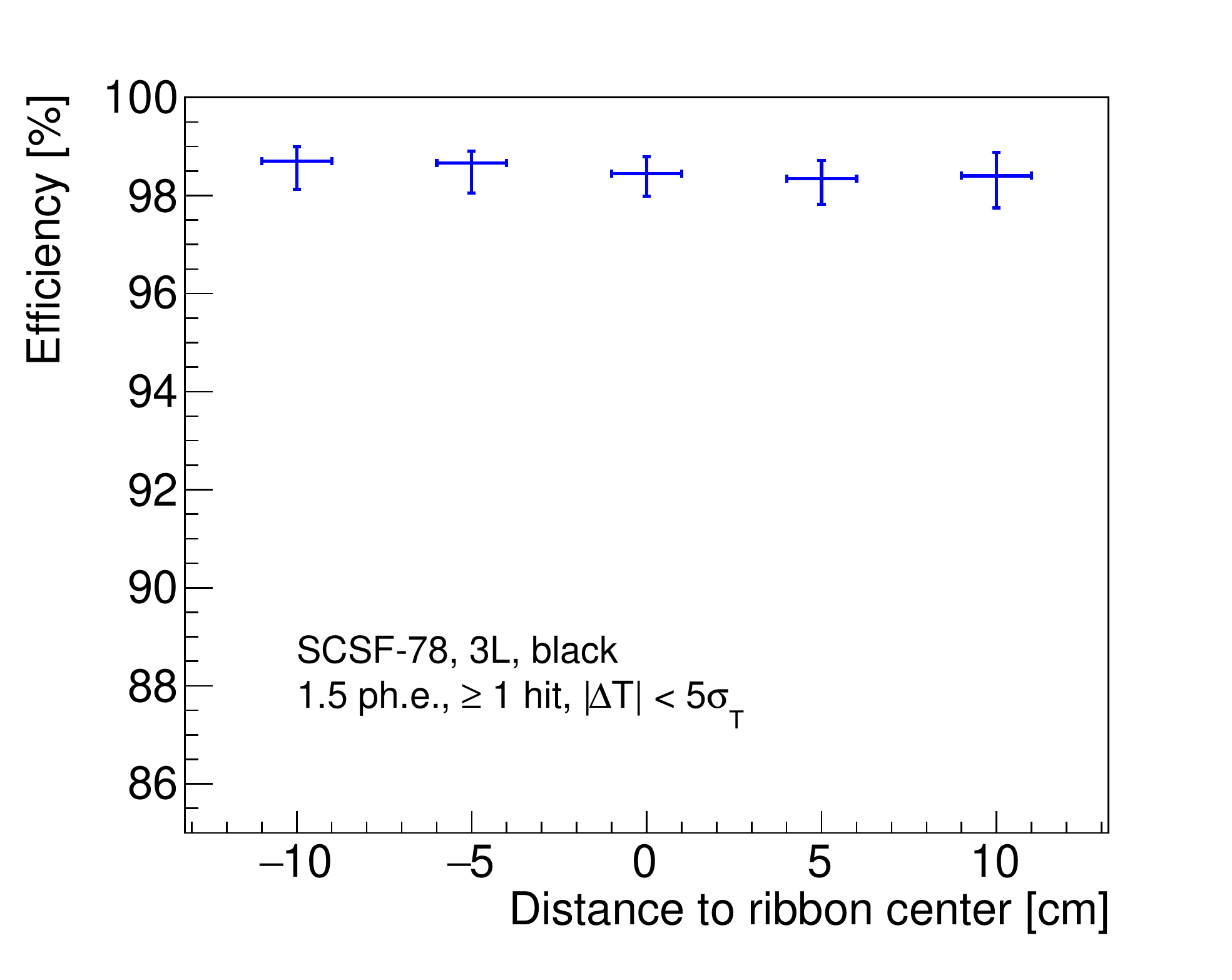}
   \includegraphics[width=0.49\textwidth]{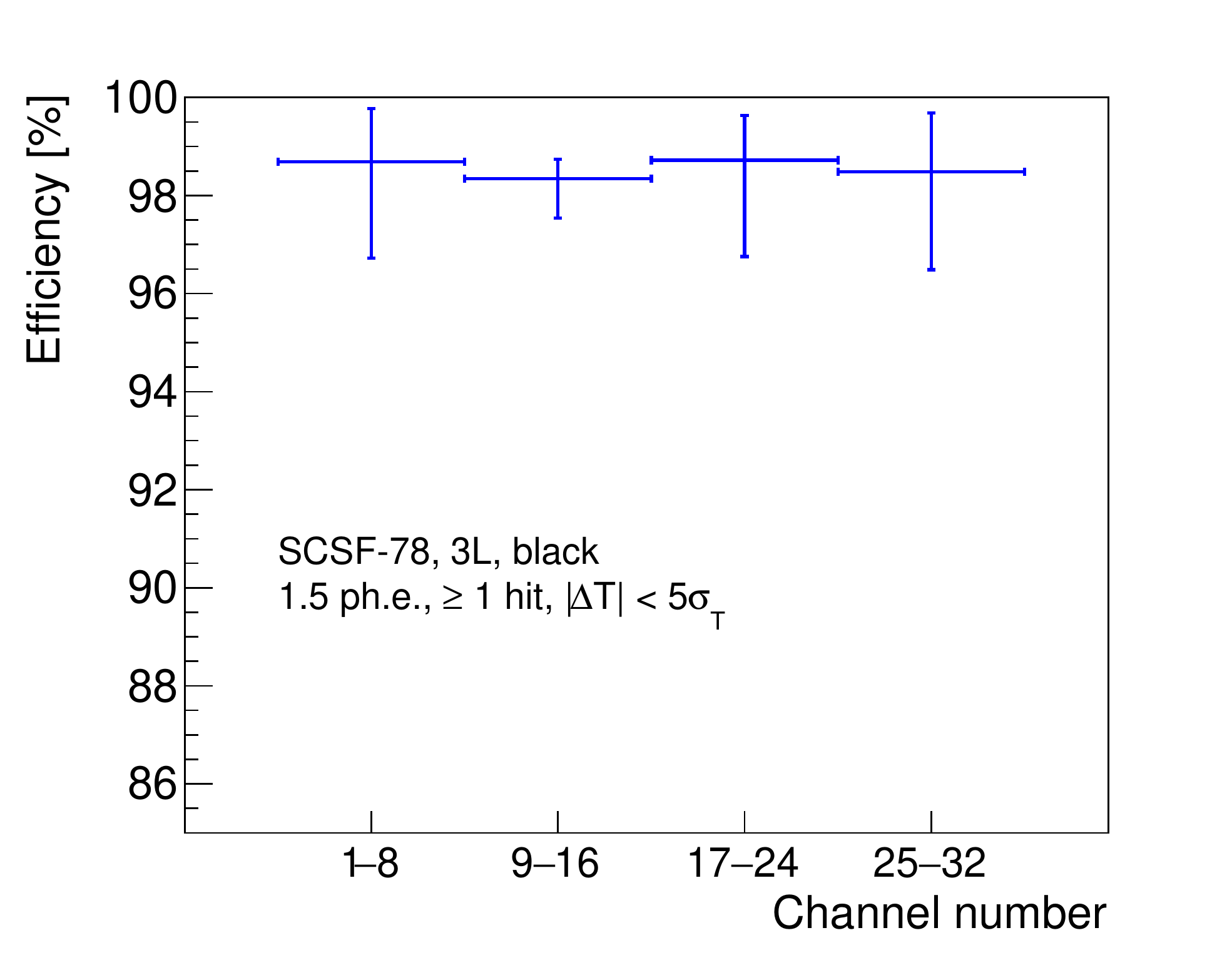}
   \caption{Cluster matching efficiency vs the beam impact point along the SciFi ribbon (horizontal scan, left)
and different impact points across the SciFi ribbon (vertical scan, right)
for a 3-layer SCSF-78 fiber ribbon prepared with black epoxy.
At least one SiPM array channel on each side of the ribbon is required to passing the 1.5 ph.e.
and a timing cut of $5 \times \sigma_{\Delta T}$ is applied.}
   \label{fig:eff_scan}
\end{figure}

The uniformity of the cluster matching efficiency along the longitudinal (horizontal) and across the transverse (vertical) directions of a SciFi ribbon
has been studied by evaluating the efficiencies for different beam impact points along the SciFi ribbon
and in consecutive groups of 8 channels of the SiPM arrays (i.e. 1~mm), respectively.
A timing cut of $5 \times \sigma_{\Delta T}$ is applied and at least one hit passing the 1.5 ph.e. threshold of 0.5 ph.e. is requested.
The efficiencies are shown in Figure~\ref{fig:eff_scan}. 
The cluster matching efficiency is quite uniform along the SciFi ribbons despite the strong variation of the light yield along the SciFi ribbon
(the light attenuation in the fibers)
(Figure~\ref{fig:clHscan}). \\
\\
\indent In summary, the cluster matching efficiency for a 3-layer SciFi ribbon is $\geq 96~\%$
for a minimum cluster multiplicity of two passing the 0.5 ph.e. threshold,
and it is $\geq 98~\%$ ($\geq 95~\%$) for a minimum cluster multiplicity of two passing the 1.5 (2.5) ph.e. threshold.
A timing cut of $5 \times \sigma_{\Delta T}$ is always applied.
For a tighter timing cut of $3 \times \sigma_{\Delta T}$ the detection efficiency drop by an additional 3\%.

\clearpage
\section{\mude SciFi Detector Implementation in \mude}
\label{sec:Mu3Eint}

In the following we discuss the implementation of the SciFi detector in the \mude detector and the readout electronics,
and how it will be operated.
The SiPM arrays will operate in a very high radiation environment of low energy electrons/positrons,
which will require the cooling of the SciFi detector below $- 10^\circ~{\rm C}$.

\subsection{Mechanics}
\label{sec:mecano}

\begin{figure}[b!]
   \centering
   \includegraphics[width=0.49\textwidth]{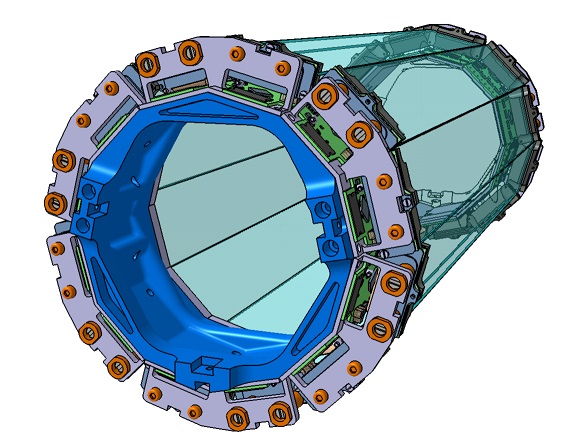}
   \hspace*{5mm}
   \includegraphics[width=0.46\textwidth]{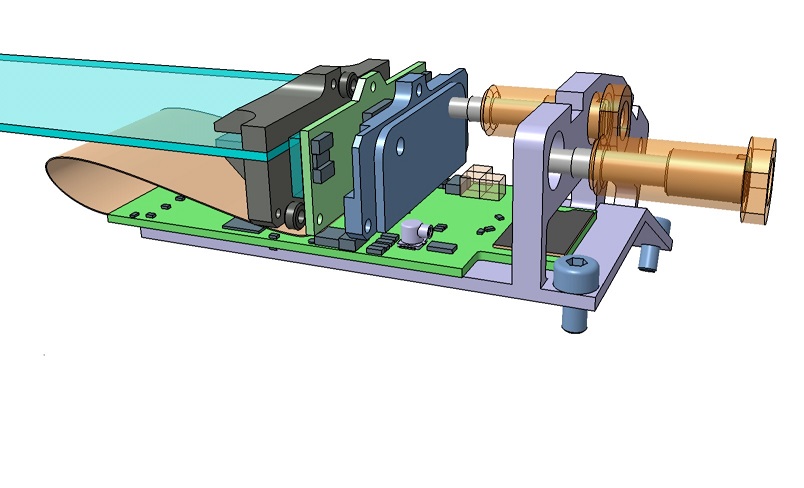}
   \caption{left) Overall structure of the scintillating fiber detector.
right) Expanded view of the SciFi support structure, showing all the elements of the detector:
SciFi ribbon, SiPM sensor, SciFi front end board and the L shape support structure.}
   \label{fig:SciFiCAD}
\end{figure}

Figure~\ref{fig:SciFiCAD} shows the overall structure of the SciFi detector with details of the principal components.
The detector is composed of 12 SciFi ribbons, 300~mm long and 32.5~mm wide.
The ribbons are staggered longitudinally by about 10~mm (Figure~\ref{fig:SciFiCAD} (left))
in order to minimize dead spaces between the ribbons
and to provide sufficient space for the spring loading of the ribbons.
To avoid sagging and to compensate for the thermal expansion (see Section~\ref{sec:scifiSag})
the ribbons are spring loaded on one side of the structure
(6 ribbons on one side and the other 6 on the other side).

To ease the sub-detector installation, the SciFi ribbons are assembled in modules.
Each module consists of two SciFi ribbons (Figure~\ref{fig:SciFiCAD} top right).
The SciFi ribbons are coupled to the SiPM arrays by simple mechanical pressure
(no grease, nor glue, nor other optical interface).
Each SiPM sensor is connected to a front-end digitizing board via a flex-print circuit.
Figure~\ref{fig:SciFiCAD} (bottom right) show an expanded view of the assembly structure:
the SciFi ribbons are attached to the SiPM arrays, which in turn are supported by stiffeners
fixed to L shaped supports,
where the assembly is also spring loaded.
The same L shapes support also the SciFi front-end boards.

The L shaped supports are fixed to a hollow dodecagonal prism shown in Figure~\ref{fig:SciFiCAD},
45~mm tall with an outer diameter of 100~mm,
which provides also the necessary cooling for the front-end electronics.
Two such cooling blocks are attached to the beam pipe on each side of the \mude detector
and connected to the pipes of the \mude liquid cooling system.
The cooling blocks are 3D-printed in aluminum with embedded piping for the circulation of the refrigerant.
Since the SiPM arrays are in thermal contact with the L shape supports, they will be cooled  by the same cooling structure.
The goal is to cool the SiPM arrays below $- 10^\circ~{\rm C}$.

\subsection{Front-End Readout Electronics}
\label{sec:MuTRiG}

The SiPMs arrays will be read out with a dedicated mixed-signal ASIC,
the MuTRiG (\underline{Mu}on \underline{T}iming \underline{R}esolver \underline{i}ncluding \underline{G}igabit-link)~\cite{MuTRiG},
capable of sustaining rates in excess of a MHz per readout channel.
The ASIC is designed in UMC 180~nm CMOS technology.
The ASIC comprises 32 input differential channels with high timing resolution,
Time-to-Digital Converters with a 50~ps bin size and a digital part to process and transfer
the data to the data acquisition system via a 1.25~Gbps LVDS serial data link with 8b/10b encoding.
Since the SiPM arrays share a common cathode, single ended input will be used.
The readout of the SiPM arrays using the MuTRiG ASIC and its performance will be discussed in detailed in~\cite{readout}.

\begin{figure}[t!]
\centering
   \includegraphics[width=0.5\textwidth]{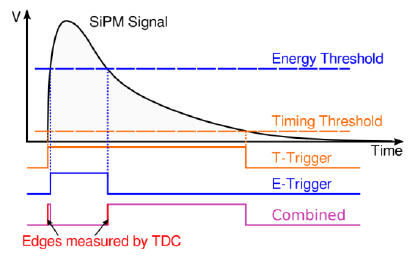}
   \caption{Principle of the MuTRiG ASIC time and energy measurements.}
   \label{fig:MuTRiG}
\end{figure}

The analog front-end is designed using fully differential structures to suppress the common mode noise
from the digital circuits in the ASIC and noise from the external sources.
The incoming SiPM signal is received by the input stage and split into a timing branch and an energy branch.
A low timing threshold extracts the time of arrival at a threshold below the SiPM single photo-electron level;
a higher energy threshold validates physical signals and provides energy information using a linearized time-over-threshold method.
The timing and energy trigger signals are combined into one single logic signal
preserving the rising edge of the timing signal and the falling edge of the energy signal,
as illustrated in Figure~\ref{fig:MuTRiG}.
The timing information is obtained from the rising edge of the timing signal
and the energy information from the time difference between the rising edge of the timing signal
and the falling edge of the energy signal.

\subsection{SiPM Irradiation}
\label{sec:rad}

\begin{figure}[t!]
  \centering
  \includegraphics[width=0.60\textwidth]{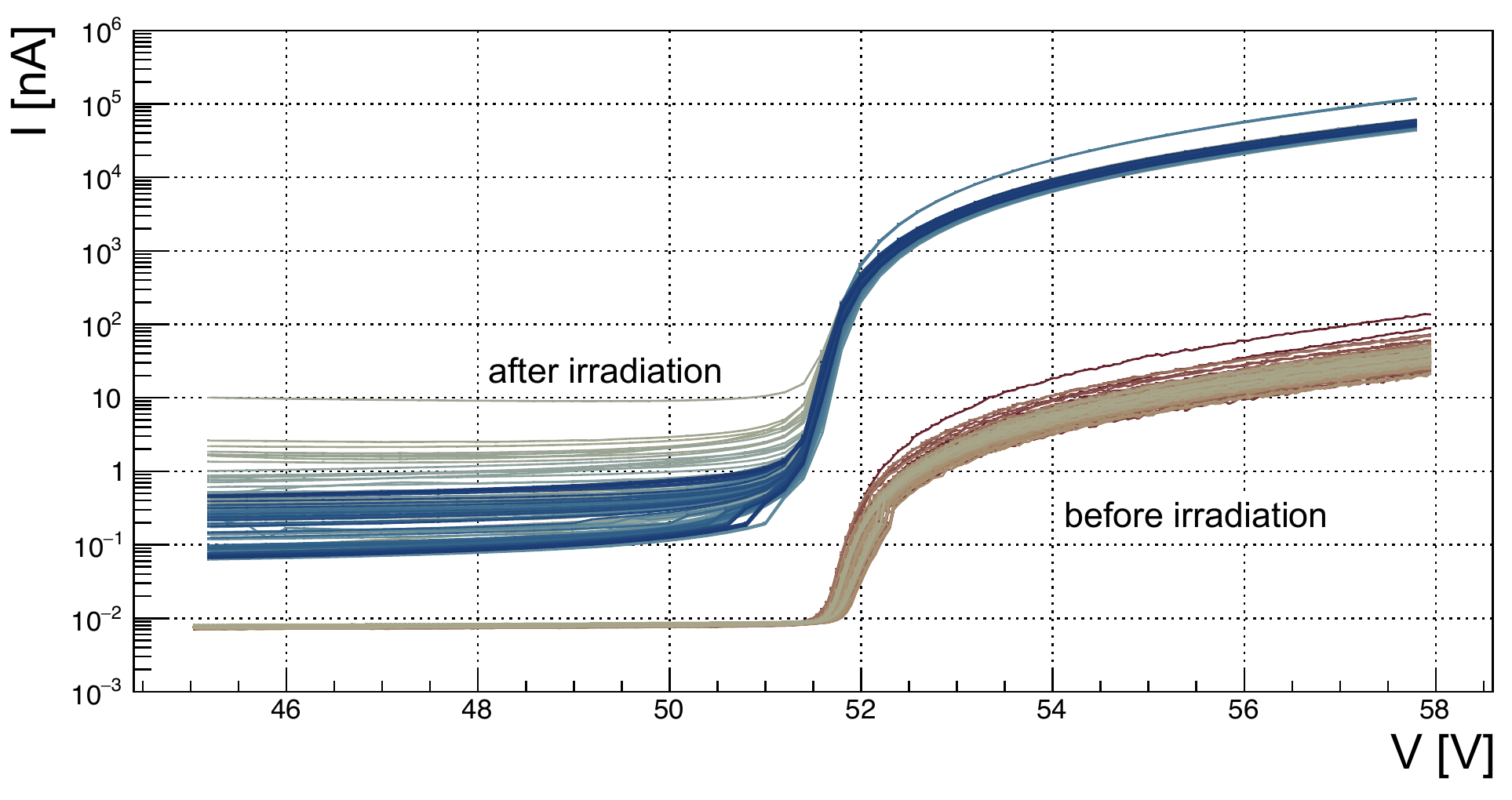}
  \caption{I--V curve for a SiPM array irradiated to $(1.5 \pm 0.3) \times 10^{12}~e^+/{\rm cm}^2$.}
\label{fig:irrad}
\end{figure}

The SiPM arrays will operate in a very high radiation environment of low energy electrons/positrons
from muon decays, which generate more damage to the sensor than m.i.p.'s, comparable to low energy neutrons.
We believe that the mechanism responsible for the damage is non-ionizing energy loss in the silicon bulk material leading to dislocations.
The literature on this is very scarce in this energy range of 10 -- 50~MeV.
Several SiPM arrays have been exposed to positrons from muon decays at rest (Michel spectrum) and irradiated at different doses
in the $\pi E5$ beamline at PSI.

\begin{figure}[b!]
  \centering
  \includegraphics[width=0.44\textwidth]{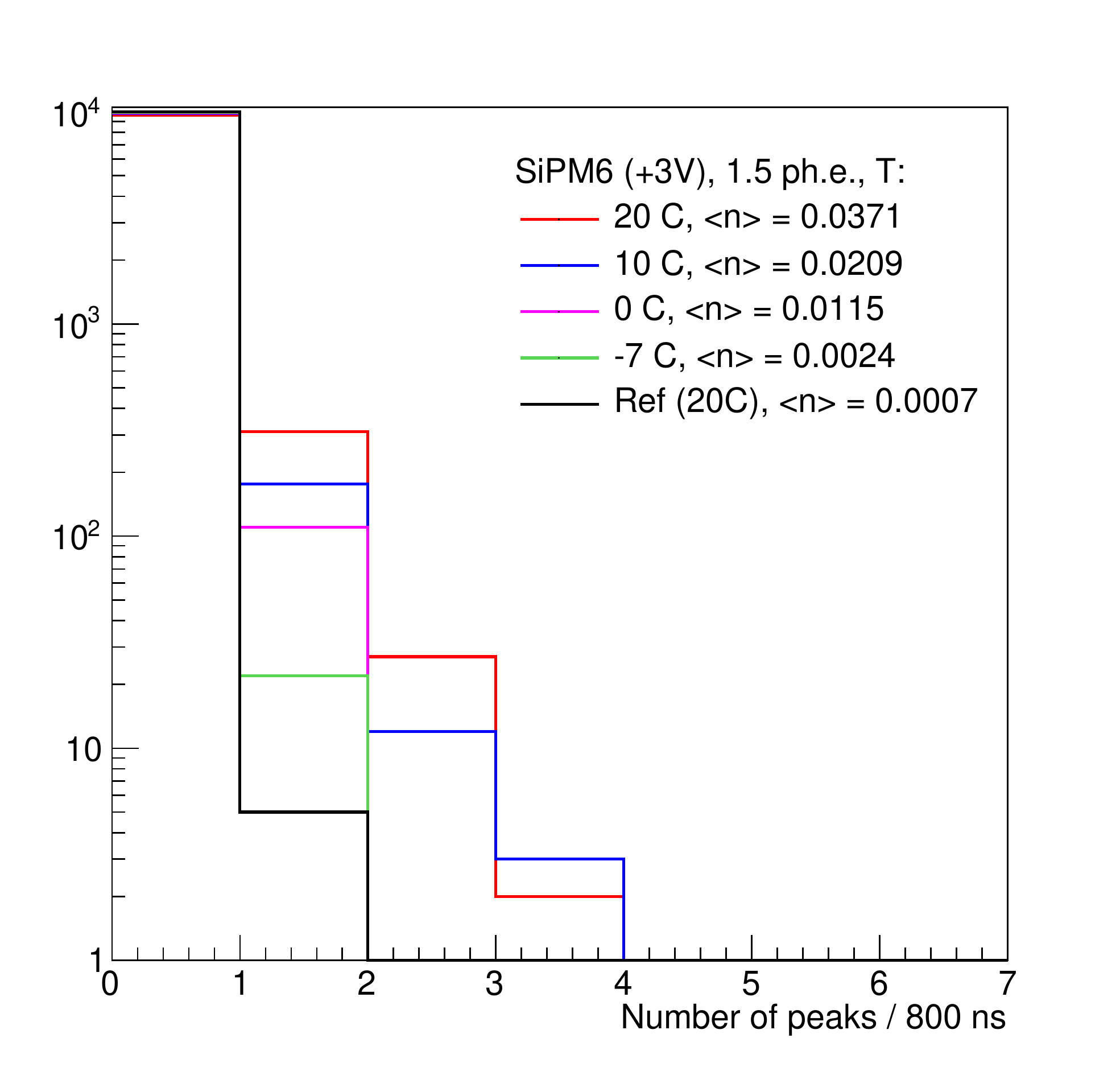}
  \includegraphics[width=0.52\textwidth]{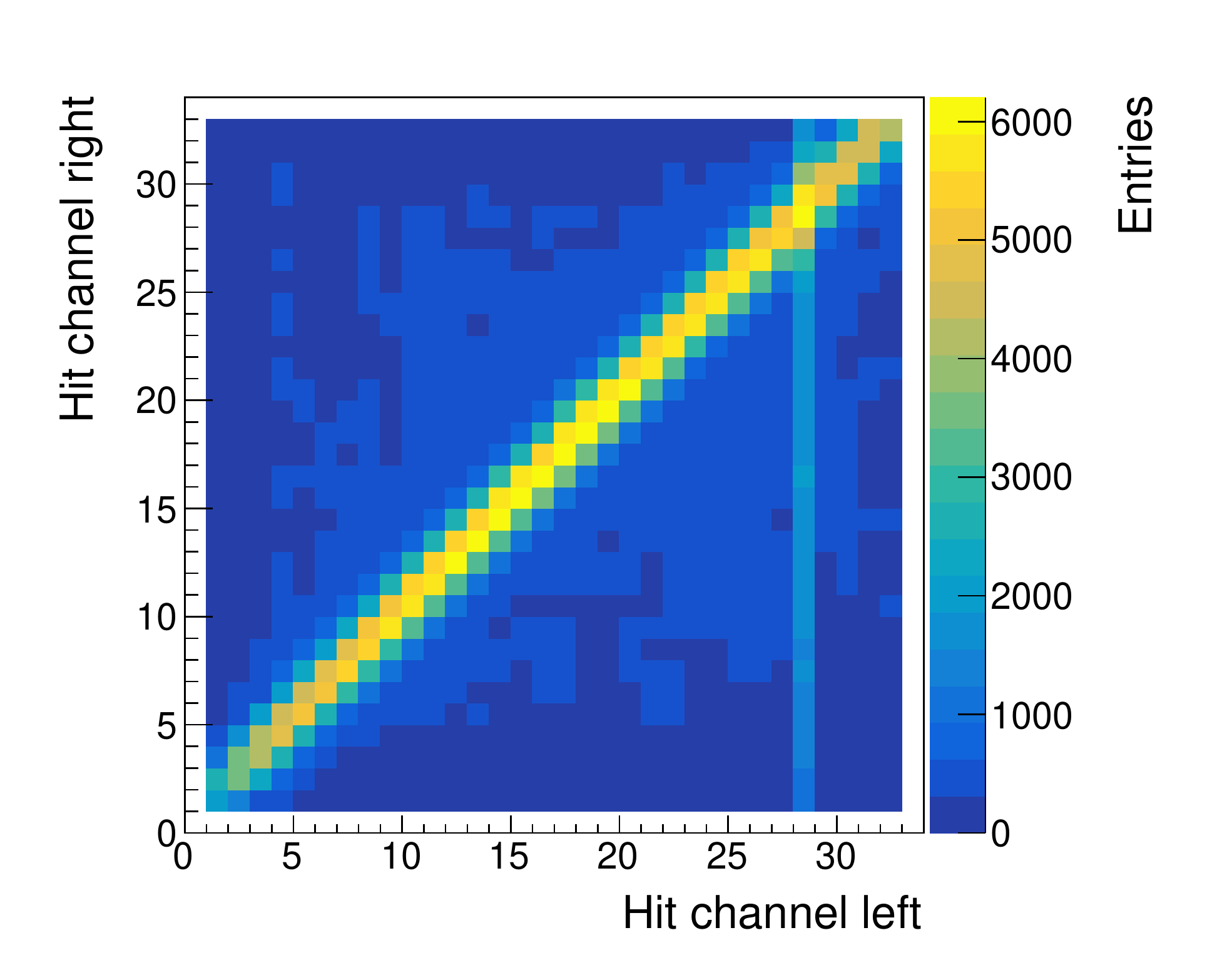}
  \caption{left) DCR for a selected channel of the same irradiated SiPM array at different temperatures operated at $V_{ob} \approx 3~{\rm V}$ for a 1.5 ph.e. threshold.
Also shown in the DCR for a non-irradiated SiPM (in black).  
right) Correlation between the two ends of the SciFi detector using the same SiPM arrays cooled to $-6^\circ~{\rm C}$.
The threshold is set at 1.5~ph.e..}
  \label{fig:correlos_irrad}
\end{figure}

For a heavily irradiated sensor, corresponding to the dose of $(1.5 \pm 0.3) \times 10^{12}~e^+/{\rm cm}^2$
(i.e. the dose expected in \mude in one year of running during the Phase~I at $10^8~{\rm muon~decays} / s$),
the dark current grows from 10~nA to $10~\mu{\rm A}$ per SiPM channel at $V_{ob} \approx 3~{\rm V}$
(from $1~\mu{\rm A}$ to 1~mA for the full sensor) as shown in Figure~\ref{fig:irrad}.
Figure~\ref{fig:correlos_irrad} (left) shows the DCR for one selected channel of the same irradiated SiPM array for a threshold above 1.5~ph.e.
at different temperatures.
The SiPM array has been cooled at different temperatures using Peltier coolers down to $-7^\circ$~C.
The rates have been measured by acquiring events with a random trigger over a 800~ns window
and by counting the number of peaks above a threshold of 1.5 ph.e..
The DCR per channel for the irradiated SiPM array at $+20^\circ~{\rm C}$ is $\sim 46~{\rm kHz}$,\footnote{To obtain the DCR,
we multiply the mean number of peaks $<n>$ in Figure~\ref{fig:correlos_irrad} (left) by $1.25 \times 10^6~{\rm s}$, i.e. we normalize $< n >$ to 1~s.}
while at $-7^\circ~{\rm C}$ the DCR drops to 3~kHz.
In \mude the SiPM arrays will be operated below $-10^\circ~{\rm C}$, which will further decrease the DCR.
Increasing the thresholds will also reduce the DCR, but at a loss of the detection efficiency.
Figure~\ref{fig:correlos_irrad} (right) shows the correlation between the two ends of the SciFi detector read out with the same irradiated SiPM array
cooled to $-6^\circ~{\rm C}$ 
with a 1.5~ph.e. threshold (compare to Figure~\ref{fig:correl} right).
From the correlations one can also deduce the average cluster width, which increases by $\sim 0.5$ channels due to the increased DCR,
which also introduces some minor background in the correlation plot.
The radiation effects will be discussed in detail in a separate work~\cite{rad}
(see also~\cite{Gerritzen}).

\subsection{Fiber Aging}

First SciFi ribbons have been produced back in 2017 and measured again after 5 years.
No change in the light yield has been observed within our sensitivity of the order of 1~photo-electron.
As an example, the vertical scan in Figure~\ref{fig:clVscan} is based on 2017 data,
while in Table~\ref{tab:MPV} measurements based on 2021 data are reported
using the same SciFi ribbons, but different SiPM arrays.

\clearpage
\section{Outlook and Conclusions}

In conclusion, we have developed a very thin scintillating fiber detector ($< 0.2\%$ of a radiation length $X_0$)
for the \mude experiment made of 3 layers of $250~\mu{\rm m}$ diameter round 
scintillating fibers and SiPM readout with a time resolution of around 250~ps, detection efficiency in excess of 96\%,
and spatial resolution of $100~\mu{\rm m}$.
In combination with the \mude HV-MAPS Si-pixel tracking detectors reaching a vertex resolution of better than $300~\mu{\rm m}$ these detectors
will allow for a full 4-dimensional reconstruction (in space and time) of the three body muon \mueee decay at a precision
sufficient to reach a sensitivity at the level of $10^{-15}$ (at 95\% CL) after a couple of years of data taking with a muon beam intensity of $10^8$ muons/s.

This technology can be used in applications requiring a very good timing, good spatial resolution, and high detection efficiency
with a very low material budget capable of operating at high rates.
To obtain very good timing resolution, the chosen detector has to feature fast rise time (i.e. fast primary signal rise time) and
low Signal-to-Noise ratio (i.e. a large primary signal amplitude and high light yield).
The shorter decay time of the NOL-11 scintillating fibers do not show a real advantage over the more conventional SCSF-78 fibers.
Only a small improvement of the timing performance of the SciFi detector is observed.
One of the limiting factors comes from the light yield of the SciFi ribbons.
The stringent \mude material budget requirements impose the use of SciFi ribbons consisting of only three staggered fiber layers.
This limits the light yield of the detector,
which in turn has an impact on the timing and the detection efficiency.
Otherwise one could use SciFi ribbons made of many more fiber layers, which would significantly increase the light yield
and improve the performance of the SciFi detector.
The light yield can be, in principle, increased by changing the doping profile of the scintillating fibers
by increasing the concentration of the {\it spectral shifter}.
That will unavoidably lead to an increased light attenuation.
However, given the short size of the SciFi ribbons the impact might not be so harmful as for longer SciFi detectors of $\mathcal{O}(1~{\rm m})$.
We plan to explore this possibility for future upgrades of the SciFi detector
by increasing the concentration of the {\it spectral shifter} by a factor of two or three.

In principle, a timing resolution much better than a few 100~ps could also be reached by a 
variety of other detector types, such as crystal scintillators (e.g. LYSO crystals), which feature intrinsic signal
rise times of few 10~ps and much larger primary signal amplitudes (some 30k photons/MeV).
However, the real challenge to be met in \mude is the requirement that the timing device has to detect electrons
of very low momenta of order 10 -- 50~MeV without affecting their momentum through multiple scattering in order not to deteriorate
the high tracking and vertexing precision.
Not only does the \mude timing detector have to be extremely thin, basically a zero-mass detector, which means also small primary signal amplitudes, 
but it also has only very limited physical space at its disposal, at the level of just a few centimeters.
Crystal scintillators that feature a few 10~ps resolution however are far too thick and bulky to be used in \mude.

Therefore, our chosen solution with thin scintillating fibers and few layers of staggered fibers has been shown to be a viable solution that
satisfies the extremely tight and challenging \mude requirements. 
We demonstrated that by operating this type of SciFi detector at a very low threshold of 0.5-1.5 photo-electrons
keeping the limited material budget, the system can sustain a high rate of the order several of several 100~kHz per SiPM channel,
reach around 250~ps time resolution at an efficiency above 96\% (for {\it left} / {\it right} coincidence signals).

However, the intrinsic limit of this scintillating fiber detector comes from the sustainable rates,
which will be exceeded in Phase~II of the Mu3e experiment
with muon rates above $10^9$ muons/s.
We anticipate that other, mostly Silicon-based technologies,
which are capable of both high tracking precision and very good timing resolutions at these very high rates,
will become available in the future.  
Such new detectors, as e.g. very thin HV-MAPS based Si-pixel detectors with $\sim 100~{\rm ps}$ time resolution, are in development at various laboratories.

\vfill
\section*{Acknowledgments}
We would like to thank the technical staff at our institutes (F.~Cadoux, Y.~Favre, S.~Debieux)
for their help in developing the tools, mechanics, and electronics for this work.
The Swiss institutes acknowledge the funding support from the Swiss National Science
Foundation grants no. 200021\_137738,
200021\_165568, 200021\_172519, 200021\_182031
and 20020\_172706.
The University of Geneva team gratefully acknowledges support from from the Ernest Boninchi Foundation
in Geneva.
This work was also supported by ETH Research Grant ETH-11 31-1
The authors thank the Paul Scherrer Institute as host laboratory for the Mu3e experiment.

\clearpage


\begin{thebibliography}{99}

\bibitem{Blondel}
\mude Collaboration, A.~Blondel {\it et al.}, \emph{Research Proposal for an Experiment to Search for the Decay $\mu \rightarrow eee$},
arXiv:1301.6113 [hep-ex].

\bibitem{TDR} \mude Collaboration, K.~Arndt, {\it Technical design of the phase I Mu3e experiment}, 
Nucl. Instr. Meth., {\bf A1014} (2021) 165679.

\bibitem{HV-MAPS} I. Peric, {\it A novel monolithic pixelated particle detector implemented in high-voltage CMOS technology},
Nucl. Instr. Meth. {\bf A582} (2007) 876; \\
Heiko's thesis or a more recent publication?

\bibitem{SiPM} LHCb Collab., {\it LHCb Tracker Upgrade TDR}, 
CERN/LHCC 2014-001 (LHCb TDR 15), 2014; \\
Hamamatsu Photonics,
\url{https://www.hamamatsu.com/eu/en/product/optical-sensors/mppc/index.html}.

\bibitem{Kuraray} Kuraray Co., \url{http://kuraraypsf.jp/index.html}.

\bibitem{MuTRiG} W.~Shen, {\it et al.}, 
{\it A silicon photomultiplier readout ASIC for time-of-flight application using a new time-of-recovery method},
IEEE Transactions on Nuclear Science, {\bf 65} (2018) 1196; \\
H.S.~Chen {\it et al.}, {\it MuTRiG: a mixed signal Silicon Photomultiplier readout ASIC with high timing resolution and gigabit data link},
J. Instrum. 12 (2017) C01043.

\bibitem{NOL11} O. Borshchev {\it et al.},
{\it Development of a New Class of Scintillating Fibres with Very Short Decay Time and High Light Yield},
J. Instrum. 12 (2017) P05013.

\bibitem{fibers} A. Bravar and Y. Demets, {\it Timing Properties of Blue-emitting Scintillating Fibers},
arXiv:2203.13322 [physics.ins-det], submitted to J. Instrum.

\bibitem{rad} {\it Radiation Damage in SiPMs by low energy electrons}, work in preparation.

\bibitem{Gerritzen} L. Gerritzen, {\it The Mu3e Scintillating Fibre Detector: Light Attenuation, Radiation Damage, and Data Acquisition},
PhD Thesis, ETH Zurich, 2022.

\bibitem{DRS} S. Ritt, R. Dinapoli, and U. Hartmann {\it Application of the DRS chip for fast waveform digitizing},
Nucl. Instrum. Meth. {\bf A~623} (2010) 486.

\bibitem{Damyanova} A.~Damyanova,
{\it Development of the Scintillating Fiber Detector for Timing Measurements in the Mu3e Experiment},
PhD Thesis, University of Geneva, 2019.

\bibitem{DamyanovaNIM} A.~Damyanova and A.~Bravar,
{\it Scintillating fiber detectors for precise time and position measurements read out with SiPMs},
Nucl. Instrum. Meth. {\bf A845} (2017) 475.

\bibitem{BravarNIM} A.~Bravar {\it et al.}, {\it The Mu3e scintillating fiber timing detector},
Nucl. Instrum. Meth. {\bf A958} (2020) 162564.

\bibitem{readout} {\it Readout of SiPM Arrays with the MuTRiG ASIC},
work in preparation.

\bibitem{Corrodi} S.~Corrodi, {\it A Timing Detector based on Scintillating Fibres for the Mu3e Experiment},
PhD Thesis, ETH Zurich, 2018.

\bibitem{PDG} Particle Data Group, P. A. Zyla {\it et al.}, {\it Review of Particle Physics}, Prog. Theor. Exp. Phys. 2020, 083C01 (2020).

\bibitem{James} F.~James, {Statistical Methods in Experimental Physics}, 2nd Ed., World Scientific Publishing, 2006.

\end{thebibliography}
\end{document}